\documentclass[11pt, a4paper]{article}
\usepackage[left=0.75in, top=0.75in, right=0.52in, bottom=1.44in]{geometry}
\usepackage{pdfpages}
\usepackage[version=4]{mhchem}
\usepackage{mathtools}
\usepackage{subfigure}
\usepackage{siunitx}
\usepackage{epigraph}
\usepackage{morefloats}
\usepackage{verbatim}
\usepackage{physics}
\usepackage{feynmp-auto}
\usepackage[section]{placeins}
\usepackage[normalem]{ulem}
\usepackage[affil-it]{authblk}
\usepackage[colorlinks,linkcolor=blue,anchorcolor=blue,citecolor=blue,urlcolor=blue]{hyperref}
\usepackage[T1]{fontenc}
\usepackage[utf8]{inputenc}
\usepackage{calc}
\usepackage{indentfirst}
\usepackage{fancyhdr}
\usepackage{graphicx,epstopdf}
\usepackage{lastpage}
\usepackage{ifthen}
\usepackage{lineno}
\usepackage{float}
\restylefloat{table}
\usepackage{amsmath}
\usepackage{amssymb} 
\usepackage{setspace}
\usepackage{enumitem}
\usepackage{mathpazo}
\usepackage{booktabs} 
\usepackage{titlesec}
\usepackage{etoolbox} 
\usepackage{tabto} 
\usepackage{xcolor, colortbl} 
\usepackage{soul} 

\usepackage{multirow}
\usepackage{microtype} 
\usepackage{tikz} 
\usepackage{totcount} 
\usepackage{changepage} 
\usepackage{paracol} 
\usepackage{attrib} 
\usepackage{upgreek} 
\usepackage{array} 
\usepackage{tabularx}
\usepackage{pbox} 
\usepackage{ragged2e} 

\usepackage[numbers, compress]{natbib}
\usepackage[bottom]{footmisc}

\begin{document}

\title{\textbf{Borexino results on neutrinos from the Sun and Earth}}
\author[1,2]{Sindhujha Kumaran}
\author[1,2]{Livia Ludhova\footnote{Correspondence: l.ludhova@fz-juelich.de}}
\author[1]{\"Omer Penek}
\author[1]{Giulio Settanta}
\affil[1]{Institute for Nuclear Physics IKP-2, Forschungszentrum J\"ulich, 52425 J\"ulich, Germany}
\affil[2]{Physics Institute IIIB, RWTH Aachen University, 52062 Aachen, Germany}
\date{}


\maketitle

\begin{abstract}
\noindent Borexino is a 280-ton liquid scintillator detector located  at the Laboratori Nazionali del Gran Sasso in Italy.  Since the start of its data taking in May 2007, it has provided several measurements of low-energy neutrinos from various sources. At the base of its success, lie unprecedented levels of radio-purity and extensive thermal stabilisation, both resulting from a years-long effort of the collaboration. Solar neutrinos, emitted in the hydrogen-to-helium fusion in the solar core,  are important for the understanding of our star, as well as neutrino properties.  Borexino is the only experiment that has performed a complete spectroscopy of the \emph{pp} chain solar neutrinos (with the exception of the \emph{hep} neutrinos contributing to the total flux at $10^{-5}$ level), through the detection of \emph{pp}, $^7$Be, \emph{pep}, and $^8$B solar neutrinos and has experimentally confirmed the existence of the CNO fusion cycle in the Sun.  Borexino has also detected geoneutrinos, antineutrinos from the decays of long-lived radioactive elements inside the Earth, that can be exploited as a new and unique tool to study our planet. This paper reviews the most recent  Borexino results on solar and geoneutrinos, from highlighting the key elements of the analyses up to the discussion and interpretation of the results for neutrino, solar, and geophysics.

\end{abstract}

\tableofcontents
\section{Introduction}
\label{sec:intro}

Neutrinos have a special position within the Standard Model (SM) of elementary particles. Having no electric charge and no colour, they interact with matter only via weak force. Thus, the probability of their interaction  is very low, and this fact has a two-fold consequence. On a positive side, neutrinos bring us unperturbed information from otherwise inaccessible locations\textemdash including the solar core and the Earth’s interior. On the other hand, this very same characteristic makes neutrino detection an experimental challenge. Large-volume detectors, constructed from specially developed materials with radioactivity levels of many orders of magnitude below the ambient values, must be placed in underground laboratories to shield them from cosmic radiation. Such measures are necessary in order to acquire high statistics of neutrino events with acceptable signal-to-background ratio. An enormous years-long effort is behind each neutrino experiment.

Borexino, a 280\,ton liquid scintillator detector located  at the Laboratori Nazionali del Gran Sasso (LNGS) in Italy, became an ``ideal scenario'' for future projects regarding the achieved radio-purity. Many years dedicated to the selection of construction materials and their treatment, involving surface cleaning or purification of the liquid scintillator, preceded the start of the data taking in May 2007. Section~\ref{sec:borex} of this paper is dedicated to the description of the basic features of the experiment and its main phases, defined according to the milestones achieved in terms of the radio-purity of the liquid scintillator and detector's thermal stability. The described arguments range from the detector setup (Section~\ref{subsec:detector-setup}), through the event reconstruction (Section~\ref{subsec:event-reco}), calibration and Monte Carlo simulation (Section~\ref{subsec:calib-MC}), background contamination (Section~\ref{subsec:bgr}), up to the interactions used to detect neutrinos and antineutrinos in Borexino (Section~\ref{subsec:detection}).

This paper reviews the most recent Borexino results on neutrinos from the Sun (Section~\ref{sec:solar}) and Earth (Section~\ref{sec:geo}). Each section introduces the field and importance of the measurement of the respective neutrinos, summarizes the details of the analysis and describes the latest results, including their interpretation and discussion.

Solar neutrino production and propagation from the solar core to our detector are discussed in Section~\ref{sec:Nu_prod_prop}. In particular, the fusion of Hydrogen to Helium proceeding via the principal \emph{pp} chain and the subdominant CNO cycle are covered in Section~\ref{sec:pp_CNO}. Section~\ref{sec:ssm_metal} instead is dedicated to the Standard Solar Model (SSM) that predicts solar neutrino fluxes as a function of the not-yet-well-understood abundances of elements heavier than Helium ("solar metallicity problem"). The neutrino-energy-dependent process that converts part of the solar neutrino flux from a pure electron flavour to a mixture of all flavours is discussed in Section~\ref{sec:nuosc_msw}. Common features of all Borexino solar neutrino analyses are synthesised in Section~\ref{sec:solar_nu_ana}. The comprehensive spectroscopy of the \emph{pp} chain solar neutrinos based on~\cite{PPchainNature} is contained in Section~\ref{sec:pp_ana}: the strategy of this particular analysis (Section~\ref{sec:ana_strat_pp}), the results on \emph{pp}, \emph{pp},$^7$Be, and $^8$B neutrinos (Section~\ref{sec:res_ppchain}) and their implications for solar and neutrino physics (Section~\ref{sec:implications}). These include: the confirmation of the origin of solar energy and of the Sun's thermal stability over the scale of 100 000 years, a mild preference towards the high-metallicity SSM with respect to the low-metallicity counterpart, and observation of the energy dependent electron neutrino survival probability for different neutrino species, excluding purely vacuum-dominated flavour conversion. The seasonal variation of the solar neutrino flux, expected due to the eccentricity of the Earth's orbit around the Sun, has been observed for  $^7$Be neutrinos~\cite{Be7mod}, as briefly discussed in Section~\ref{sec:be7_seasonal}. About 1\% of the solar energy is expected to be produced in the CNO cycle, a fact confirmed by the recent Borexino discovery~\cite{CNOpap}. Section~\ref{sec:CNO_ana} is dedicated to this observation of solar neutrinos from the CNO fusion with the analysis strategy summarised in Sections~\ref{sec:CNO_strategy} and~\ref{sec:CNO_LPoF}. The expected sensitivity (\cite{CNOsens} and Section~\ref{CNO_sens_corr}) is in agreement with the results (Section~\ref{sec:CNO_results}) obtained from both a spectral fit and a counting analysis. Precise measurement of solar neutrinos can lead to constraining of the parameter space of suggested non-standard neutrino interactions (Section~\ref{sec:bxnsi} based on~\cite{Bx-NSI20}) or placing strong limits on the effective neutrino magnetic moment (Section~\ref{sec:bxnmm} based on~\cite{Bx-NMM17}).

The part of this paper dedicated to geoneutrinos is introduced by providing essential information about the Earth's structure and heat budget (Section~\ref{subsec:earth}), Bulk Silicate Earth models (Section~\ref{subsec:bse}) that predict the composition of the primitive Earth and consequently also the abundances of $^{238}$U and $^{232}$Th, which produce geoneutrinos, a new tool to probe our planet (Section~\ref{subsec:geonu-intro}). The latest Borexino geoneutrino analysis~\cite{Agostini:2019dbs} is discussed in Section~\ref{sec:geo-analysis}, including the expected levels of geoneutrino, and other antineutrino signals, as well as non-antineutrino backgrounds, event selection cuts, spectral fit, and sources of systematic errors in Sections~\ref{subsec:exp-signal} to~\ref{subsec:syst-geo}, respectively. The results and their interpretation in terms of measured geoneutrino signal at LNGS, extracted geoneutrino signal from the Earth's mantle and the corresponding radiogenic heat, as well as imposed limits on the power of a hypothetical georeactor at different locations inside the Earth are discussed in Sections~\ref{subsec:results-geo-signal} to~\ref{subsec:georea}, respectively.

The final concluding remarks about the Borexino solar and geoneutrino analyses are given in Section~\ref{sec:conclusions}.

\section{The Borexino experiment}
\label{sec:borex}  

Borexino is an ultra-pure liquid scintillator detector located in Hall C of the LNGS in central Italy, at a depth of 3800 meters water equivalent. It is the radio-purest large-scale neutrino experiment ever built~\cite{BxDetector}. The laboratory has been designed to use the Gran Sasso mountain as a shielding against the cosmic muon radiation, which is suppressed at LNGS by a factor of $\sim$$10^6$. Thus, the laboratory represents an ideal place to probe low-energy neutrino physics with a high signal-to-noise ratio~\cite{LongPaperPhaseI}.

The Borexino data-taking started with the so-called \emph{Phase-I}, which extended from May 16\textsuperscript{th} 2007 until May 16\textsuperscript{th} 2010. During and right after the end of Phase-I, several detector calibration campaigns with radioactive sources were performed, in the period between November 2008 and July 2010. The calibrations were followed by a dedicated scintillator purification campaign between May 2010 and October 2011 with the goal of further reducing several contaminants, in order to improve the sensitivity to solar neutrinos. This extensive campaign with 6 cycles of closed-loop water extraction led to a significant reduction of the backgrounds, namely, $^{85}$Kr, $^{210}$Bi, $^{238}$U, and $^{232}$Th. The $^{238}$U and $^{232}$Th contamination reached the levels of $< 9.5\times10^{-20}$g/g (95\% C.L.) and $< 5.7\times10^{-19}$g/g (95\% C.L.), respectively~\cite{LongPaperPhaseI,NusolAfterNature}. Furthermore, the $^{210}$Bi and $^{85}$Kr contents were reduced by a factor of about 2.3 and 4.6, respectively. The so-called \emph{Phase-II} extended from December 11\textsuperscript{th} 2011 until May 22\textsuperscript{nd} 2016. During Phase-II, dynamical mixing of the scintillator was observed, resulting from convective currents due to temperature gradients present in the detector, caused by both human activities in the laboratory and seasonal temperature variations. The radioactive component heavily affected by these convective currents was $^{210}$Po, which was brought from the vessel holding the scintillator into the central parts of the detector. As a consequence, the Borexino collaboration decided to thermally stabilise the detector through a thermal insulation campaign. From May to December 2015, 900\,m$^2$ of mineral wool was installed on the surface of the detector. In order to track the temperature changes, the detector was equipped with 66 probes of an active temperature control system, surrounding the whole apparatus. This lead to the stabilisation of the temperature profile from the bottom to the top of the detector, ranging from 7.5 to 15.8$^\circ$C with a gradient of $\Delta T/\Delta z\approx0.5^\circ$C per meter. Thermal stability of the detector is the key element for the observation of CNO neutrino~\cite{CNOpap} and the main characteristic of the Borexino \emph{Phase-III}, that started on July 17\textsuperscript{th} 2016 and is still ongoing.

This Section, dedicated to the description of the Borexino experiment, is divided into five parts. Section~\ref{subsec:detector-setup} describes the detector setup. Section~\ref{subsec:event-reco} discusses the event reconstruction algorithms in Borexino, focusing on the position and energy reconstruction as well as particle identification techniques. Section~\ref{subsec:calib-MC} provides details about the calibration campaigns and the Monte Carlo simulation that was tuned on these. The levels of detector backgrounds are discussed in Section~\ref{subsec:bgr}.
The neutrino and antineutrino detection principles in Borexino are given in final Section~\ref{subsec:detection}.

\subsection{Detector setup}
\label{subsec:detector-setup}

\begin{figure}[t]
\centering
    \includegraphics[width=0.6\textwidth]{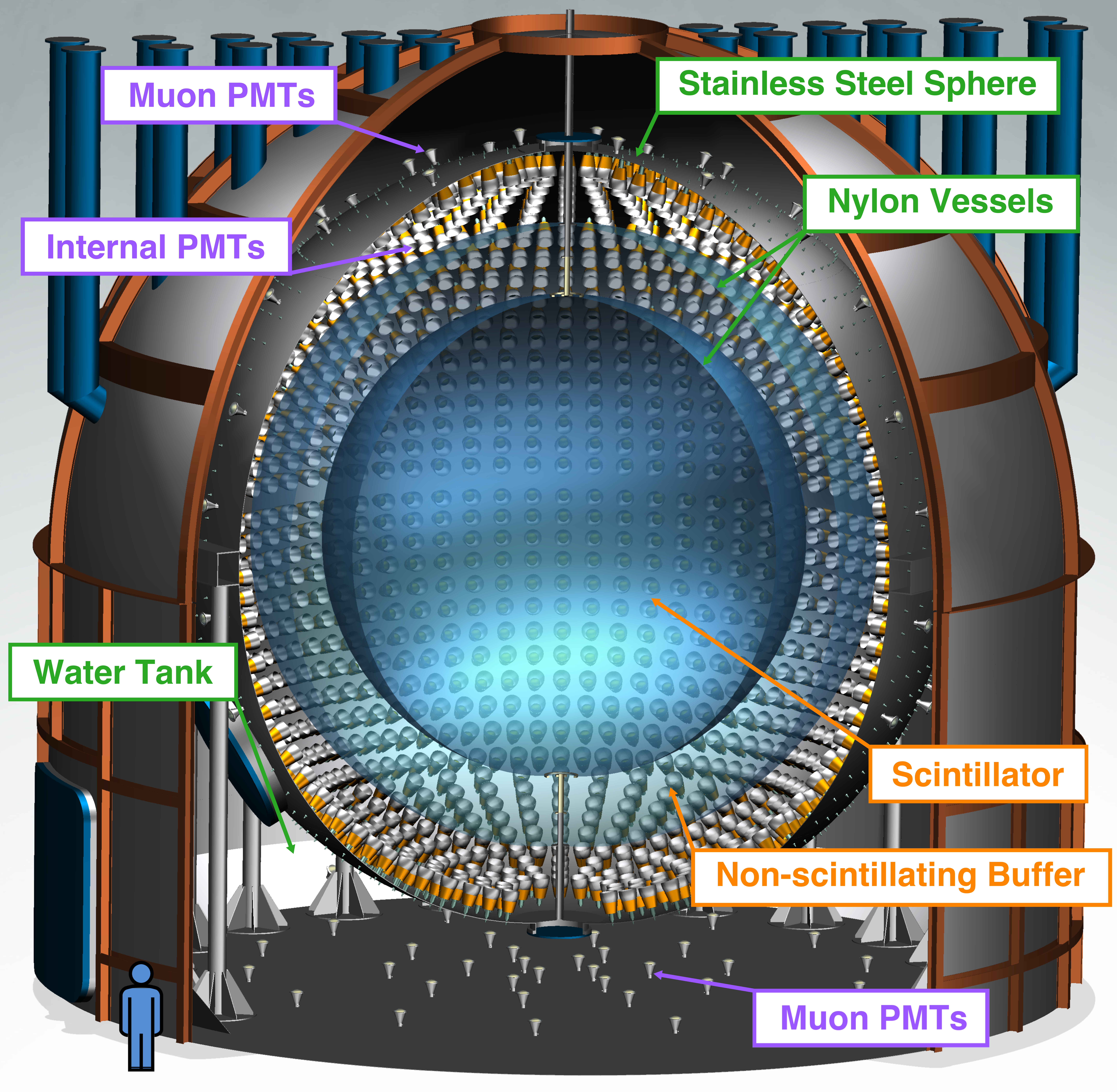}
    \caption{Schematic view of the Borexino detector. From inside to outside: the liquid scintillator contained in the nylon inner vessel, nylon outer vessel, stainless steel sphere, and the water tank for the Cherenkov muon veto. From~\cite{PPchainNature}. 
    \label{fig:BxDetector}} 
\end{figure}

The Borexino detector has an onion-like structure with the radio-purity of materials increasing towards the centre. A schematic of the detector is shown in Figure~\ref{fig:BxDetector}. The main neutrino target is 280\,ton of liquid scinitillator (LS) located in the core of the detector. Pseudocumene (PC, 1,2,4-trimethylbenzene) is used as a solvent with 1.5\,grams per litre of PPO (2,5-diphenyloxazole) as a solute. The scintillator density is $(0.878\pm0.004)$\,g\,cm$^{-3}$~\cite{Agostini:2019dbs}, where the error considers the changes due to the temperature variations during the whole data-taking period. The scintillator is contained in a thin spherical nylon inner vessel (IV) with a radius of 4.25\,m. The LS is surrounded by a non-scintillating buffer liquid (inner buffer) made up of PC doped with dimethylphthalate as a quencher. The shape of the IV changes with time, because of a leak of the LS from the IV to the buffer region which started around April 2008~\cite{LongPaperPhaseI,Agostini:2019dbs}. The leak was identified by reconstructing a large number of events outside the IV. The IV shape is reconstructed based on contaminants on its surface selected between 0.8-0.9\,MeV, dominated by external backgrounds such as $^{40}$K, $^{214}$Bi, and $^{208}$Tl (see Section~\ref{subsec:bgr}).
The density of the inner buffer is almost the same as for the scintillator material. This buffer region is held by a nylon outer vessel (OV) with a radius of 5.50\,m, followed by a second outer buffer region, which in turn is surrounded by a stainless steel sphere (SSS) with a radius of 6.85\,m, which holds 2218 8-inch photomultiplier tubes (PMTs), facing inwards. The 2.6\,m thick buffer region shields the inner volume against external radioactivity from the PMTs and the SSS. Moreover, the OV serves as a shielding against inward-diffusing Radon. The inner components contained inside the SSS are called the \emph{inner detector} (ID). The SSS is enclosed in a cylindrical tank filled with high-purity water, additionally endowed with 208 external PMTs, which define the \emph{outer detector} (OD). This water tank serves as an extra shielding against external gammas and neutrons, and as an active Cherenkov veto for residual cosmic muons passing through the detector. 

Charged particles interacting with molecules of the LS produce scintillation light, the amount of which is roughly proportional to the deposited energy. The exact amount of the emitted light depends on the particle type. In addition, the energy scale is to some extent intrinsically non-linear due to the ionisation quenching~\cite{Birks} and emission of the Cherenkov light. Particles causing high ionisation density experience high levels of quenching. Due to this, the visible energy of $\alpha$s' is quenched in the LS with a factor of about 10 with respect to electrons of the same energy.

The PMTs in Borexino convert the light to photoelectrons (p.e.), defined as the electrons removed from the photocathode of the PMT through incident photons.
In Borexino, the effective light yield is about 500\,p.e. per 1\,MeV of electron equivalent. The PMTs have random dark-noise coincidences with a rate of about 1 hit per 1\,$\mu$s in the whole detector. The average number of working PMT channels in the ID varies in time and is 1576 and 1238 for Phase-II and Phase-III, respectively~\cite{NusolAfterNature,CNOpap}.

The ID PMTs are coupled to an analog front-end (FE board, FEB) that amplifies the signal. Further processing differs for the two data acquisition (DAQ) systems~\cite{BxDetector}: the main DAQ and a semi-independent flash analog-to-digital converter (FADC) sub-system commissioned in November 2009. The two systems run independently and have different triggers. The main DAQ treats every PMT information separately, has a threshold of about 50\,keV, and a read-out window of 16\,$\mu$s. The FADC system instead processes an integrated signal of 24 FEB outputs, has an energy threshold of 1\,MeV, and 1.28\,$\mu$s read out window.

In the main DAQ, the FEB signal is fed to a digital circuit (Laben board, LB). A fast amplified timing signal can trigger a threshold discriminator, set to about 1\% of the p.e., defining a PMT hit. The FEB also integrates the PMT current and provides the second input to the digital LB. This second signal provides the charge of the hit, integrated in 80\,ns, that is proportional to the number of photons hitting the PMT during the integration time. The photons eventually falling on the same PMT in the time interval from 80 to 180\,ns are not detected.

\subsection{Event reconstruction}
\label{subsec:event-reco}

In this Section, we describe the event reconstruction algorithms applied to the triggers of the main DAQ system. Firstly, the \emph{clustering algorithm} identifies an accumulation of hits in the 16\,$\mu$s DAQ gate that correspond to a single physical event. Typically, there is just one cluster present, but multi-cluster triggers do exist. Higher level event reconstruction algorithms such as position, energy, and particle identification are then applied on each identified cluster. The main variables used in the solar and geoneutrino analyses presented in this paper are obtained from the data of the main DAQ system. The FADC system is optimised for multi-MeV events~\cite{GRBs,ImprovedB8} and was also successfully used to improve the muon tagging efficiency and to identify noise events~\cite{Agostini:2019dbs}.

\paragraph{\emph{Position reconstruction}}

The position reconstruction is based on the time-of-flight (TOF) technique, that subtracts from each measured hit time $t_i$ a position-dependent time-of-flight $T_{i}^{\mathrm{TOF}}$ from the point $\vec{r}_0$ of particle interaction at time $t_0$ to the PMT at position $\vec{r}_{i}$, where the $i{^\mathrm{th}}$ hit was detected:
\begin{equation}
    T_{i}^{\mathrm{TOF}}=(t_{i}-t_{0}) = \frac{n_{LS}}{c_{0}}\abs{\vec{r}_{i}-\vec{r}_{0}}.
\end{equation}
Here, $n_{LS}=1.68$ is the effective refraction index of the LS determined using calibration data~\cite{BxCalibPaper} and $c_{0}$ the speed of light in vacuum. The algorithm determines the most likely vertex ($t_0$, $\vec{r_{0}}$) of the interaction, using the arrival times $t_{i}$ of the detected hits on each PMT and the position vectors $\vec{r_{i}}$ of the corresponding PMTs. The likelihood maximization uses the probability density functions (PDFs) of the hit detection as a function of the time elapsed from the emission of scintillation light due to the interaction of an electron. The shape of the PDFs changes according to the charge of each hit~\cite{LongPaperPhaseI}. The position resolution is about 10\,cm at 1\,MeV at the centre of the detector~\cite{LongPaperPhaseI}. For other positions with larger radii, the resolution decreases on average by a few centimetres. 
 
\paragraph{\emph{Energy reconstruction}}

The visible energy in Borexino is different for different particle types. The detected light is proportional to the deposited energy, up to the leading order. There are intrinsic non-linearities of the energy scale due to the particle dependent ionisation quenching~\cite{Birks} and a small contribution from Cherenkov radiation~\cite{NusolAfterNature}.
The deposited energy is parameterized via the following energy estimators:
\begin{itemize}
	\item $N_{h}$: number of PMT hits, including multiple hits on a single PMT.
	\item $N_{p}$: number of triggered PMTs, ignoring multiple hits on the same PMT.
	\item $N_{p}^{dt_{1(2)}}$: a variant of $N_{p}$, restricting the considered time interval to 230(400)\,ns after the cluster start time. 
	\item $N_{pe}$: charge expressed in number of photoelectrons collected in all PMT hits.
\end{itemize}
The energy estimators are normalised to 2000 working PMTs, since not all PMTs are active during the data-taking. In addition, the energy estimators can also be geometrically normalised, considering the relative weight of each PMT to be proportional to the solid angle with respect to the reconstructed position of the event. This normalisation takes into account the fact that the amount of light seen by each PMT depends on the distance of this PMT to the event. An electron with kinetic energy of 1\,MeV produces approximately 500 photoelectrons in the Borexino detector. This results in $5\%/\sqrt{E~(\text{MeV})}$ energy resolution.

\paragraph{$\alpha / \beta$\emph{ discrimination}}
 
\begin{figure}[t]
\centering
\subfigure[]{\includegraphics[width = 0.45\textwidth]{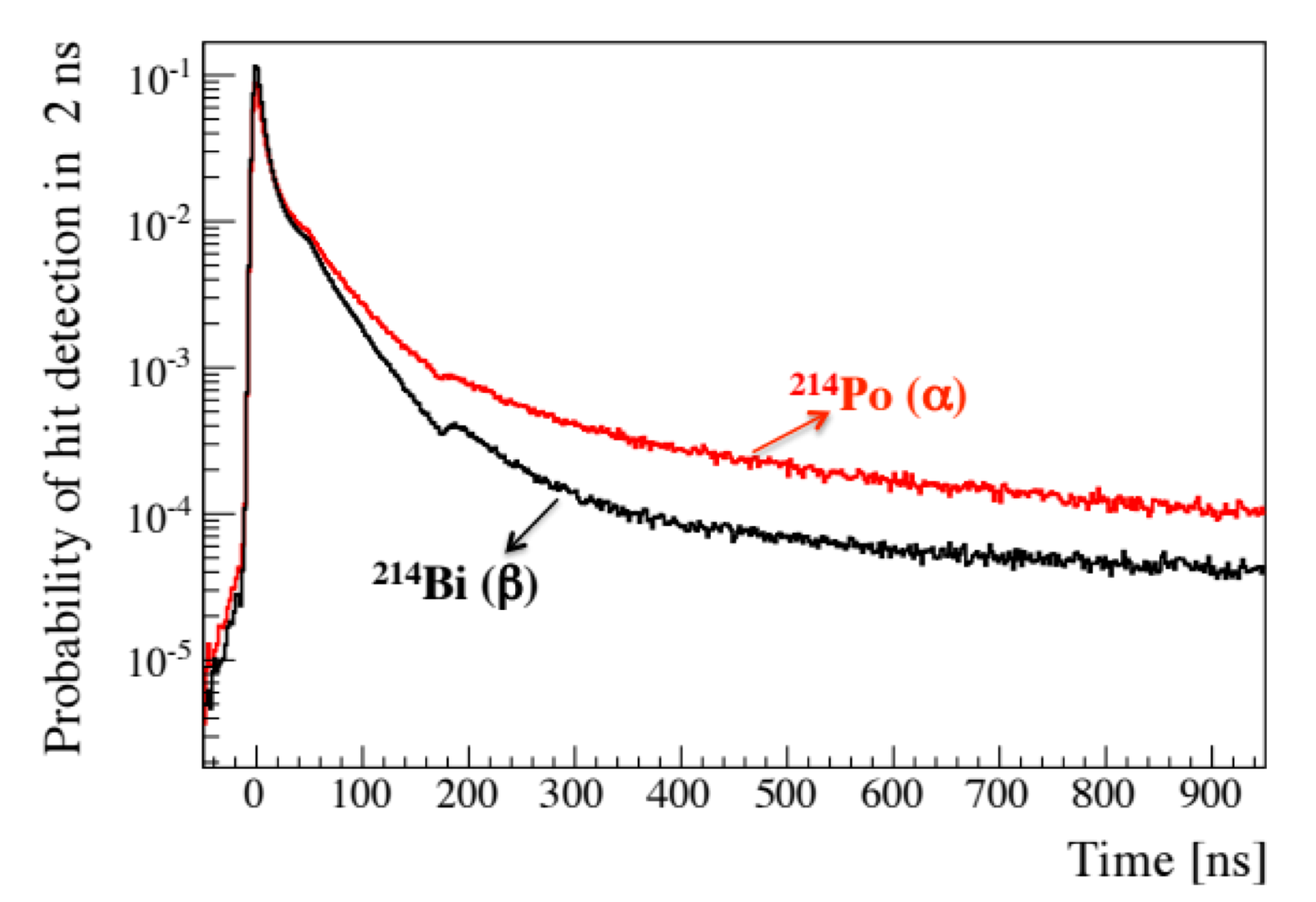}\label{fig:ab_hit_time_profile}}
\subfigure[]{\includegraphics[width = 0.45\textwidth]{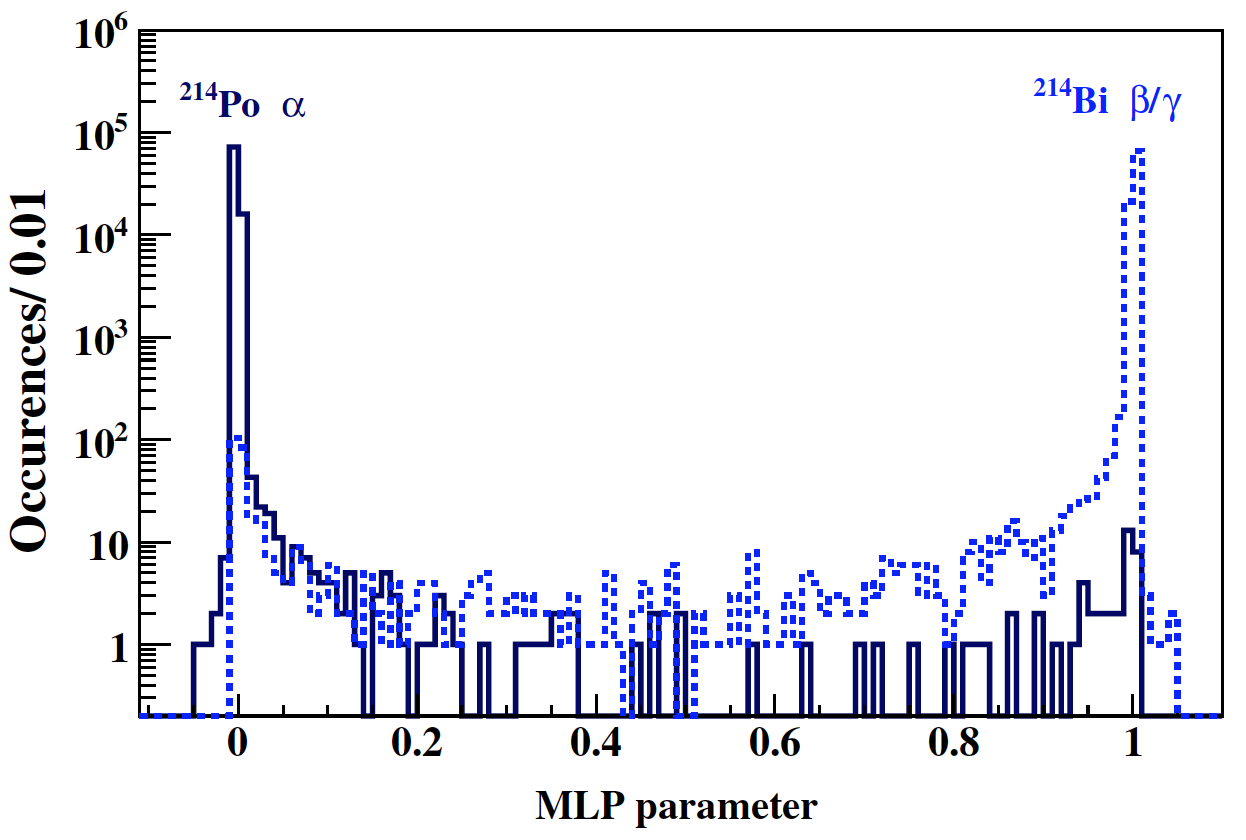}\label{fig:mlp_geo}}
\caption{$\alpha$ and $\beta$ separation in Borexino. (a) Hit time profiles for $^{214}$Po-$\alpha$s' (in red) and $^{214}$Bi-$\beta$s' (in black) after TOF subtraction. From~\cite{LongPaperPhaseI}. (b) The \emph{MLP} variable, shown for the same events, characterised by a high separation power between $\alpha$ and $\beta/\gamma$ like events. From~\cite{Agostini:2019dbs}.
\label{fig:AB_time_vs_MLP}} 
\end{figure}

The discrimination of $\alpha$ and $\beta$ particles is based on the different types of interactions of these particles. $\alpha$s' have a high ionisation quenching compared to $\beta$s', leading to different hit time profiles, as shown for $^{214}$Bi-$\beta$s' and $^{214}$Po-$\alpha$s' in Figure~\ref{fig:ab_hit_time_profile}. In Borexino, $\alpha$/$\beta$  discrimination is performed on an event-by-event basis. The algorithm is tuned based on $^{222}$Rn-correlated $^{214}$Bi($\beta^{-}$)-$^{214}$Po($\alpha$) fast coincidences introduced into the detector by a small air leak during the water extraction cycles performed between June 2010 and August 2011. The $\alpha/\beta$ discrimination variable currently used in Borexino is the so-called MultiLayer Perceptron (\emph{MLP}), while other variables have also been used in the past, as described in~\cite{LongPaperPhaseI}. The \emph{MLP} variable is based on neural networks and, in Borexino, it consists of one input layer, one hidden layer, and an output layer. It can distinguish between two classes of events by training on the acquired data. Thus, it can exploit not only the time profiles but also the pulse shape variables such as mean times, variances, skewness, and kurtosis associated with the given training data sample~\cite{Agostini:2019dbs}. The $\alpha$ particles in Borexino have an \emph{MLP} value around 0, while the $\beta/\gamma$ particles have a value around 1, as shown in Figure~\ref{fig:mlp_geo}. 

\paragraph{\emph{$\beta^{+} / \beta^{-}$  discrimination}}

\begin{figure}[t]
\centering
\subfigure[]{\includegraphics[width = 0.45\textwidth]{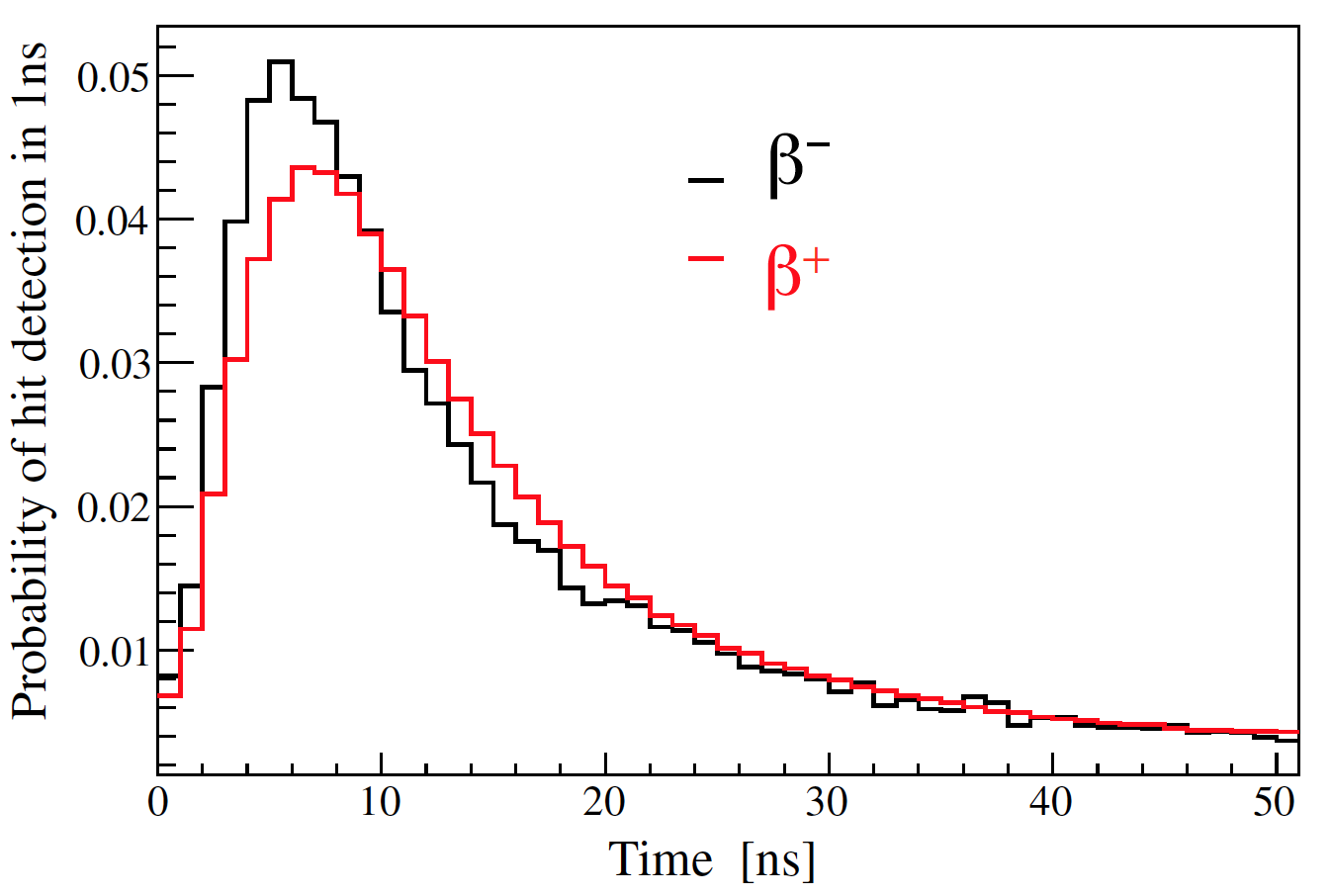}\label{fig:eminus_eplus_hit_time_only}}
\subfigure[]{\includegraphics[width = 0.478\textwidth]{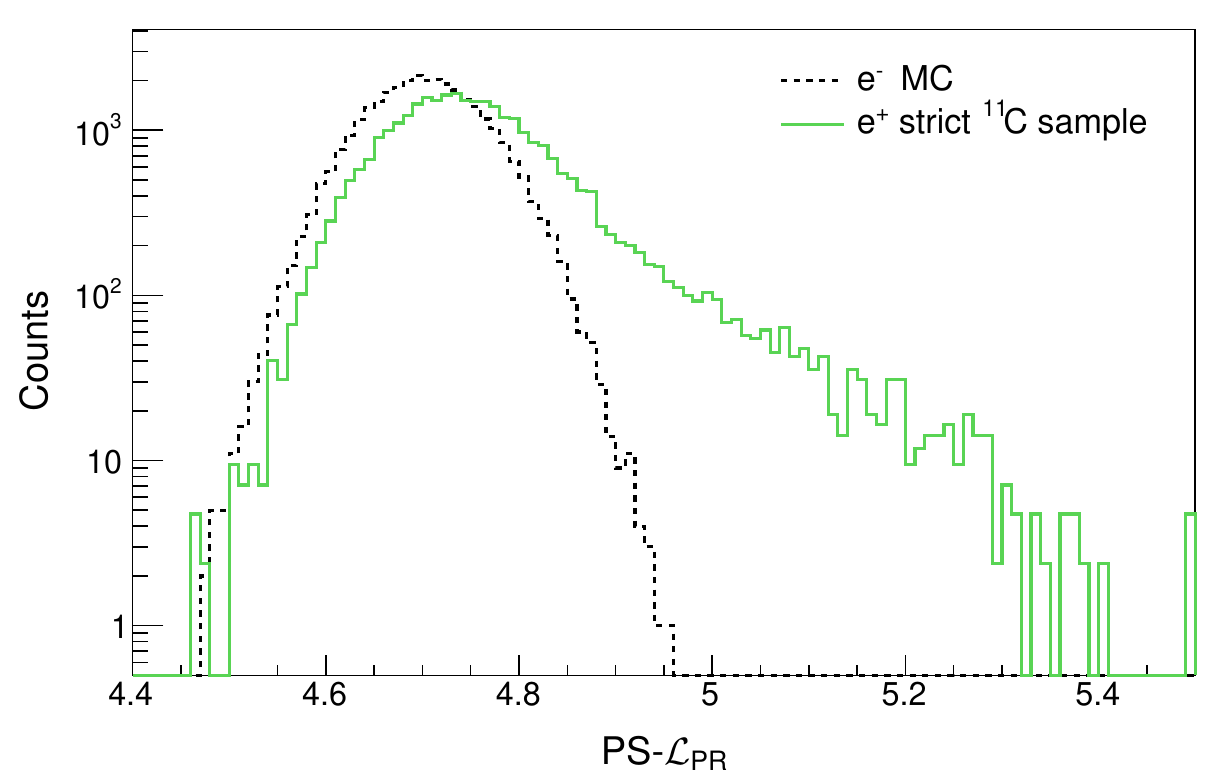}\label{fig:eminus_eplus_user_only}}
\caption{$\beta^-$ (electron) and $\beta^+$ (positron) separation in Borexino. $\beta^-$ distributions are based on Monte Carlo simulations. $\beta^+$ distributions represent a pure sample of $^{11}$C cosmogenic events selected from data. (a) Hit time profiles of $\beta^-$s' and $\beta^+$s' after TOF subtraction. The two distributions show a similar behavior. From~\cite{LongPaperPhaseI}. (a) PS-$\mathcal{L}_{\text{PR}}$ variable in the energy window from 400 to 650 $N_{h}$. From~\cite{NusolAfterNature}. \label{fig:user_eminus_eplus}} 
\end{figure}

The hit time profiles of electrons and positrons look very similar in the liquid scintillator, making their separation very challenging. Therefore, the discrimination is done on a statistical basis and not on an event-by-event basis. Positrons emitted in the LS build ortho-positronium in 50\% of the cases, as discussed in~\cite{Perasso_2014}. This formation leads to a delay of the $e^{+}e^{-}\rightarrow\gamma\gamma$ annihilation process, with a typical lifetime of $\sim$3\,ns. The lifetime of para-positronium is about 125\,ps in vacuum, making its contribution indistinguishable from the prompt light emission caused by the positron~\cite{LongPaperPhaseI,Perasso_2014}. The pulse shape discrimination of $\beta^{-}$s' and $\beta^{+}$s' in Borexino is based on the likelihood of the position reconstruction, normalised to the number of PMTs $N_{p}$. This parameter is called PS-$\mathcal{L}_{\text{PR}}$. The position reconstruction is based on the emission profiles for electrons, as discussed in the \emph{Position reconstruction} paragraph. While the spatial position reconstruction for $\beta^{-}$s' and $\beta^{+}$s' is precise within the position reconstruction resolution, the likelihood value is worse for positrons than for electrons. This causes the difference in the PS-$\mathcal{L}_{\text{PR}}$ variable, making pulse shape discrimination possible. Figure~\ref{fig:user_eminus_eplus} shows the $\beta^{-}$ and $\beta^{+}$ hit time distributions (Figure~\ref{fig:eminus_eplus_hit_time_only}) and the PS-$\mathcal{L}_{\text{PR}}$ variable distributions (Figure~\ref{fig:eminus_eplus_user_only}) in the region from 400 to 650 $N_{h}$, corresponding to approximately 1.0 to 1.8\,MeV. The PS-$\mathcal{L}_{\text{PR}}$ variable for $\beta^{-}$s' is taken from Monte Carlo simulations, while for $\beta^{+}$s' it is taken from a pure sample of the $^{11}$C($\beta^{+}$) cosmogenic background, selected via the \emph{three-fold coincidence} algorithm (see Sections~\ref{subsec:bgr} and~\ref{sec:solar_nu_ana}).

\subsection{Calibration and Monte Carlo simulation}
\label{subsec:calib-MC}

In order to understand the whole Borexino detector and validate the physics model adopted for the description of the light emission, propagation, and detection by PMTs, a dedicated Geant4-based Monte Carlo (MC) simulation code has been developed by the Borexino collaboration~\cite{BxMCpaper}. The MC code has been tuned based on calibration of the detector with radioactive sources and various laboratory measurements~\cite{BxCalibPaper}. The Borexino calibration campaigns were performed in November 2008, January 2009, June-July 2009, and July 2010 using different types of radioactive sources as extensively discussed in~\cite{BxCalibPaper} and~\cite{BxMCpaper}.

\paragraph{\emph{Calibrations}}

The goal of the calibration campaign was to (1) determine the accuracy and resolution of the position reconstruction, (2) measure the absolute energy scale and resolution, (3) estimate the energy response and non-uniformity depending on the energy and position of an event, (4) tune the MC simulation framework. The different calibration sources were used to study the responses of $\alpha$s', $\beta$s', $\gamma$s', and neutrons, covering an energy range of 0.122-10\,MeV. These sources were deployed in 295 locations. The energy scale was determined through the usage of different monochromatic $\gamma$ sources ranging from 0.122 to 2.615\,MeV located at the centre of the detector and at few positions along the vertical detector axis. To study the uniformity of the trigger efficiency, some of these $\gamma$ sources were also deployed at different positions and at larger distances from the centre. $^{222}$Rn was used as a $\alpha/\beta$-source, while $^{241}$Am-$^9$Be was used as a neutron-source (see Sections~\ref{sec:solar_nu_ana} and~\ref{sec:geo-analysis}). The calibration sources used to calibrate the event position reconstruction were $^{222}$Rn and $^{241}$Am-$^9$Be, which have been placed in 182 and 29 positions in the scintillator, respectively~\cite{BxCalibPaper}. The external calibration was done using $^{228}$Th whose daughter nuclide $^{208}$Tl is a strong gamma source. This helped in studying the exact determination of the IV shape, the external $\gamma$ background near the IV, and the asymmetries in the energy response near the IV. The source positions have been measured with a charge coupled device (CCD) camera system~\footnote{Kodak DC290 2.4 megapixel consumer grade digital cameras, each equipped with a Nikon FC-E8 fisheye lens.}~\cite{BxCalibPaper}. In total, 7 CCD cameras were mounted on the SSS. The differences (nominal-to-reconstructed) in the $(x,y,z)$ coordinates have been determined with a precision better than 2\,cm~\cite{LongPaperPhaseI}. In addition to these dedicated calibration campaigns, there are constant offline checks of the detector's stability and regular online PMTs' charge and timing calibration~\cite{LongPaperPhaseI}, in order to monitor the quality of the acquired data.

\paragraph{\emph{Monte Carlo Simulations}}

The Borexino Monte-Carlo (MC) simulation~\cite{BxMCpaper} is able to simulate all the processes after the interaction of a particle in the detector, including the knowledge of detector effects. It is based on the GEANT4 software (v4.10.5). The code is able to generate the physical event and track the light production, propagation and detection. Furthermore, the electronics and the trigger responses are fully simulated. The framework monitors the detector evolution in time, based on the electronics status, trigger settings, and active PMTs. The outputs from the MC simulation and the real data are treated in exactly the same way. The energy response and position for the sources placed at the detector centre have been reproduced with a precision better than 0.8\% and 1\%, respectively~\cite{BxMCpaper}. The MC simulations are needed for every Borexino analysis and are especially relevant for the evaluation of systematic uncertainties (see Sections~\ref{sec:solar} and~\ref{sec:geo}.) The overall agreement of data and MC in the energy region below 3\,MeV is within the order of 1\% while above 3\,MeV it is of the order of 1.9\%~\cite{ImprovedB8}.

\subsection{Background levels}
\label{subsec:bgr}

In Borexino, the backgrounds can be classified into internal, external, and cosmogenic~\cite{LongPaperPhaseI,PPchainNature,Bellini:2013cosmo}. The \emph{internal background} isotopes, namely, $^{238}$U and $^{232}$Th chains, $^{14}$C, $^{85}$Kr, $^{210}$Pb, $^{210}$Bi, $^{210}$Po, and $^{208}$Tl are contaminants of the liquid scintillator itself. 
The \emph{external background} components originate from materials outside of the LS and are typically represented by $\gamma$s' that are able to reach the LS volume. The \emph{cosmogenic background} consists of muons and consecutive events created by muon spallation in the detector and the surrounding rock. The backgrounds are summarised in~Tables~\ref{tab:bkg_ext-int} and~\ref{tab:cosmo_bckg} and are discussed below. 

\paragraph{\emph{Internal background}}

\begin{itemize}
    \item[] $^{238}$U chain\,($\tau$ = 6.4$\times10^9$\,years): a primordial long-lived radioactive isotope with 99.3\% abundance in natural Uranium. The chain contains eight $\alpha$ and six $\beta$ decays. The chain contains the fast $^{214}$Bi($\beta^{-}$, \emph{Q}\,=\,3.272\,MeV)-$^{214}$Po($\alpha$, \emph{Q}\,=\,7.686\,MeV) decay sequence with $\tau$ = 238\,$\mu$s, allowing a coincidence tagging assuming secular equilibrium. This chain is highly suppressed by the water extraction campaign as a consequence of the LS purification, leading to an upper limit of $<9.5\times10^{-20}$\,g/g\,(95\% C.L.) on the whole chain.
    \item[]
    \item[] $^{232}$Th chain\,($\tau$ = 20.2$\times10^9$\,years): a primordial long-lived radioactive isotope with 100\% abundance in natural Thorium. The chain has six $\alpha$ and four $\beta$ decays. Assuming secular equilibrium, it is possible to determine its content through the fast $^{212}$Bi($\beta^{-}$, \emph{Q}\,=\,2.252\,MeV)-$^{212}$Po($\alpha$, \emph{Q}\,=\,8.955\,MeV) decay sequence with $\tau$ = 433\,ns. The content of this chain after the LS purification campaign reached an upper limit of $<5.7\times10^{-19}$\,g/g\,(95\% C.L.). From this chain, $^{208}$Tl\,(\emph{Q}\,=\,4.999\,MeV, $\tau$ = 4.4\,min), which simultaneously emits an electron and gamma rays, has high relevance for the $^8$B analysis (see Section~\ref{sec:solar_nu_ana}). Since there is also $^{208}$Tl background originating from the IV contamination, the component coming from the $^{232}$Th contamination of the LS is specifically called \emph{Bulk} $^{208}$Tl.
    
    \item[]
    \item[] $^{14}$C\,($\beta^-$ decay, \emph{Q}\,=\,0.156\,MeV, $\tau$ = 8270\,years): this isotope is a natural component of the organic liquid scintillator and is chemically identical to the stable isotope $^{12}$C.  Therefore, it cannot be removed through purification. It dominates the low-energies, relevant for the \emph{pp}-$\nu$ analysis, and dictates the trigger rate which is about 25\,Hz at 50\,keV threshold. Its rate is stable in time and determined as $(40\pm2)$\,Bq/100\,ton. In Borexino, the relative ratio of $^{14}$C to $^{12}$C is $\approx10^{-18}$\,g/g~\cite{C14paper}.
    \item[]
    \item[] $^{85}$Kr\,($\beta^-$ decay, \emph{Q}\,=\,0.687\,MeV, $\tau$ = 15.4\,years): this isotope occurs in the air due to nuclear explosions. Its decay rate can be determined by following the procedure described in~\cite{LongPaperPhaseI}, which exploits the $^{85}$Kr-$^{85m}$Rb fast delayed coincidence decay with a branching ratio of 0.43\%. It is a major background of $^7$Be solar neutrinos (Section~\ref{sec:solar_nu_ana}) and has been suppressed by a factor of about 4.6 after the LS purification campaign. 
    \item[]
    \item[] $^{210}$Pb\,($\beta^-$ decay, \emph{Q}\,=\,0.064\,MeV, $\tau$ = 32.2\,years) and $^{210}$Bi\,($\beta^-$ decay, \emph{Q}\,=\,1.162\,MeV, $\tau$ = 7.2\,days): $^{210}$Pb is an isotope which is contained in the LS. Its very low \emph{Q} value is below the Borexino analysis threshold. It has a long lifetime, so it can be considered stable in time during few-years long analysis periods. The $^{210}$Pb in the LS is assumed to be in secular equilibrium with $^{210}$Bi, its short-lived daughter nuclei. The content of $^{210}$Bi has been reduced by a factor of about 2.3 after the water extraction period. It is an important background for $^7$Be solar neutrinos (Section~\ref{sec:pp_ana}) and a major challenge for CNO solar neutrinos (Section~\ref{sec:CNO_ana}). While $^{210}$Pb is most probably present also on the surface of the IV, there is no evidence that it would be leaching to the inside of the scintillator, causing an additional source of $^{210}$Bi and its daughter, $^{210}$Po events. 
    \item[]
    \item[] $^{210}$Po\,($\alpha$ decay, \emph{Q}\,=\,5.304\,MeV, $\tau$ = 200\,days): the $^{210}$Po contamination follows a more complicated history in Borexino. The visible energy of $\alpha$s' is highly quenched in the LS and the peak of $\alpha$s' from $^{210}$Po decays occurs in the region around 0.4\,MeV of the electron equivalent energy scale.
    $^{210}$Po can be produced along the decay of $^{210}$Pb:
    \begin{equation}
     \textrm{$^{210}$Pb}\xrightarrow[32\,y]{\beta^-}\textrm{$^{210}$Bi} \xrightarrow[7.2\,d]{\beta^-}\textrm{$^{210}$Po} \xrightarrow[199.6\,d]{\alpha}\textrm{$^{206}$Pb}\text{ (stable)}.
     \label{eq:Pb210chain}
    \end{equation}
    Assuming an equilibrium state of the above chain, the $^{210}$Po and $^{210}$Bi rates are equal. We call this term, originating from the $^{210}$Pb/$^{210}$Bi contaminating the scintillator, as supported $^{210}$Po$^\mathrm{S}$~\cite{CNOpap,CNOsens}. 
    However, there are two additional $^{210}$Po components that are not linked to the local $^{210}$Bi and are thus breaking the secular equilibrium condition: (1) vessel $^{210}$Po$^\mathrm{V}$ originating from the IV and (2) unsupported $^{210}$Po$^\mathrm{U}$, which is a residual component introduced during some cycles of the water extraction phase of the LS purification campaign. The latter, $^{210}$Po$^\mathrm{U}$ component has decayed over time, reaching asymptotically a value of zero in Phase-III. The $^{210}$Po$^\mathrm{V}$ component detaches from the IV and moves into the scintillator, effectively driven by the slow convective currents, triggered by the seasonal variation of the temperatures. Thermal stabilization of the detector performed before the start of the Phase-III period, discussed in the beginning of this Section, has helped in reducing this convective component. The precise determination of the $^{210}$Po$^\mathrm{S}$ content is fundamental to obtain information about the $^{210}$Bi contamination of the LS, which is highly relevant for the CNO-$\nu$ analysis (see Section~\ref{sec:CNO_LPoF}). The $^{210}$Po background is also important for the geoneutrino analysis, since it can trigger ($\alpha, \textrm{n}$) interactions which can mimic geoneutrino signals (see Section~\ref{subsec:exp-nonanti-bckg}). In addition, mono-energetic  $^{210}$Po events, that can be selected on an event-by-event basis using the MLP variable, are an important "standard candle" to follow the stability of the detector response over time.
\end{itemize}

\begin{table}[t] 
\centering
\caption{Summary of the internal and external backgrounds in Borexino. The columns (from left to right) specify the isotope and its decay type, \emph{Q} value, lifetime, details of the analysis (data-taking phase, where the abbreviation p indicates that only a part of the corresponding phase was considered (see also Table~\ref{tab:SolAnal}), fiducial mas, and energy range) where the background was relevant, the estimated rate, and the corresponding reference. The MLP in the fourth column stands for $^{210}$Po events, which have been selected with the MLP variable (see Section~\ref{subsec:event-reco}). The rates for the internal backgrounds are given for zero threshold, while the rates of the external backgrounds are given for the energy range of the analysis. The rates of the external backgrounds do not scale linearly with the volume and were normalised to 100\,ton just for the sake of comparison.}
\label{tab:bkg_ext-int}
\scalebox{0.92}{
\begin{tabular}{lcccccc}
\toprule
Type  &	$Q$ & $\tau$ & Phase (FV, Energy) & Rate  & Ref  \\ 
     & [MeV] &   &   [ton,\,MeV] &  & \\ 
\midrule
Internal 	& &		&  &	   \\ 
\midrule
  &   &   &    & [Bq/100\,ton]  & \\ 
  $^{14}$C\,($\beta^-$) & 0.156 & 8270\,y & II (71.3,0.19-2.93) & $40 \pm 2$ & \cite{PPchainNature,NusolAfterNature}  \\
    &   &   &    & [cpd/100\,ton]  & \\ 
$^{85}$Kr\,($\beta^-$) &  0.687 & 15.4\,y& II (71.3,0.19-2.93) &  $6.8 \pm 1.8 $ & \cite{PPchainNature,NusolAfterNature} 	  \\
$^{210}$Bi\,($\beta^-$)& 1.162 & 7.2\,d & II(71.3,0.19-2.93) & $17.5 \pm 1.9$ & \cite{PPchainNature,NusolAfterNature}  	 \\
       &   &  & III(71.3,0.19-2.93) & $\leq(11.5 \pm 1.3)$&\cite{CNOpap}	 	 \\
$^{210}$Po\,($\alpha$) & 5.304 & 199.6\,d & II(71.3, 0.19-2.93) & $260.0\pm 3.0$ & \cite{PPchainNature,NusolAfterNature}  \\
       &   &  & I(p)+II+III(p)(245.8, MLP) & $1275.0\pm 8.0$ & \cite{Agostini:2019dbs}\\
$^{208}$Tl\,($\beta^-\gamma$) (bulk) & 4.999 & 4.4\,min & I(p)+II+III(p)(227.8,3.2-5.7) &  $0.018 \pm 0.004 $ & \cite{PPchainNature, BxB82010}  	 \\
\midrule
External	& &	&	&  &	 &  \\ 
\midrule
$^{40}$K\,($\gamma$) & 1.461 & 1.8$\times10^9$\,y&II(71.3,0.19-2.93)& $1.0 \pm 0.6 $& \cite{PPchainNature,NusolAfterNature}		\\
$^{214}$Bi\,($\gamma$)	& 2.448 & 28.7 min& II(71.3,0.19-2.93) & $1.9 \pm 0.3 $ & \cite{PPchainNature,NusolAfterNature}		\\
$^{208}$Tl\,($\gamma$)& 2.615 & 4.4\,min& II(71.3,0.19-2.93) & $3.3 \pm 0.1 $ & \cite{PPchainNature,NusolAfterNature}	\\
$^{208}$Tl\,($\beta^-\gamma$) (emanated) & 4.999 & 4.4\,min & I(p)+II+III(p)(227.8,3.2-5.7) &  $0.206 \pm 0.028 $ & \cite{PPchainNature, ImprovedB8}  	 \\
$^{208}$Tl\,($\beta^-\gamma$) (surface) & 4.999 & 4.4\,min & I(p)+II+III(p)(227.8,3.2-5.7) &  $0.478 \pm 0.020 $ & \cite{PPchainNature, ImprovedB8}  	 \\

$[(\alpha,\,\textrm{n}),X]\rightarrow \gamma$ & - & - & I(p)+II+III(p)(227.8,3.2-5.7) & $0.098 \pm 0.034$&\cite{PPchainNature,BxB82010}		\\
$[(\alpha,\,\textrm{n}),X]\rightarrow \gamma$ & - & - & I(p)+II+III(p)(266.0,5.7-16.0) & $0.090 \pm 0.008$ & \cite{PPchainNature,BxB82010}		\\
\midrule
\bottomrule
\end{tabular}
}
\end{table}

\begin{table}[t] 
\centering
\caption{Summary of the relevant muon-induced cosmogenic background in Borexino. The columns (from left to right) specify the isotope and its decay type, \emph{Q} value, lifetime, the estimated rate expressed in counts per day per 100\,ton, and the corresponding reference. The rates with statistical and systematic errors summed in quadrature and 3$\sigma$ upper limits are extracted from the Borexino data through a dedicated analysis, in a scintillator mass of 99.6\,ton~\cite{Bellini:2013cosmo}. The exception is the $^{11}$C rate that is taken from the Phase-II results on solar neutrinos, measured in 71.3\,ton~\cite{PPchainNature}. The n-capture time in Borexino was measured using the $^{241}$Am-$^{9}$Be neutron calibration source~\cite{Bellini:2011yd}.}
\label{tab:cosmo_bckg}
\begin{tabular}{lcccc}
\toprule
Type	& $Q$    &$\tau$ &  Rate            & Ref \\ 
        & [MeV] &              & [cpd/100\,ton]  & \\
\midrule
Cosmogenic 		&	&  	 	  & &\\ 
\midrule
(n,p or $^{12}$C)$\rightarrow \gamma$ &  2.23 or 4.95 & (254.5$\pm$1.8)\,$\mu$s~\cite{Bellini:2011yd}	&  $90.2 \pm 3.1$ & \cite{Bellini:2013cosmo}\\
$^{12}$N\,($\beta^+$)  & 17.3 & 15.9\,ms&  $\textless$0.03 (3$\sigma$) & \cite{Bellini:2013cosmo} \\
$^{12}$B\,($\beta^-$)  & 13.4 & 29.1\,ms &   $1.62 \pm 0.09$	&	 \cite{Bellini:2013cosmo}\\
$^{8}$He\,($\beta^{-}n$)  & 10.7 & 171.7\,ms &  $\textless$0.042 (3$\sigma$) & \cite{Bellini:2013cosmo} \\
$^{9}$C\,($\beta^+$)  	  & 16.5 & 182.5\,ms &  $\textless$0.47 (3$\sigma$) & \cite{Bellini:2013cosmo}\\
$^{9}$Li\,($\beta^{-}n$)  	 	 & 13.6 & 257.2 ms	& $0.083 \pm 0.009$	& \cite{Bellini:2013cosmo}\\
$^{8}$B\,($\beta^{+}$)   & 18.0 & 1.11 s	& $0.410 \pm 0.163$	& \cite{Bellini:2013cosmo}\\
$^{6}$He\,($\beta^-$)  	& 3.51 & 1.16 s	& $1.110 \pm 0.452$		 &  \cite{Bellini:2013cosmo}\\
$^{8}$Li\,($\beta^-$)  	 & 16.0 & 1.21 s	& $0.210 \pm 0.191$ & \cite{Bellini:2013cosmo}\\
$^{11}$Be\,($\beta^-$)   & 11.5 & 19.9 s	& $\textless$0.20 (3$\sigma$) & \cite{Bellini:2013cosmo}\\
$^{10}$C\,($\beta^+$)  	 & 3.65 & 27.8 s	& $0.52 \pm 0.13$		& \cite{Bellini:2013cosmo}\\
$^{11}$C\,($\beta^+$) 	  & 1.98 & 29.4 min	& $26.8 \pm 0.2$ & \cite{PPchainNature,NusolAfterNature}		 \\
\bottomrule
\end{tabular}
\end{table}

\paragraph{\emph{External background}}

External background is represented by the particles created outside of the scintillator but reaching the fiducial volume of the analysis. Typically, only $\gamma$-rays represent external background and other particles, as $\alpha$s' and $\beta$s' produced in external materials, are absorbed before being able to enter the central parts of the LS. The contamination levels of the detector's construction materials, such as PMTs, SSS, or IV, are extensively discussed in~\cite{Alimonti:2000xc}. The external background can be divided into three categories: (1) $\gamma$s' from $^{40}$K, $^{208}$Tl (from the $^{232}$Th contamination), and $^{214}$Bi (from the $^{238}$U contamination) relevant at energies below 3\,MeV, (2) $^{208}$Tl background from the $^{232}$Th contamination of the IV relevant for the $^8$B solar neutrino analysis~\cite{ImprovedB8}, and (3) a recently identified source of high-energy $\gamma$s' from the captures of neutrons produced by $(\alpha,\,\textrm{n})$ reactions, that themsleves are triggered by $\alpha$s' from the decays of Uranium and Thorium present in the SSS and the PMTs' glass (see Section~\ref{sec:pp_ana})~\cite{ImprovedB8}. The different external backgrounds are further discussed below.

\begin{itemize}
    \item[] $^{40}$K\,($\gamma$ decay, \emph{Q}\,=\,1.461\,MeV, $\tau$ = 1.8$\times10^9$\,years): this primordial nuclide has an electron capture reaction with 10.7\% probability, leading to the emission of a 1.461\,MeV $\gamma$. This $\gamma$ has the highest probability of occurrence when compared to other decay branches. The most important source of this background is the glass of the PMTs.
    \item[]
    \item[] $^{214}$Bi\,($\gamma$ decay, \emph{Q}\,=\,2.448\,MeV, $\tau$ = 28.7\,min): this isotope originating from the $^{238}$U contamination of the construction materials (mostly SSS and PMTs) has a 99.98\% probability to decay via $\beta^-$-emission to an excited state at 2.448\,MeV, which emits a $\gamma$ with a branching fraction of 1.5\%. This decay occurs with highest probability compared to the other excited $\gamma$ states.
    \item[]
    \item[] $^{208}$Tl($\beta^-\gamma$ decay, \emph{Q}\,=\,4.999\,MeV, $\tau$ = 4.4\,min): this isotope originates from the $^{232}$Th contamination and is a direct decay product of $^{212}$Bi with 36\% branching ratio. $^{208}$Tl emits simultaneously an electron and gamma during its decay. Therefore, as previosuly mentioned, $^{208}$Tl gives rise to two kinds of external backgrounds. From the $^{208}$Tl decays in the SSS and PMTs, only 2.6\,MeV $\gamma$-rays can penetrate the LS, making it an external background for the low energy solar neutrino analysis below 3\,MeV (Section~\ref{sec:pp_ana}). However, when the source of contamination is the IV, there is a chance that the emitted electron also deposits its energy in the LS, which effectively increases the visible energy of this background above 3\,MeV. This kind of background is important for the $^{8}$B solar neutrino analysis~\cite{ImprovedB8}, which uses peripheral areas of the IV for detection.  When $^{208}$Tl decays, it can be located within the nylon membrane (\emph{surface} $^{208}$Tl) or in the fluid in a close proximity to the IV (\emph{emanated} $^{208}$Tl). The latter component can be caused, for example, by $^{220}$Rn, a volatile progenitor that can diffuse out of the IV.
    \item[]
    \item[] $[(\alpha,\,\textrm{n}),X]\rightarrow \gamma$: this background is represented by high energy gammas produced in captures of radiogenic neutrons. The latter are produced by ($\alpha$, n) interactions triggered by $\alpha$s' from decays of $^{238}$U/$^{235}$U and $^{232}$Th chains occurring in the SSS and the PMTs' glass. The MC simulation shows that these neutrons are mainly captured on the SSS iron and on the protons and $^{12}$C in the 80\,cm buffer layer adjacent to the SSS. The energy of the gammas from these neutron captures extends upto 10\,MeV. This background is considered in the energy ranges from 3.2 to 5.7\,MeV and from 5.7 to 16.0\,MeV, targeted for the $^8$B neutrino analysis (see Section~\ref{sec:solar_nu_ana}).
    
\end{itemize}

\paragraph{\emph{Cosmogenic background}}

Cosmogenic backgrounds in Borexino can be divided into three main categories: cosmic muons, cosmogenic neutrons, and cosmogenic radioisotopes~\cite{Bellini:2013cosmo}. 
Cosmic muons are created due to the interaction of high energy primary cosmic rays with the nuclei in the atmosphere. Cosmogenic fast neutrons can arise from the spallation of muons passing either through the OD and/or the ID, or the surrounding rocks and can penetrate through the detector materials, due to their high energy and no charge. The cosmogenic radioisotopes are created due to the spallation of cosmic muons on detector materials. In contrast to cosmogenic neutrons, the charged ions of radioactive isotopes have low penetration ability and act as backgrounds only when produced inside the liquid scintillator. The three categories of cosmogenic backgrounds in Borexino are discussed below. The detection and measurement of cosmogenic background are extensively discussed in~\cite{Bellini:2011yd,Bellini:2013cosmo,Agostini:2019dbs}.

\begin{itemize}
    \item[]Cosmic muons: The primary cosmic muon flux arriving at the Earth's surface is about 6.5$\times10^5$ m$^{-2}$\,h$^{-1}$ and is attenuated by a factor of $\sim$10$^{6}$ at LNGS due to the mountains above and this corresponds to a measured flux of (3.432$\pm$0.003) $\cdot$ 10$^{-4}$\,m$^{-2}s^{-1}$~\cite{Agostini:2018fnx}. The mean energy of muons at LNGS is about 280\,GeV, compared to about 1\,GeV at the Earth's surface, since the lower energy muons incident at the surface are absorbed, and the flux steeply falls as a function of, abou energy. Muons in Borexino are classified into three types: \emph{internal}, \emph{external} and \emph{special} muons, which are extensively discussed in~\cite{Agostini:2019dbs}. The \emph{internal muons}, about 4300 per day, are the ones crossing both the OD and the ID. They are identified using three flags: either by a special electronics flag called the \emph{Muon Trigger Flag}, which means that the Water Cherenkov OD triggered; or through the software reconstruction algorithm called the \emph{Muon Cluster Flag}, that identifies clusters of hits among those detected by the OD; or via the \emph{Inner Detector Flag}, that uses different cluster variables for the reconstruction of muon pulse-shape information in the ID. The dead time applied after these muons differ for different analyses, depending on their relevant cosmogenic backgrounds. \emph{External muons} cross only the OD and do not form clusters in the ID. They are detected by either the Muon Trigger Flag or the Muon Cluster Flag with an overall rate similar to that of internal muons. A 2\,ms dead time, nearly 8 times the neutron capture time, is applied after all external muons to eliminate fast neutrons crossing the LS after these muons. \emph{Special muon} flags are designed to tag a very small category of muons that cross the detector, typically in times when the detector was in a special state. These special categories of muons include also noise events and are particularly important in the geoneutrino analysis (see Section~\ref{sec:geo-analysis}), where the signal rate is extremely low. Therefore, depending on the analysis, special muons can also be treated as internal muons~\cite{Agostini:2019dbs}. In addition to the different muon tagging methods mentioned above, the FADC-sub system allows for an accurate pulse shape discrimination of muons. It plays a key role in analyses where the muon tagging is extremely important such as the geoneutrino analysis~\cite{Agostini:2019dbs} and the \emph{hep} solar neutrino analysis~\cite{ImprovedB8}. The combined muon tagging efficiency of the main DAQ and the FADC sub-system in Borexino is 99.9969\%~\cite{Agostini:2019dbs}. The measurement of muons using 10\,years of Borexino data, and their seasonal and annual modulations are discussed in detail in~\cite{Agostini:2018fnx}. 
    \item[]
    \item[] Cosmogenic neutrons: Cosmic muons in Borexino can lead to the creation of cosmogenic neutrons, due to different spallation processes on carbon nuclei. Neutrons in the Borexino LS are captured with a lifetime of (254.5 $\pm$ 1.8)\,$\mu$s (measured using $^{241}$Am-$^{9}$Be neutron calibration source~\cite{Bellini:2011yd}) and create a 2.2\,MeV $\gamma$ when captured on a proton, or a 4.95\,MeV $\gamma$ when captured on a $^{12}$C nucleus. The 2.2\,MeV $\gamma$ is not relevant for the solar neutrino analysis, but the 4.95\,MeV $\gamma$ is important for the $^{8}$B solar neutrino analysis~\cite{ImprovedB8}. A 1.28\,ms gate is opened after each internal muon to guarantee high detection efficiency of cosmogenic neutrons. The above-mentioned 2\,ms dead time ($\sim$8 times the n-capture time) is usually enough to suppress these fast neutrons arising from the passage of muons. Fast neutrons from muons crossing the water tank and from undetected muons crossing the surrounding rocks are relevant backgrounds for the geoneutrino analysis. The neutrons from the water tank are estimated through the analysis of the acquired data, while a dedicated Monte-Carlo simulation is required to estimate the contribution from the surrounding rocks. This is discussed in detail in~\cite{Agostini:2019dbs}.
    \item[]
    \item[] Cosmogenic radioisotopes: The spallation of cosmic muons on $^{12}$C nuclei leads to the creation of $^{11}$C isotope ($\beta^+$ decay, \emph{Q}\,=\,0.960\,MeV, $\tau$ = 29.4\,min), an important background for \emph{pep} solar neutrinos. Due to its long lifetime of 29.4\,min, $^{11}$C cannot be suppressed with a simple time veto. It is tagged through the \emph{three-fold coincidence} (TFC) algorithm, which exploits the fact that $^{11}$C is mostly produced in time coincidence with neutrons, and further divides the data into TFC-enriched and TFC-depleted spectra for the solar neutrino analysis, which is discussed in more detail in Section~\ref{sec:solar_nu_ana}. In addition, since the decay mode of this isotope is via $\beta^{+}$, it can be distinguished, on a statistical level, from the solar neutrino signal ($\beta^{-}$) through pulse-shape discrimination, as discussed in Section~\ref{subsec:event-reco}. The other cosmogenic isotopes relevant for the solar and geoneutrino analyses are listed in detail in Table~\ref{tab:cosmo_bckg}. All these backgrounds are relevant for the $^{8}$B solar neutrino analysis. They are suppressed using a long dead time of 6.5\,s after a muon signal, with the exception of $^{10}$C and $^{11}$Be. $^{10}$C requires a longer time veto and an additional space veto, while the $^{11}$Be background is treated using a multivariate fit and its residual rate is found to be compatible with zero~\cite{ImprovedB8}. $^{9}$Li represents an important non-antineutrino background for geoneutrinos due to its ($\beta^{-} + n$) decay mode, which imitates the geoneutrino signal (Section~\ref{subsec:detection}). They are suppressed using sophisticated time and spatial vetoes~\cite{Agostini:2019dbs}. Other isotopes such as $^{8}$He and $^{12}$B are are also relevant for the geoneutrino analysis and are discussed in~\cite{Agostini:2019dbs}, but their contribution is negligible when compared to the $^{9}$Li isotope.

\end{itemize}

\subsection{Neutrino and antineutrino detection}
\label{subsec:detection}

The detection principles of neutrinos and antineutrinos are significantly different in Borexino and are discussed in this Section.

\paragraph{\emph{Neutrino detection}}

Neutrinos $\nu$ of all flavours are detected via the neutrino-electron elastic scattering process:
\begin{equation}
 \nu_{e,\mu,\tau}+e\rightarrow\nu_{e,\mu,\tau}+e,
 \label{eq:nu_detection}
\end{equation}
in which neutrinos interact with the electrons present in the LS, which has a density of $(3.307\pm0.015)\times10^{31}$ electrons per 100\,ton~\cite{CNOpap}. In this process, a fraction of the neutrino energy is transferred to the electron, which is finally responsible for the generation of scintillation light in the detector. The electron recoil spectrum, continuous even in the case of mono-energetic neutrinos, extends up to a maximum energy $T_e^{max}$ given by:
\begin{equation}
T_e^{max}= \frac {E_\nu } {1+ \frac {m_e c^2} {2 E_\nu}},
\label{eq:Tmax}
\end{equation}
where $m_{e}$ is the electron mass and $E_{\nu}$ the neutrino energy. The elastic scattering process has no threshold. The cross section for $\nu_{e}$s' is in the order of $10^{-45}$ to $10^{-43}$ cm$^2$ for solar neutrino energies, i.e. below 20\,MeV~\cite{BahcallRadiative}. It is about 5 times larger with respect to the $\nu_{\mu,\tau}-e$ scattering process. This is due to the fact that the latter interact only through neutral current (NC) interactions, while $\nu_{e}$s' additionally interact via charged current (CC) interactions. In Borexino, however, the scattering process induced by $\nu_{\mu}$s' and $\nu_{\tau}$s' cannot be distinguished from each other with the current amount of data. The cross sections used for the solar neutrino analysis consider leading order radiative corrections and are taken from~\cite{BahcallRadiative}, with improved measurements taken from~\cite{Nakamura_2010}.

\paragraph{\emph{Antineutrino detection}}

Electron antineutrinos $\bar{\nu}_{e}$ are detected via the \emph{Inverse Beta Decay} (IBD) reaction:
\begin{equation}
 \bar{\nu}_{e}+\text{p}\rightarrow\text{n}+e^{+},
 \label{eq:antinu_detection}
\end{equation}
in which an electron flavour antineutrino is captured on a free proton (Hydrogen nucleus), producing a neutron and a positron. Hereby, the LS has a density of $(6.007\pm0.001)\times10^{30}$ protons per 100\,ton~\cite{Agostini:2019dbs}. The IBD has a kinematic threshold of 1.806\,MeV due to the larger mass of the neutron compared to the proton. The positron first deposits its kinetic energy and then annihilates, producing two gammas with  $E_{\gamma}=0.511$\,MeV each. These two processes cannot be distinguished and lead to the creation of a \emph{prompt event}. The energy of the antineutrino is mostly transferred to the positron and thus the visible energy of a prompt event $E_{vis}^{e^{+}}$ can be directly connected with the energy of the incident antineutrino $E_{\bar{\nu}_{e}}$: 
\begin{eqnarray}
E_{vis}^{e^{+}} = 2 \cdot E_{\gamma} + E_{\bar{\nu}_{e}} - 1.806\, \textrm{MeV} = E_{\bar{\nu}_{e}} - 0.784\,\textrm{MeV}.
\label{eq:Evis_anti_nu}
\end{eqnarray}
After its creation, the neutron is thermalised and then it is captured after a typical time of (254.5 $\pm$ 1.8)\,$\mu$s~\cite{Bellini:2011yd}. The capture is accompanied by a de-excitation gamma. In the majority of cases, the neutron is captured on a proton and the energy of the gamma is 2.2\,MeV. With a probability of 1.1\%~\cite{Agostini:2019dbs}, a 4.95\,MeV $\gamma$ is emitted after the neutron capture on $^{12}$C. Each gamma of this energy range deposits its energy in the scintillator predominantly by multiple Compton scatterings. Several Compton electrons are then detected as a single \emph{delayed event}. At MeV energies, the IBD cross section is in the order of $10^{-42}$ cm$^2$~\cite{CrossSectionVissani}, which is about 100 times larger compared to neutrino-electron elastic scattering. Antineutrinos are able to interact also via elastic scattering, but it is much more convenient to detect them via the IBD process, providing a golden channel to identify the rare interactions and significantly suppress backgrounds, exploiting the fast prompt-delayed coincidence signal.

\section{Solar neutrinos}
\label{sec:solar}

The Sun is a strong natural source of neutrinos, and the emitted flux of solar neutrinos is of the order of $10^{10}$cm$^{-2}$s$^{-1}$, with their energy spectrum extending up to about 15\,MeV. They are produced in the electron flavor ($\nu_e$) along the nuclear fusion processes that occur in the core of the Sun. Differently from photons, also produced in these interactions, neutrinos are able to travel directly from the production site to the Earth, without being deflected or absorbed. Therefore, solar neutrinos are a direct probe to the Sun’s interior. Indeed, they are being extensively used to understand the fundamental processes powering our star since decades~\cite{GALLEX99,GNO05,GALLEX10,SAGE09,SNO13,LongPaperPhaseI,SK16_solar}. Historically, solar neutrino measurements led to the experimental evidence of neutrino flavor transformation~\cite{SNO_1,SNO_2}. Even today, they are at the base of the most precise determination of the $\theta_{12}$ mixing angle~\cite{PDG2020}. More recently, they are being used in searches for physics beyond the Standard Model~\cite{Bx12_NSI,Bx-NSI20} and are among the goals of future experiments~\cite{JUNOB8,Jinping17,Theia20,LAr16}.
The vast majority (about 99\%) of the energy produced in the Sun comes from a series of reactions fusing Hydrogen to Helium, called \emph{pp} chain. The associated neutrino flux is generated by various sub-processes, and includes the so-called \emph{pp}, \emph{pep}, $^7$Be, $^8$B, and \emph{hep} neutrinos. The remaining small fraction of solar energy is produced in the so-called CNO cycle, in which the Hydrogen-to-Helium fusion is catalysed by the presence of Carbon, Nitrogen, and Oxygen. More details about the production mechanism and propagation of solar neutrinos from the Sun to the Earth are given in Section~\ref{sec:Nu_prod_prop}. In particular, Section~\ref{sec:pp_CNO} reports about the \emph{pp} chain and CNO cycle. Section~\ref{sec:ssm_metal} discusses the so-called \emph{Standard Solar Model} (SSM) that predicts the fluxes of different species of neutrinos, that depend on the so-called \emph{metallicity}, i.e., the abundance of elements heavier than Helium. Solar neutrinos arrive on the Earth as a mixture of all flavours. The process of the flavour conversion of solar neutrinos, maximised for neutrinos with energy greater than $\sim$2\,MeV by the presence of the dense solar matter via the Mikheyev-Smirnov-Wolfenstein (MSW) effect~\cite{PhysRevD.17.2369, Mikheev:1986gs} is briefly discussed in Section~\ref{sec:nuosc_msw}.
Borexino is the only experiment that has performed a complete spectroscopy of solar neutrinos. Section~\ref{sec:solar_nu_ana} presents the basic principles of Borexino solar neutrino analysis and underlines the features common to various analysis aimed to extract rates of different solar neutrino species. The following sections then discuss the particularities of different analyses, their results, and physics implications. Section~\ref{sec:pp_ana} describes specifically the measurement of \emph{pp} chain~\cite{PPchainNature}, while Section~\ref{sec:CNO_ana} is devoted to the discovery of CNO neutrinos~\cite{CNOpap}. Finally, Section~\ref{sec:bsm} gives a brief overview of Beyond-the-Standard-Model physics probed by Borexino. Searches for spectral deformation of electrons scattered off $^7$Be solar neutrinos due to eventual \emph{Non-Standard Interactions} (NSI) are described in Section~\ref{sec:bxnsi}, while the tight limits set on \emph{Neutrino Magnetic Moment} (NMM) are discussed in Section~\ref{sec:bxnmm}.

\subsection{Solar neutrinos production and propagation}
\label{sec:Nu_prod_prop}

\subsubsection{Hydrogen to Helium fusion: \emph{pp} chain and CNO cycle}
\label{sec:pp_CNO}

\begin{figure}[t]
\centering
    \includegraphics[width=0.65\textwidth]{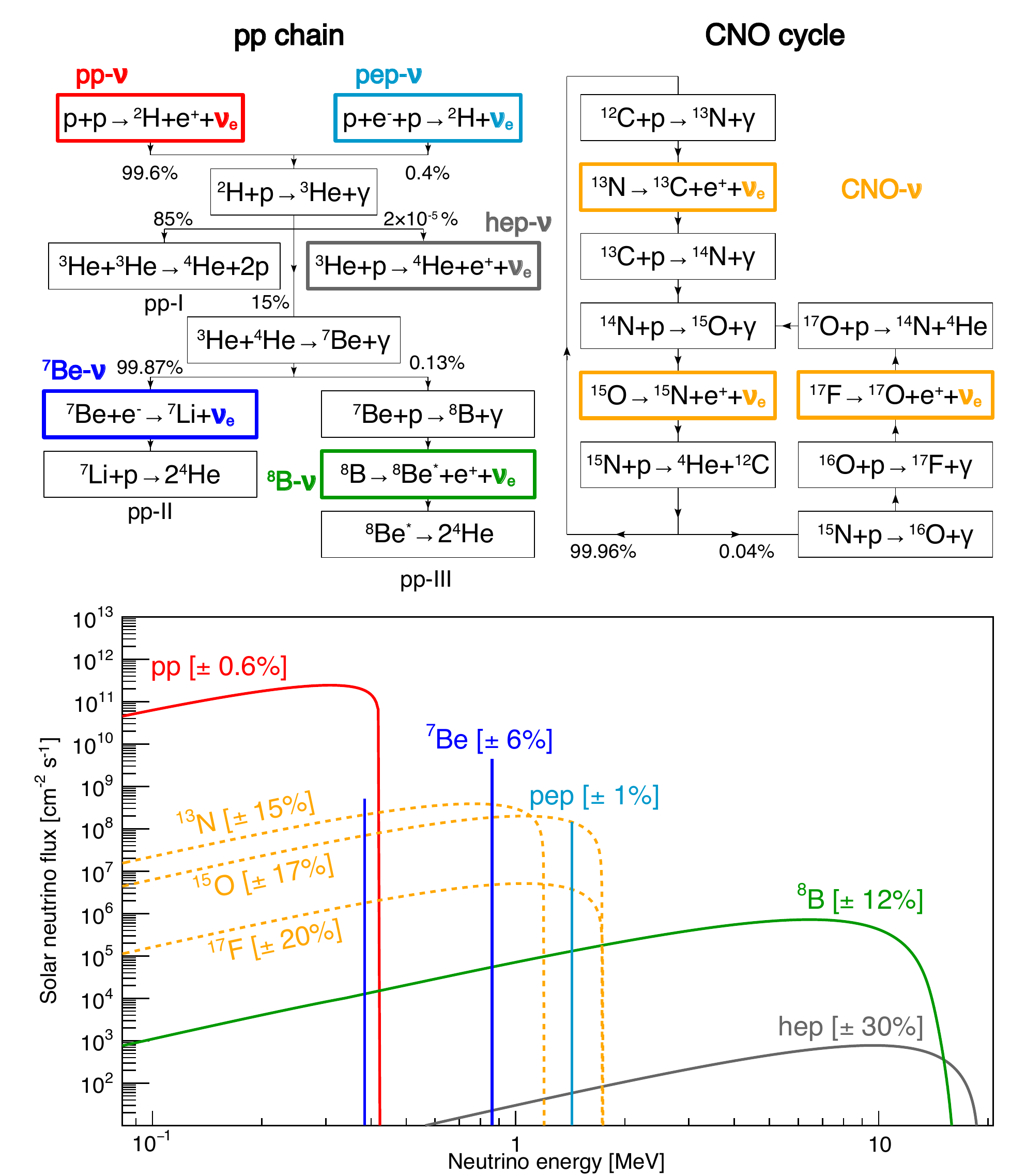}
    \caption{Top: Schematic view of the \emph{pp} (left) and CNO (right) nuclear fusion sequences. Bottom: The solar-neutrino energy spectrum obtained from \url{http://www.sns.ias.edu/~jnb/}, using the updated fluxes taken from~\cite{B16SSM}. The flux (vertical scale) is given in units of cm$^{-2}$\,s$^{-1}$\,MeV$^{-1}$ for continuum sources and in cm$^{-2}$\,s$^{-1}$ for mono-energetic sources. The numbers in brackets represent the relative flux uncertainties for HZ-SSM prediction as quoted in Table~\ref{tab:SolarFluxes} after constraining the solar luminosity as presented in~\cite{Bergstrom_2016}.}
    \label{fig:pp_cno_sketch}
\end{figure}

Solar neutrinos are emitted during the fusion of protons to Helium nuclei taking place in the solar core:
\begin{equation}
    4p + 2 e^- \to ^4{\rm \!\!He} + 2 \nu_e + 26.73 \, {\rm\,MeV}.
    \label{eq:HtoHe}
\end{equation}
 The dominant fusion process is the \emph{pp} chain, while a subdominant fraction of solar energy is produced in the so-called CNO cycle, in which the fusion is catalysed by the presence of heavier elements, namely Carbon, Oxygen, and Nitrogen. Thus, the CNO contribution to the Sun’s fusion processes depends directly on the core’s metallicity, i.e. the mass abundance of elements heavier than Helium. In addition, the net contribution of the CNO cycle is strongly dependent on the core’s temperature. The Sun is a relatively small and cold star in the Universe and the CNO contribution to the luminosity budget is around 1\%. For heavier stars, having a mass greater than $\sim$1.3 solar masses, the CNO cycle is instead believed to be the dominant process which burns Hydrogen into Helium. It is therefore considered as the main nuclear fusion process occurring in the Universe. The precise CNO contribution in the Sun, however, is unknown, since the Sun’s metallicity is not known with precision (see Section~\ref{sec:ssm_metal}).

The top part of Figure~\ref{fig:pp_cno_sketch} shows the schemes of the \emph{pp} chain and of the CNO cycle, while its bottom part shows the energy spectrum of solar neutrinos. The fluxes, and thus the normalisation of the energy spectra, are predicted by the SSM~\cite{B16SSM} as discussed in the following Section. The \emph{pp} chain consists of three branches, indicated as pp-I, pp-II, and pp-III, each terminated by the production of $^4$He. The flux of solar neutrinos is dominated by the \emph{pp} neutrinos (order of $10^{10}$\,s$^{-1}$ cm$^{-2}$) having a continuous energy spectrum with 0.420\,MeV endpoint. In the \emph{pp} chain, also mono-energetic $^7$Be ($10\%$ branching at $0.384$\,MeV (excited state) and $90\%$ branching at $0.862$\,MeV (ground state)) and \emph{pep} (1.44\,MeV) neutrinos are produced, as well as $^8$B neutrinos characterised by lower flux (order of $10^{6}$ s$^{-1}$ cm$^{-2}$) and a continuous energy spectrum extending up to about 16.3\,MeV. The $hep$ neutrinos, with an extremely low flux and $18.784$\,MeV endpoint energy, has been not experimentally observed yet. 
The CNO cycle is dominated by the $^{13}$N and $^{15}$O decays, while $^{17}$F decays contribute only at the $1\%$ level. All three components are continuous spectra of similar shapes with endpoints below 1.8\,MeV. In the Borexino analysis, we call the CNO spectrum the weighted sum of all three components. 

\subsubsection{Standard Solar Model and the metallicity problem}
\label{sec:ssm_metal}

The so-called Standard Solar Model (SSM) is a spherically symmetric quasi-static model of a star in hydrostatic equilibrium with one solar mass $M_{\odot}$, including several differential equations derived from basic physical principles. SSM assumes that the Sun was initially chemically homogeneous and that the mass loss is negligible during the whole 4.57\,Gyr of its existence~\cite{B16SSM}. The calibration is done by adjusting the mixing length parameter and the initial Helium and metal~\footnote{In solar astrophysics, elements heavier than Helium are called metals.} mass fractions in order to satisfy the constraints imposed by the present-day solar luminosity $L_{\odot}$, radius $R_{\odot}$, and surface metal-to-hydrogen abundance ratio (so-called \emph{metallicity}, $(Z/X)_{\odot}$)~\cite{B16SSM}. SSM assumes that solar energy is generated by the \emph{pp} chain and the CNO cycle.

\begin{table}[t] 
\centering
\caption{Solar neutrino fluxes predicted by Standard Solar Models B16-GS98 (High Metallicity, HZ-SSM) and  B16-AGSS09met (Low Metallicity, LZ-SSM)~\cite{B16SSM} in units of cm$^{-2}$ s$^{-1}$ with the exponential factors given in the last column. The relative difference between the two model predictions is also quoted.}
\label{tab:SolarFluxes}
\scalebox{0.95}{
\begin{tabular}{lcccc}
\toprule
Solar $\nu$  & B16-GS98 (HZ) & B16-AGSS09met (LZ)&$(\textrm{HZ}-\textrm{LZ})/\textrm{HZ}$ $[\%]$ & Exp \\   
\midrule
\emph{pp}-cycle &  & & & \\ 
\midrule
\emph{pp} & $5.98 (1.0 \pm 0.006)$ & $6.03 (1.0 \pm 0.005)$ & $-0.8$ & $\times10^{10}$ \\ 
[4pt] $^7$Be & $4.93 (1.0 \pm 0.06)$ & $4.50 (1.0 \pm 0.06) $ & $8.9$ & $\times10^{9}$ \\
[4pt] \emph{pep} & $1.44 (1.0 \pm 0.01)$ & $1.46 (1.0 \pm 0.009)$ & $-1.4$ & $\times10^{8}$ \\
[4pt] $^8$B & $5.46 (1.0 \pm 0.12) $ & $4.50 (1.0 \pm 0.12)$ & $-17.6$ & $\times10^{6}$\\
[4pt] $hep$ & $7.98 (1.0 \pm 0.30)$ & $8.25 (1.0 \pm 0.12)$ & $-3.4$ & $\times10^{3}$ \\
[4pt] \midrule CNO-cycle &  & & & \\ 
\midrule$^{13}$N & $2.78 (1.0 \pm 0.15)$ & $2.04 (1.0 \pm 0.14)$ & $26.6$ & $\times10^{8}$\\
[4pt] $^{15}$O & $2.05 (1.0 \pm 0.17)$ & $1.44 (1.0 \pm 0.16)$ & $29.7$ & $\times10^{8}$\\
[4pt] $^{17}$F & $5.29 (1.0 \pm 0.20)$ & $3.26 (1.0 \pm 0.18)$ & $38.3$ & $\times10^{6}$\\
[4pt] CNO& $4.88 (1.0 \pm 0.11)$ & $3.51 (1.0 \pm 0.10)$ & $ 28.1$ & $\times10^{8}$\\ 
\bottomrule
\end{tabular}
}
\end{table}

Among the outputs of the SSM are the neutrino fluxes, summarised in Table~\ref{tab:SolarFluxes} for the new generation of SSM called B16~\cite{B16SSM}, that includes updates on nuclear reaction rates, more consistent treatment of the equation of state, and a novel treatment of opacity uncertainties. The prediction is given separately for the two canonical sets of solar abundances. So-called \emph{low metallicity (LZ or AGSS09met-LZ)} scenario~\cite{AGSS15a,AGSS15b} represents the most recent revision of solar abundances based on development of three-dimensional hydro-dynamical models of the solar atmosphere, of techniques to study line formation, and the improvements of atomic properties such as transition strengths. The other solar metal composition scenario, so-called \emph{high metallicity (HZ or GS98-HZ)} scenario~\cite{GS98}, is based on older one-dimensional modeling of the solar atmosphere and predicts higher metal abundances. In this regard, emerges the so-called \emph{metallicity puzzle}. The fact is that newer LZ-SSM spoil the earlier agreement between the observed sounds speed profile (helioseismology data) and the corresponding SSM predictions. The origin of this discrepancy is currently not understood. However, the SSM prediction of neutrino fluxes depend on the metallicity: the metallicity influences the opacity of the Sun and consequently also the temperature in the core and the fusion rates. There is a sizeable difference of 8.9\% and 17.6\% between the HZ and LZ SSM predictions of $^7$Be and $^8$B fluxes, respectively (Table~\ref{tab:SolarFluxes}). The largest difference between the fluxes predicted by the LZ and HZ SSM results for the CNO cycle and amounts to about 32\%. The metallicity indeed directly influences the efficiency of the CNO cycle, since the “metals”, Carbon, Nitrogen, and Oxygen are the elements which catalyze the process. The CNO neutrino flux is then directly dependent upon the core’s metallicity, which keeps memory of the Sun’s elemental composition at the time of formation. Since the metal abundance in the core is believed to not be influenced by the surface, CNO neutrinos preserve the core’s information in its pristine conditions. Thus, neutrinos produced in the CNO cycle are a unique probe to the Sun’s primordial composition. In summary, precise measurements of the solar neutrino fluxes can provide important boundary conditions for the future development of the SSMs and our understanding of the stars in general.

\subsection{Solar neutrino analysis in a nutshell}
\label{sec:solar_nu_ana}
\begin{table}[t!]
\centering
\caption{Details of the solar neutrino analysis of the Borexino Phase II(+I) and Phase III data. The first column specifies the solar neutrino species extracted by the fit of the data using the exposure mentioned in the second column. The third column shows the variables used in the fit: global variable $E$, standing for various energy estimators (Section~\ref{subsec:event-reco}) or $R$, reconstructed radius. The parameters in square brackets identify the variables used in the multivariate part of the fit in the LER: pulse shape variable used to constrain residual $^{11}$C($e^+$) and radial distribution constraining external background. The fourth column gives the energy range of the global variable, while the last column highlights the species whose rates were constrained in the fit. The source of information used to extract these independent constraints is given in brackets, while more details are given in text. UL = Upper Limit; Cosm. = cosmogenic background; indep. data = independent data; LER/HER = low/high energy region; LPoF = Low Polonium Field.} \label{tab:SolAnal}
\scalebox{0.92} {
    \begin{tabular}{lllll}
    \toprule
     Solar $\nu$  & Exposure & Fit   & Energy  & Constraints   \\
                 & [day $\times$ ton] & Variable          & [MeV]   & \\[3pt]
     \midrule
     Phase-II~\cite{PPchainNature,NusolAfterNature} &  &  & &    \\[3.8pt]
      12/2011 - 05/2016   &  &  & &    \\[3.8pt]
    \midrule
    \emph{pp}, $^7$Be, \emph{pep}  & 1291 $\times$ 71.3  & $E$($N_h$, $N_{p}^{dt1,2}$)  & LER & CNO [SSM-HZ(LZ)] \\ [3.8pt]
                  &      & +[e$^{+/-}$, $R$] & 0.19-2.93  & $^{14}C$ [indep. data]  \\[3.8pt]
    CNO(UL)& 1291 $\times$ 71.3 & $E$($N_h$, $N_{p}^{dt1,2}$) & LER & \emph{pp}/\emph{pep} [SSM-HZ(LZ)] \\[3.8pt]
              &      & +[e$^{+/-}$, $R$] & 0.19-2.93 & $^{14}C$ [indep. data]   \\[3.8pt]
    \midrule
     Phase-I (part) + II + III (part)~\cite{PPchainNature,ImprovedB8} &  &  & &    \\[3.8pt]
     01/2008 - 12/2016   &  &  & &    \\[3.8pt]
    \midrule
    $^8$B & 2062 $\times$ 227.8 & $R$  & HER-I & $^{208}$Tl:bulk($^{212}$Bi-$^{212}$Po) \\ [3.8pt]
           &      & & 3.2 - 5.7   &  $^{214}$Bi($^{214}$Bi-$^{214}$Po)  \\[3.8pt]
           &      &   & & Cosm. [indep. data]   \\[3.8pt]
    $^8$B & 2062 $\times$ 266.0 & $R$  & HER-II  &\\ [3.8pt]
      &      &   & 5.7 - 16.0  & Cosm. [indep. data]   \\[3.8pt]
     \midrule
     Phase-I (part) + II + III (part)~\cite{ImprovedB8} &  &  & &    \\[3.8pt]
     11/2009 - 10/2017   &  &  & &    \\[3.8pt]
     \midrule
    \emph{hep} & 1259 $\times$ 216.0 & $E$($N_{pe}$)  & 11.0 - 20.0 & \\[3.8pt]
    \midrule
     Phase-III~\cite{CNOpap} &  &  &  & \\
     07/2016 - 02/2020   &  &  & &    \\[3.8pt]
     \midrule
    CNO  & 1072 $\times$ 71.3 & $E$($N_h$) & LER & UL($^{210}$Bi)[LPoF]   \\[3.8pt]
                 &      & +[$R$] & 0.32 - 2.64 & \emph{pep} [SSM+solar]   \\[3.8pt]
    \bottomrule
    \end{tabular}
    }
\end{table}

\subsubsection{Neutrino flavour conversion and matter effects}
\label{sec:nuosc_msw}

In the standard three-flavour neutrino framework, the electron neutrino survival probability, i.e., the probability to measure solar neutrino in the same flavour as it was produced, can be cast in the form~\cite{Capozzi2018}:
\begin{equation}
    P_{3\nu}(\nu_{e}\rightarrow\nu_{e}) \simeq \cos^{4} \theta_{13} P_{2\nu}(\nu_{e}\rightarrow\nu_{e}) + \sin^4 \theta_{13},
    \label{eq:Pee_3nu}
\end{equation}
where $P_{2\nu}(\nu_{e}\rightarrow\nu_{e})$ corresponds to the 2$\nu$ probability for $\theta_{13} = 0$, which depends on the solar mixing angle $\theta_{12}$ and the mass splitting $\Delta m^2_{12}$ only. 

Should the oscillation happen in vacuum, this survival probability could be approximated by~\cite{PPchainNature}:
\begin{equation}
    P_{3\nu}^{\mathrm{vac}}(\nu_{e}\rightarrow\nu_{e}) =   P_{ee}^{\mathrm{vac}} \simeq \cos^{4} \theta_{13} \left(1 - 0.5 \sin^2 2 \theta_{12}\right) + \sin^4 \theta_{13},
    \label{eq:Pee_vac}
\end{equation}
that does not depend on energy and has an approximate value of 0.54. In reality, solar neutrinos are crossing the dense solar matter. The electron flavour neutrinos experience an extra potential due to the charge-current interaction with the electrons present in the Sun. This affects the neutrino oscillation probability that changes with respect to a pure vacuum oscillation scenario. This effect is called Mikheyev-Smirnov-Wolfenstein (MSW) effect~\cite{PhysRevD.17.2369, Mikheev:1986gs}. Thus, the survival probability $P_{ee}^{\mathrm{MSW}}$ depends not only on the oscillation parameters, but also on the neutrino-energy-dependent potential. Assuming an adiabatic decrease of the electron density with radius, it can be expressed as follows~\cite{Bahcall_2004,PPchainNature}:
\begin{equation}
    P_{3\nu}^{\mathrm{MSW}}(\nu_{e}\rightarrow\nu_{e}) =   P_{ee}^{\mathrm{MSW}} \simeq 0.5\cos^{4} \theta_{13} \left(1 + \cos 2 \theta_{12}^{\mathrm M} \cos 2\theta_{12}\right),
    \label{eq:Pee_msw}
\end{equation}
where
\begin{equation}
     \cos 2 \theta_{12}^{\mathrm M} = \frac {\cos 2\theta_{12} - \beta}{\sqrt{(\cos 2\theta_{12} - \beta)^2 +  \sin^2 2\theta_{12} }}
    \label{eq:cos-msw}
\end{equation}
and
\begin{equation}
     \beta = \frac {2\sqrt{2} G_F \cos^2\theta_{13} n_{e} E_{\nu}}{\Delta m _{12}},
    \label{eq:beta_msw}
\end{equation}
where $\theta^{{\mathrm{M}}}$ is the mixing angle in matter, $E_{\nu}$ the neutrino energy, $n_e$ the electron density in matter, and $G_F$ the Fermi coupling constant. This MSW survival probability stays at the vacuum value at low energies typical for \emph{pp} neutrinos (\emph{vacuum dominated region}), while for high energy $^8$B neutrinos it decreases to about 0.32 (\emph{matter dominated region}). The situation is further complicated by the fact, that different neutrino species have their production regions at different radii~\cite{Bahcall_2006}, and thus, propagate through regions of different electron densities. Calculation of the survival probabilities considering non-adiabatic corrections and averaging over the production region for each solar neutrino species has been presented in~\cite{deHolanda:2004fd}. Finally, the exact form of the \emph{transition region} between the vacuum and matter dominated regions might be sensitive to different models of the non-standard neutrino interactions and thus, is a point of interest for searches for physics beyond the Standard Model~\cite{Capozzi2018}.

Solar neutrinos are detected via the elastic scattering of electrons, as it was discussed in Section~\ref{subsec:detection}. Therefore, even for mono-energetic neutrinos, as $^7$Be or \emph{pep} neutrinos, the spectrum of scattered electrons is a continuous one, characterised only by the Compton-like edge, corresponding to a maximal energy of the scattered electrons (Equation~\ref{eq:Tmax}). It is impossible to distinguish the electrons scattered off by solar neutrinos from the background components, on an event-by-event basis~\footnote{As discussed in Section~\ref{subsec:detection}, it is only possible to perform $\alpha/\beta$ separation on an event-by-event basis.}. Therefore, the analysis proceeds in two steps: (1) the event selection, with a set of cuts to maximize the signal-to-background ratio, and (2) the extraction of the neutrino and residual background rates through a combined fit of the distributions of global quantities of the events surviving the cuts. 

Table~\ref{tab:SolAnal} summarizes some basic details of the solar neutrino analysis discussed in this paper such as the extracted solar species, analyzed periods, exposure, fit variable, energy range, and constraints used in the fit. Measurement of the \emph{pp} chain neutrinos is discussed in Section~\ref{sec:pp_ana}, based on~\cite{PPchainNature,NusolAfterNature,ImprovedB8}. The interaction rates of the \emph{pp}, $^7$Be, and \emph{pep} neutrinos were obtained through a spectral fit of the Phase-II data in the so-called LER (\emph{Low Energy Region}) below 3\,MeV. The measurement of $^8$B neutrinos  was performed on the combined Phase-I+II data, through a fit of the radial distribution. This strategy avoids any assumption on the spectral shape, i.e., on the shape of the survival probability in the transition region, defining the so called "upturn", the energy interval sensitive to possible NSI (Section~\ref{sec:nuosc_msw}). This analysis is performed in the HER (\emph{High Energy Region}) above 3.2\,MeV and below 16\,MeV. This energy interval is divided into two sub-parts namely, HER-I (below 5.7\,MeV) and HER-II (above 5.7\,MeV), each characterised by different backgrounds. The experimental confirmation of the existence of the CNO fusion in the Sun~\cite{CNOpap} was performed using the Phase-III data in the energy interval similar to LER, with an increased energy threshold, due to the worsened resolution associated  with the loss of PMTs. The CNO analysis is discussed in more detail in Section~\ref{sec:CNO_ana}.

The main event selection criteria are conceptually similar for all solar neutrino analyses and are conceived to reject cosmic muons surviving the mountain shield, reduce the cosmogenic background, select an optimal spatial region of the scintillator (the fiducial volume, FV), and remove eventual noise events. 

The muon detection combines the information of the external
Cherenkov veto with that of the inner detector, including the pulse shape analysis. Cosmogenic background is reduced
by applying a veto following each muon. A 2\,ms veto suppresses neutron captures from \emph{external muons}, which cross only the water buffer. A vast majority of captures occur on protons, emitting 2.2\,MeV$\gamma$s', while around 1\% of the captures occur on $^{12}$C nuclei, emitting $4.95$\,MeV $\gamma$s'. For the \emph{internal muons} crossing the scintillator, the veto time differs. In the LER, a time cut of 300\,ms is used
to suppress the saturation effects of electronics. In the HER, a much longer veto of 6.5\,s is applied, in order to suppress
$^{12}$B, $^{8}$He, $^{9}$C, $^{9}$Li, $^{8}$B, $^{6}$He, and $^{8}$Li decays. In addition, the presence of untagged $^{11}$Be (\emph{Q}\,=\,11.5\,MeV, $\beta^-$, $\tau$\,=\,19.9\,s) is estimated through a multivariate approach and found to be compatible with zero. In HER, an additional spherical cut of 0.8\,m radius is also applied for 120\,s around the capture position of cosmogenic neutrons to remove $^{10}$C(\emph{Q} = 3.6\,MeV, $\beta^{+}$, $\tau$ = 27.8\,s) background. 

 Specifically in the LER analysis, the cosmogenic $^{11}$C(\emph{Q} = 0.96\,MeV, $\beta^{+}$, $\tau$ = 29.4\,min) background requires a dedicated treatment. Due to its relatively long life-time, it cannot be removed by a simple veto. The so-called \emph{three-fold coincidence} (TFC) algorithm~\cite{LongPaperPhaseI, NusolAfterNature} uses the fact that $^{11}$C, created by the spallation of muons on $^{12}$C, is mostly produced together with neutrons:
\begin{equation}
	\mu + {^{12}\mathrm{C}} \rightarrow \mu + {^{11}\mathrm{C}} + \textrm{n}.
\end{equation}
Based on the characteristics and mutual configuration of the muon and cosmogenic neutrons, the TFC algorithm identifies the space-time regions with increased probability to create $^{11}$C and/or assigns to each event a probability to be $^{11}$C. Based on this, all events are divided into the {\emph TFC-subtracted} and {\emph TFC-enriched} categories. The TFC-subtracted spectrum preserves about $64\%$ of the total exposure, while the $^{11}$C suppression preserves more than $90\%$.

In both, LER and HER, $^{214}$Bi and $^{214}$Po events are removed with about 90\% efficiency, through the space-time correlation of their fast $\beta+\alpha$ delayed coincidence. 

The low- and high-energy analyses use different fiducial volumes (FV). In the LER, the FV represents the central region of 71.3\,ton, selected to maximally suppress external $\gamma$s' from $^{40}$K, $^{214}$Bi, and $^{208}$Tl, originating from the materials surrounding the scintillator. It is contained within the radius $R<2.8$\,m and the vertical coordinate -1.8\,m\,$< z < $\,2.2\,m. The HER is above the energy of the aforementioned external background. The analysis in HER-I requires only a $z<$\,2.5\,m cut to suppress background events related to a small pinhole in the inner vessel that causes liquid scintillator to leak into the buffer region. The total selected mass in this case is 227.8\,ton. In contrast, the analysis in HER-II uses the entire scintillator volume of 266\,ton, since the above-mentioned external background does not affect this energy window.

Backgrounds which survive the selection cuts are treated in a Poissonian binned likelihood fit to disentangle them from solar neutrino signals. In the HER-I and HER-II, a fit of the radial distribution of events is performed to separate the $^8$B neutrino signal (uniformly distributed in the scintillator) from the external background. The LER analysis follows a multivariate approach, which simultaneously fits the two (TFC-subtracted and TFC-tagged) energy spectra, the radial, and in Phase-II, also the pulse-shape estimator distributions. The radial fit helps to constrain the external background. The pulse shape distribution in Phase-II is used to constrain residual $^{11}$C($e^+$) in the TFC-subtracted spectrum. The likelihood function is constructed by the multiplication of the likelihoods corresponding to the listed distributions:
\begin{equation}
	\mathcal{L}_{\mathrm{MV}} = 
	\mathcal{L}_{\mathrm{E,sub}}^{\textrm{TFC}}\cdot
	\mathcal{L}_{\mathrm{E,tag}}^{\textrm{TFC}}\cdot
	\mathcal{L}_{\mathrm{R}}\cdot
	\mathcal{L}_{\mathrm{PS}},
	\label{eq:likelihood}
\end{equation}
where the last term was used only in Phase-II. The free parameters of the fit are the rates of solar neutrinos and background components from zero threshold. When needed, some of these can be constrained by multiplicative pull terms (Gaussian or semi-Gaussian) to break the correlation among species having similar spectral shapes. This is, of course, only possible, when there is an independent way to evaluate these rates. This will be discussed in the following Section~\ref{sec:pp_ana} for Phase II and Section~\ref{sec:CNO_ana} for Phase-III analyses. A summary of these conditions is shown in the last column of Table~\ref{tab:SolAnal}. 

The probability distribution functions (PDFs) used in the fit for signal and backgrounds are typically obtained by the Geant-4 based MC simulation (Section~\ref{subsec:calib-MC})~\footnote{The only exception is the pulse shape PDF of positrons used in Phase-II analysis that is based on a very pure data sample of $^{11}$C selected with the TFC method tuned for this purpose.}. This MC-based method has the advantage that the detector response is automatically taken into account by the simulation. The main disadvantage is that an extensive analysis campaign is needed in order to evaluate the systematic uncertainties related to the imprecision of the MC-based PDFs.  In addition, this approach cannot adjust for eventual changes that might appear in the detector. An alternative is the so-called analytical fit of the energy spectra used in the Phase-II analysis~\cite{NusolAfterNature}. Here, the detector response is represented by analytical functions, with parameters such as the effective light yield, non-linearity of the energy scale, resolution and non-uniformity parameters. In the fit, some of these parameters are kept free and some are constrained or fixed. Thus, this approach is more flexible to adjust for changes in the detector and does not require to evaluate systematic uncertainty towards the parameters left free in the fit. The disadvantage of this approach is the usage of several fit parameters, which makes this approach generally more prone to correlations.

\subsection{Spectroscopy of \emph{pp}-chain neutrinos}
\label{sec:pp_ana}

This section is dedicated to the latest Borexino analysis of \emph{pp} chain solar neutrinos~\cite{PPchainNature,ImprovedB8,NusolAfterNature}. After the discussion of the main principles of this analysis in the previous section, the particularities of the analysis strategy of \emph{pp} chain solar neutrinos is presented in Section~\ref{sec:ana_strat_pp} and the results in Section~\ref{sec:res_ppchain}. The discussion of their implications is presented in Section~\ref{sec:implications}. Observation of the seasonal modulation of $^7$Be neutrino rate -- a direct evidence of its solar origin, is briefly presented in Section~\ref{sec:be7_seasonal}. 

\subsubsection{Analysis strategy}
\label{sec:ana_strat_pp}

 As mentioned in the previous section, the LER and HER follow different analysis strategies. These are explained below.
 
\paragraph{\emph{LER Analysis}}

\begin{figure}[t]
\centering
\subfigure{\includegraphics[width = 0.46\textwidth]{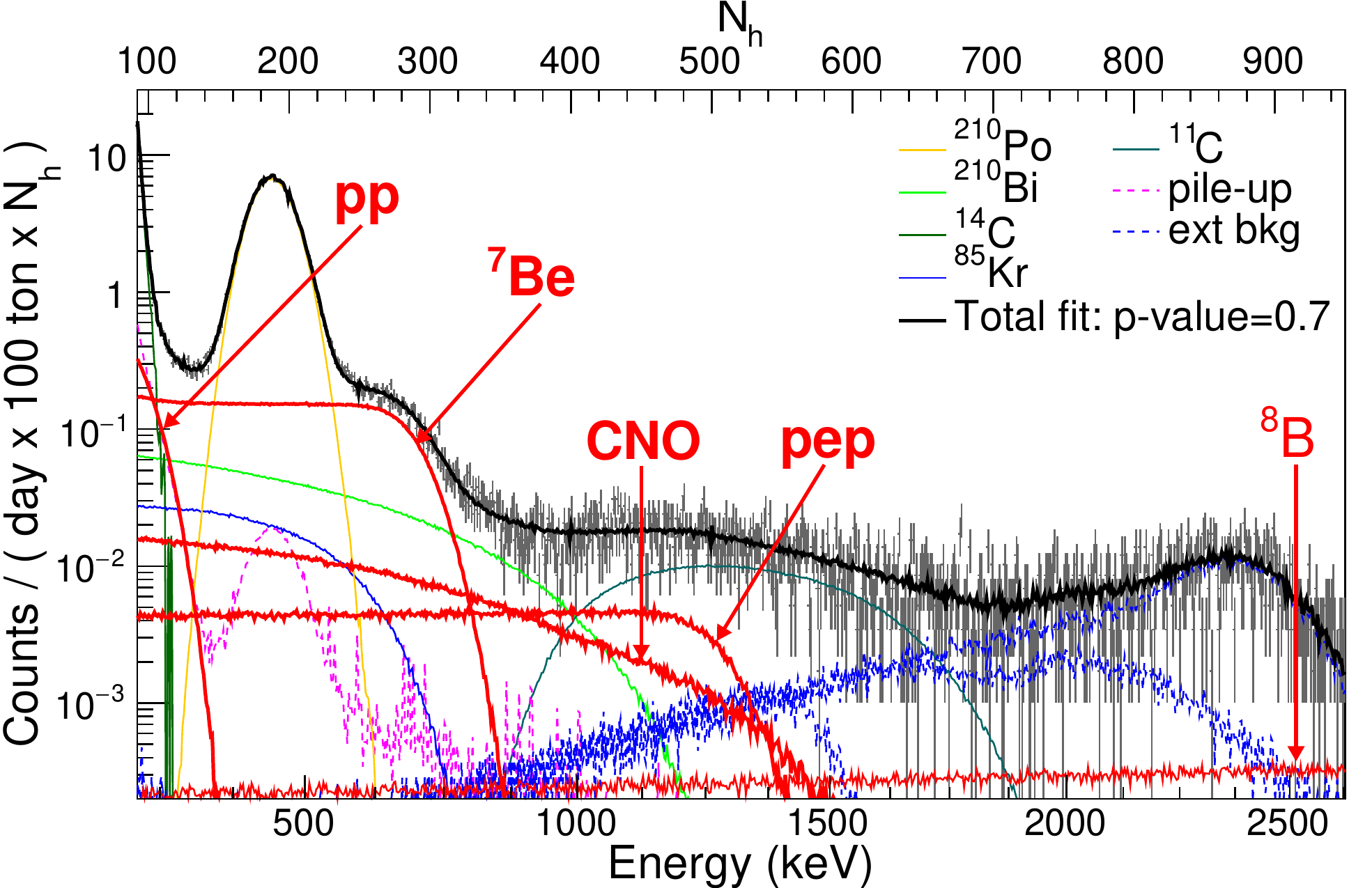}}
\subfigure{\includegraphics[width = 0.46\textwidth]{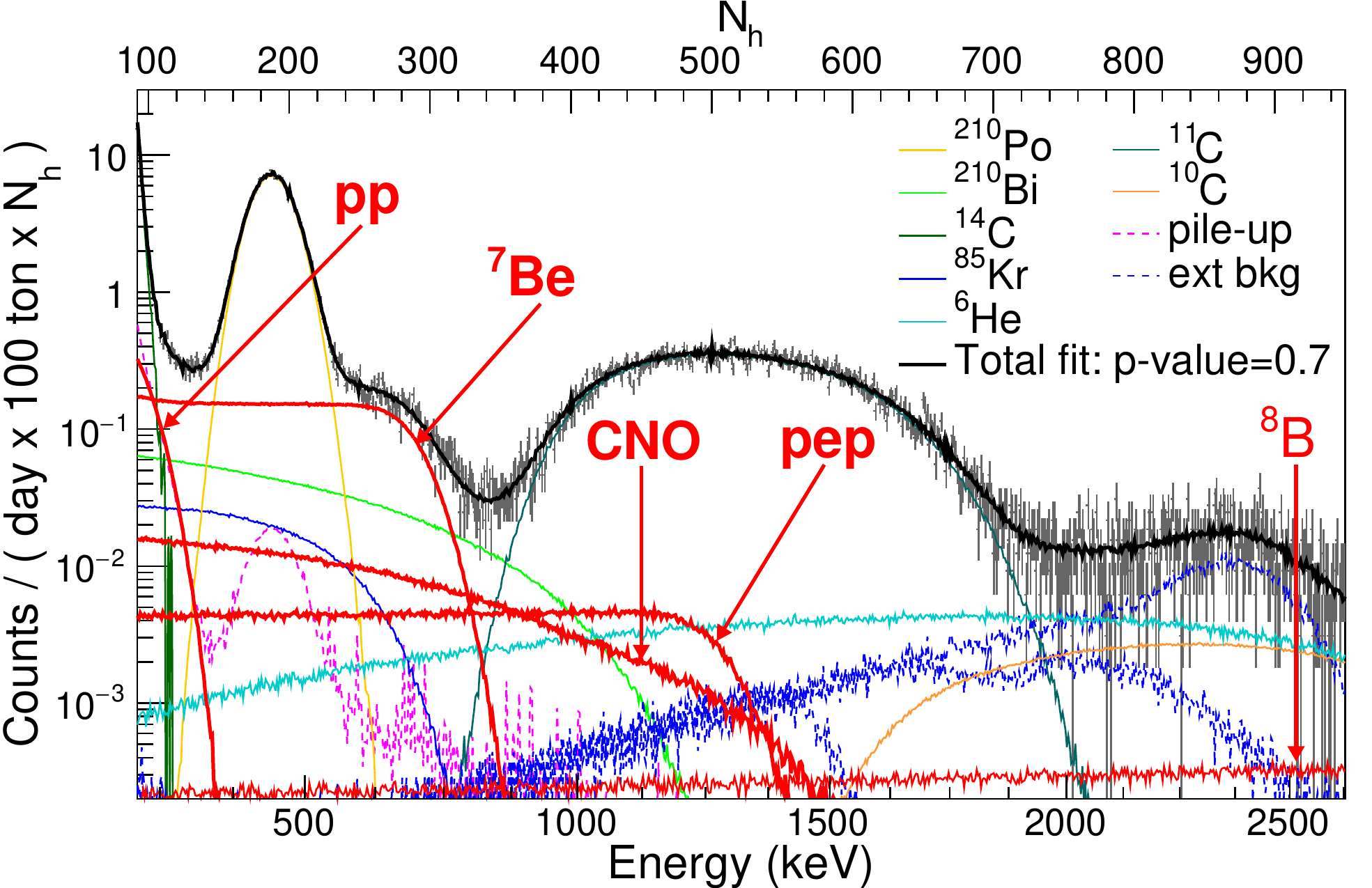}}
\subfigure[]{\includegraphics[width = 0.46\textwidth]{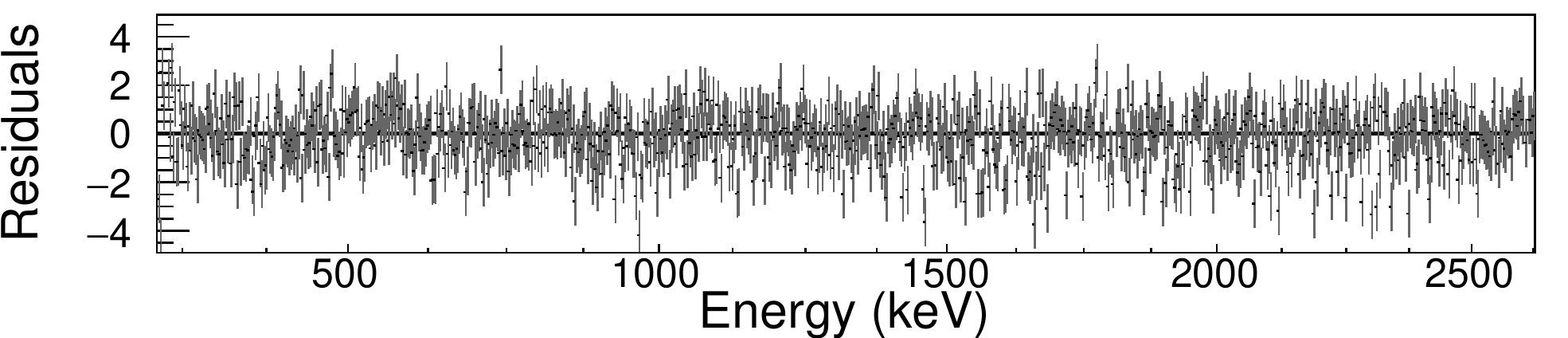}}
\subfigure[]{\includegraphics[width = 0.46\textwidth]{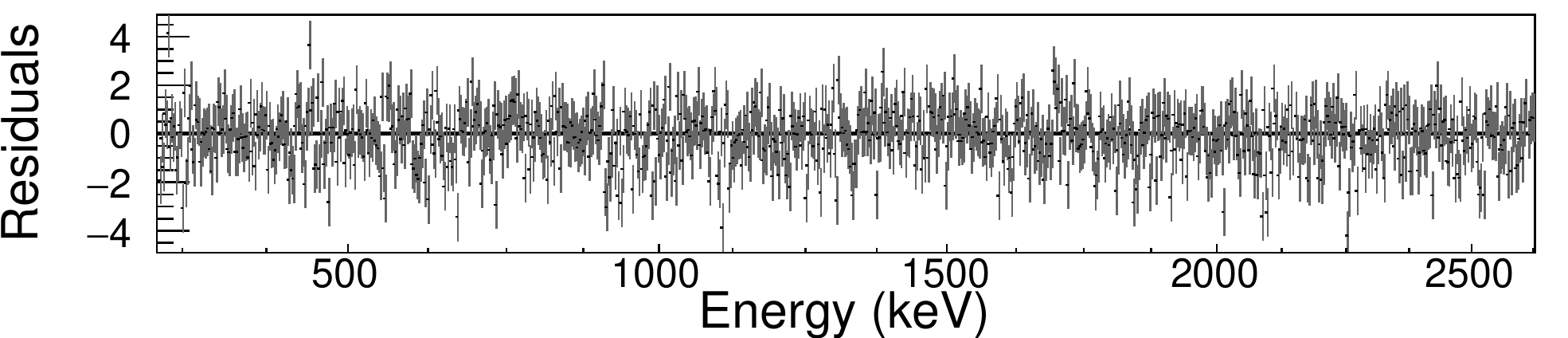}}
\subfigure[]{\includegraphics[width = 0.46\textwidth]{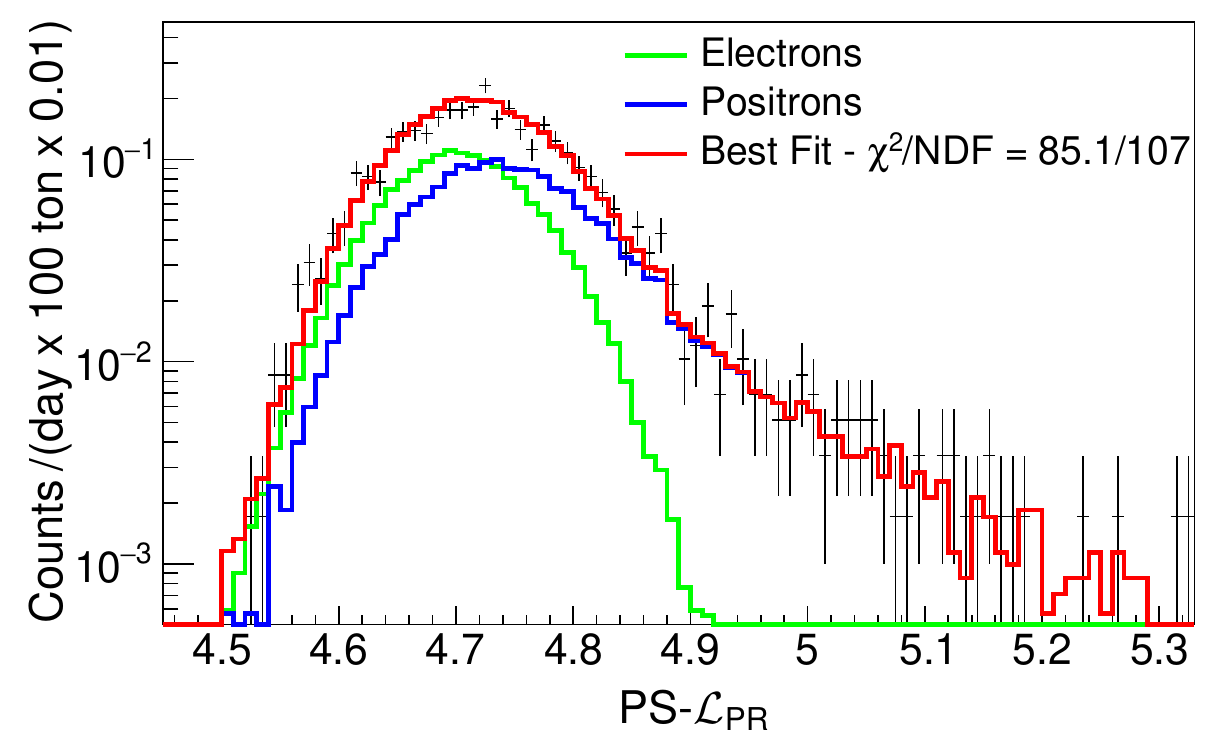}}
\subfigure[]{\includegraphics[width = 0.46\textwidth]{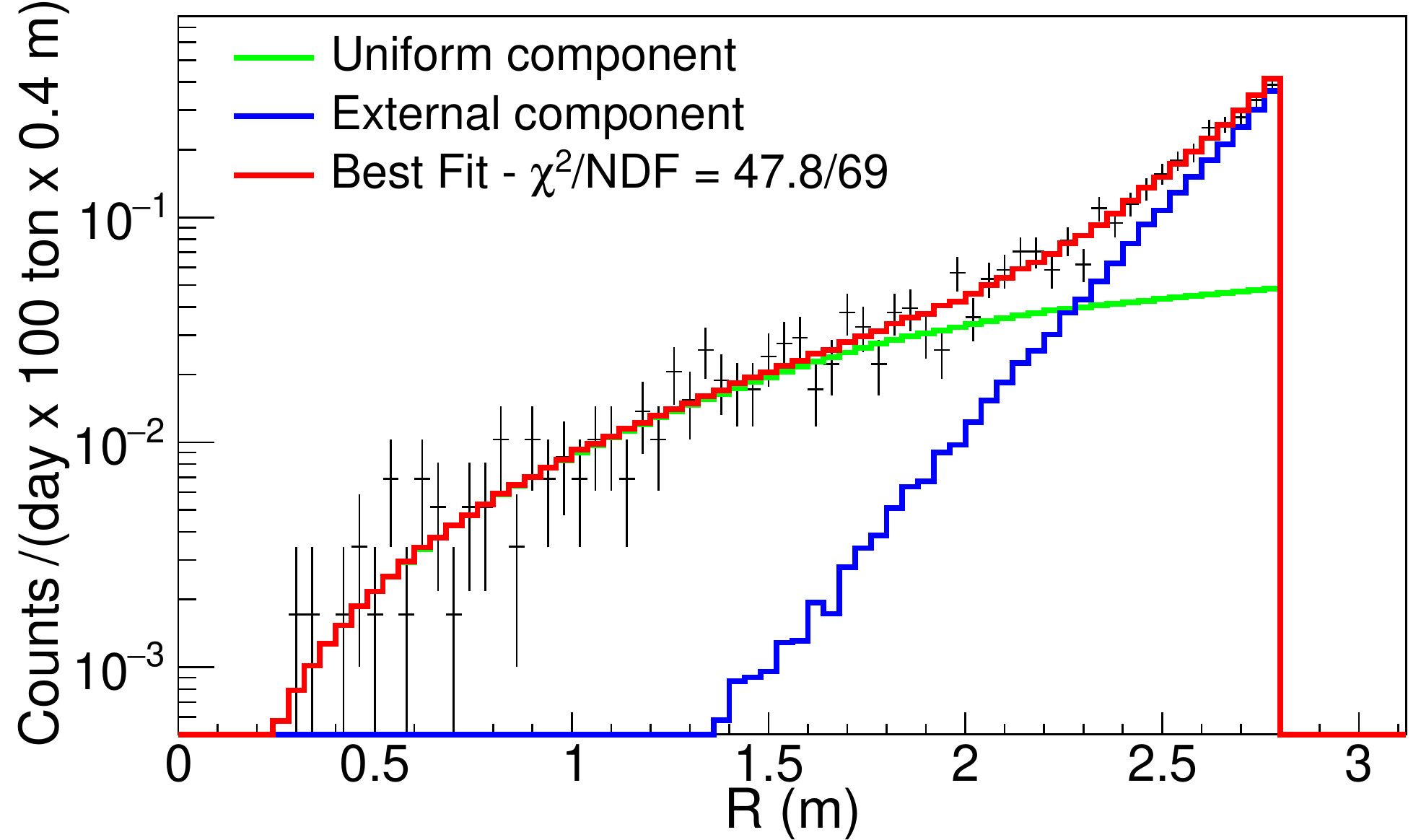}}
\caption{Multivariate fit of Borexino Phase-II data in LER, using MC-based method and $N_{h}$ energy estimator. The solar neutrino species are shown in red. The total fit $p$-value is $0.7$. (a) Spectral fit of the TFC-subtracted  ($^{11}C$ depleted) energy spectrum. (b) Spectral fit of the TFC-tagged ($^{11}C$ enriched) energy spectrum. (c) Fit of the pulse shape variable. (d) Fit of the radial distribution of events. From~\cite{NusolAfterNature}. \label{fig:fit_LER_Phase2}} 
\end{figure} 

The goal of this analysis is to measure \emph{pp}, $^{7}$Be, and \emph{pep} neutrino interaction rates with the same multivariate fit in the energy interval from 0.19 to 2.93\,MeV. Due to the different energy interval of these three neutrino species, the extraction of each species faces different challenges. In order to make it easier to follow this section, we exemplify the spectral fit in Figure~\ref{fig:fit_LER_Phase2}.

For low energy \emph{pp} neutrinos, there is a high correlation with the irreducible $^{14}$C background. Nevertheless, the $^{14}$C rate can be independently determined from the fit of the energy spectrum of the second clusters~\cite{BxPP2014}, i.e., events randomly falling in the last part of the 16\,$\mu$s long data acquisition window triggered by the preceding event (first cluster). The obtained $^{14}$C rate is $(40\pm2)$\,Bq/100\,ton and this value was used to constrain it in the fit. 

The other challenge for \emph{pp} solar neutrinos are pile-up events, i.e., events happening so close in time to each other, that they are reconstructed as single events. In this case, events occurring far away from each other in space, even out of the FV, can be erroneously reconstructed in the FV. Pile-up is dominated by the overlap of $^{14}$C+$^{14}$C, but non-negligible contributions arise also from the pileup between external background with either $^{14}$C or $^{210}$Po. Pile-up is treated using the following two methods described in~\cite{BxPP2014,BxMCpaper}. In one case, {\emph{synthetic pile-up spectrum}} is constructed, starting from real data or MC, that is used as an additional spectral component, fully constrained  in shape and rate during the fit. In the other case, all spectral components are convoluted with a randomly acquired spectrum, i.e. with events acquired with a solicited, external trigger. Thus, in this approach, all energy PDFs are slightly deformed (with the dominant change only on $^{14}$C spectrum) and no additional component is added in the spectral fit.

The Compton-like edge of the $^{7}$Be neutrino signal is a clearly visible feature in the spectrum, see Figure~\ref{fig:fit_LER_Phase2}. This spectral feature makes the fit relatively easy, even if there is a correlation with $^{85}$Kr and $^{210}$Bi backgrounds.

The measurement of \emph{pep} neutrinos is complicated by the presence of $^{11}$C background, that is treated by the TFC technique discussed in Section~\ref{sec:solar_nu_ana}. It is the measurement of \emph{pep} neutrinos that required the multivariate fit approach. In addition, there is a strong correlation between \emph{pep}, $^{210}$Bi, and CNO spectral shapes. Thus, in order to break this degeneracy, the CNO rate is constrained in the fit to the SSM prediction including MSW-LMA oscillations.
The analysis is repeated for both HZ-SSM and LZ-SSM, with the expected CNO rate of $(4.92\pm 0.55)$\,cpd/100\,ton and $(3.52\pm 0.37)$\,cpd/100\,ton, respectively. In case of different results, these are quoted separately.

The $^{8}$B neutrino rate does not affect the LER analysis and is always fixed to its SSM prediction, while the \emph{hep} neutrino rate is simply neglected due to its fully negligible expected rate.

\paragraph{\emph{HER Analysis}}

\begin{figure}[t]
\centering
\subfigure[]{\includegraphics[width = 0.45\textwidth]{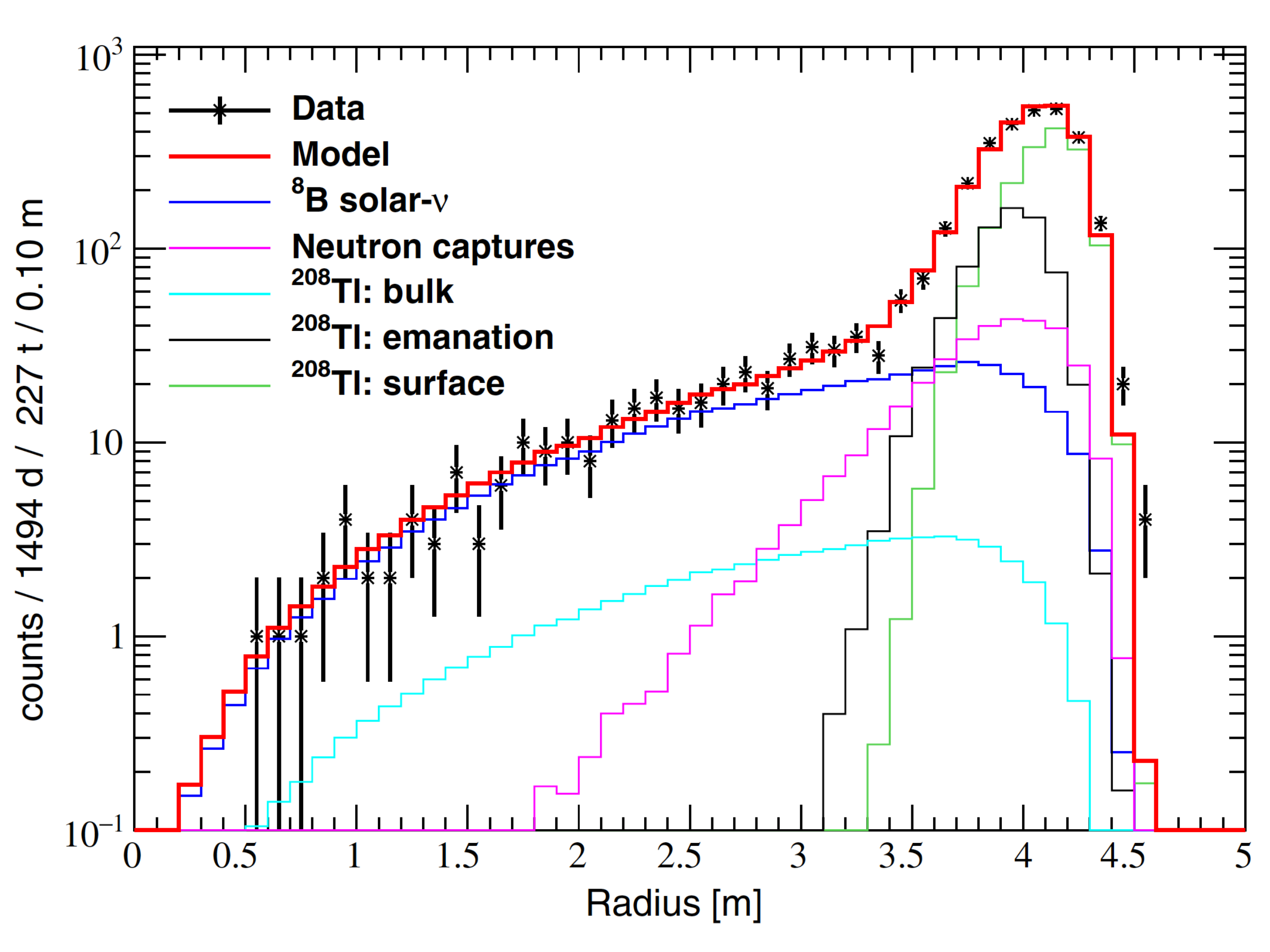}\label{fig:her1_p2}}
\subfigure[]{\includegraphics[width = 0.45\textwidth]{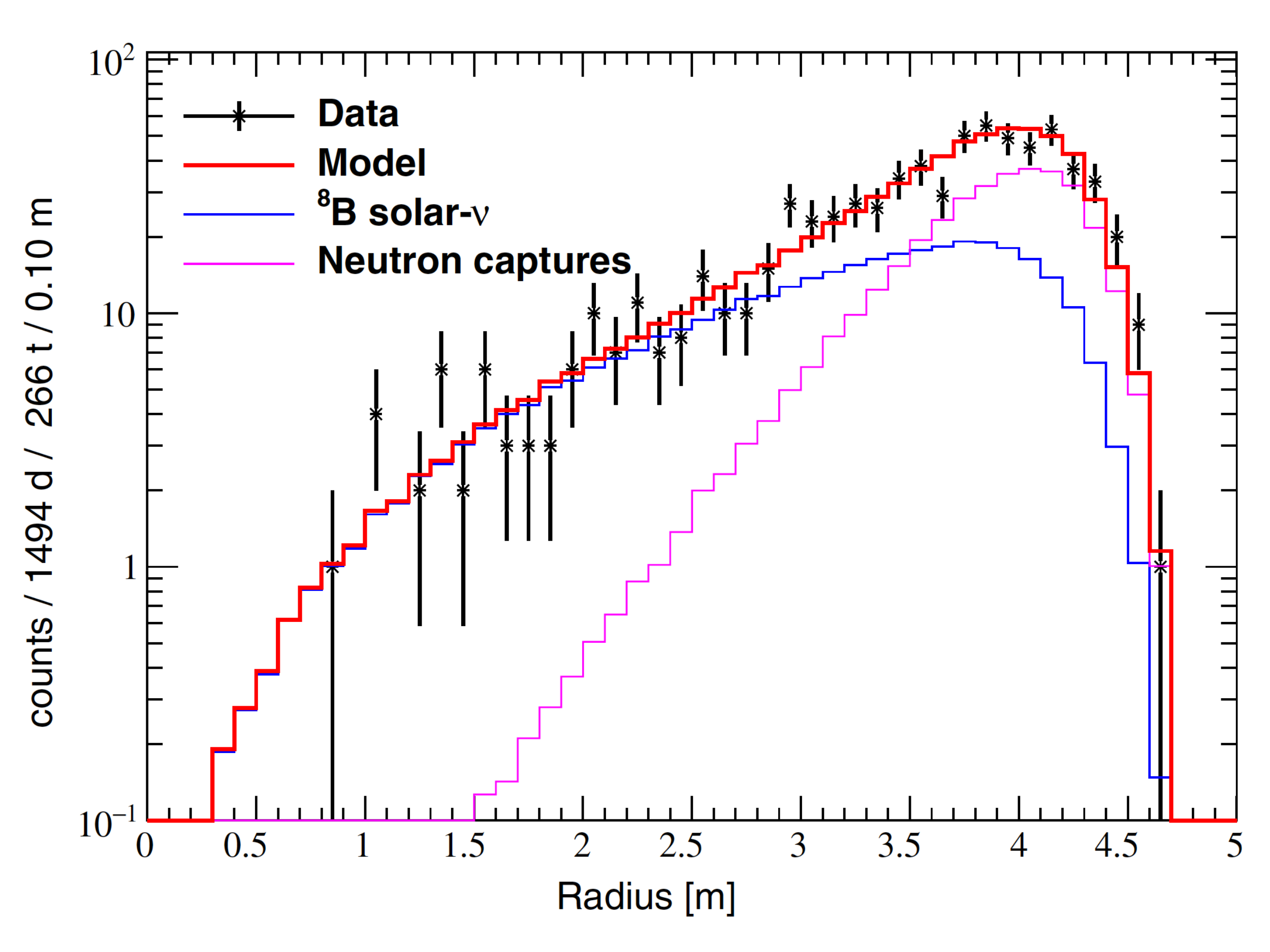}\label{fig:her2_p2}}
\caption{ (a) Fit of the event radial distribution in the HER-I range,
[1650, 2950] p.e., corresponding to [3.2,5.7]\,MeV. (b) Fit of the event radial distribution in the HER-II range, [2950, 8500] p.e., corresponding to [5.7,16.0]\,MeV. From \cite{ImprovedB8}.\label{fig:b8_radial_fits}} 
\end{figure} 

The strategy to extract the $^8$B solar neutrino rate is based on  radial fits in the HER-I and HER-II, that are shown in Figure~\ref{fig:b8_radial_fits}. In both fits, the dominant uniform contribution is from $^8$B solar neutrinos. Some residual uniform background due to muons, cosmogenic isotopes, and $^{214}$Bi decays surviving the cuts is very small. This is estimated following the procedure in~\cite{BxB82010}, and constrained in the fit. Only in HER-I there is an additional uniform background from bulk $^{208}$Tl(\emph{Q}\,=\,5\,MeV, $\beta^-, \gamma$), which comes from the residual $^{232}$Th contamination of the liquid scintillator and is constrained in the fit to the value based on the $^{212}$Bi–$^{212}$Po ($\beta$ + $\alpha$) fast delayed coincidences. External $^{208}$Tl contamination contributes to the HER-I with two distinct components: one from contamination directly on the inner vessel surface, and another from decays of nuclei that have recoiled from the inner vessel into the liquid scintillator or originated from the volatile progenitor of $^{208}$Tl, $^{220}$Rn, which has emanated from the nylon. The rates of both components are left free to vary in the radial fit. Finally, HER-I and HER-II are also polluted by $\gamma$-rays following the capture of radiogenic neutrons produced via ($\alpha$, n) or spontaneous fission reactions of $^{238}$U, $^{235}$U, and $^{232}$Th in the Stainless Steel Sphere (SSS) and PMTs. This rate is also a free parameter in the fit. The neutron captures are the only background in the HER-II analysis, as there are no naturally long-lived isotopes above $5$\,MeV.

\subsubsection{Results on \emph{pp}, \emph{pep}, $^{7}\textrm{Be}$, and $^{8}\textrm{B}$ neutrinos} 
\label{sec:res_ppchain}

\begin{table}[t!]
\centering
\caption{Results of the Borexino Phase-II~\cite{PPchainNature} and Phase-III~\cite{CNOpap} solar neutrino analyses. The rates and fluxes are integral values without any threshold; the first error is statistical, the second systematic. The rate-to-flux conversion assumes neutrino flavour conversion~\cite{deHolanda:2004fd} with the neutrino oscillation parameters from~\cite{Capozzi2018}. The last column shows the fluxes as predicted by the HZ- and LZ-SSM~\cite{B16SSM} (see Table~\ref{tab:SolarFluxes}). The result for $^7$Be neutrinos contains the ground and excited state lines as shown in Figure~\ref{fig:pp_cno_sketch}. The fluxes of \emph{pp}, $^7$Be, \emph{pep}, CNO, $^8$B, and \emph{hep} neutrinos are normalised to $10^{10}$, $10^{9}$, $10^{8}$, $10^{8}$, $10^{6}$, and $10^{3}$ (see Exp factor definitions in Table~\ref{tab:SolarFluxes}), respectively.} \label{tab:phase2-3-res}
\scalebox{0.92}{
    \begin{tabular}{llll}
    \toprule
     Solar $\nu$ & Rate [cpd/100\,ton] & Flux (cm$^{-2}$s$^{-1}$) & SSM Flux (cm$^{-2}$s$^{-1}$) \\
     \midrule
     Phase-II~\cite{PPchainNature,NusolAfterNature} &  &  &  \\
      12/2011 - 05/2016   &  &  &  \\
    \midrule
    \emph{pp} & 134 $\pm$ 10$^{+6}_{-10}$ & (6.1 $\pm$ 0.5$^{+0.3}_{-0.5}$)  & 5.98(1.0 $\pm$ 0.006)  \text{(HZ)} \\ [3pt]
    & & & 6.03(1.0 $\pm$ 0.006)  \text{(LZ)} \\ [3pt]
    $^{7}$Be & 48.3 $\pm$ 1.1$^{+0.4}_{-0.7}$ & (4.99 $\pm$ 0.11$^{+0.06}_{-0.08}$) & 4.93(1.0 $\pm$ 0.06) \text{(HZ)} \\ [3pt]
    & & & 4.50(1.0 $\pm$ 0.06)  \text{(LZ)} \\ [3pt]
     \emph{pep} (HZ) & 2.43 $\pm$ 0.36$^{+0.15}_{-0.22}$ & (1.27 $\pm$ 0.19$^{+0.08}_{-0.12}$) & 1.44(1.0 $\pm$ 0.01) \text{(HZ)} \\ [3pt]
    & & & 1.46(1.0 $\pm$ 0.009)  \text{(LZ)} \\ [3pt]
    \emph{pep} (LZ) & 2.65 $\pm$ 0.36$^{+0.15}_{-0.24}$ & (1.39 $\pm$ 0.19$^{+0.08}_{-0.13}$)  & 1.44(1.0 $\pm$ 0.01)  \text{(HZ)} \\ [3pt]
    & & & 1.46(1.0 $\pm$ 0.009)  \text{(LZ)} \\ [3pt]
    CNO & $\textless$8.1 (95\% C.L.) & $\textless$7.9  (95\% C.L.) & 4.88(1.0 $\pm$ 0.11)  \text{(HZ)} \\ [3pt]
    & & & 3.51(1.0 $\pm$ 0.10)  \text{(LZ)} \\ [3pt]
    \midrule
    Phase-I + II~\cite{PPchainNature,ImprovedB8} &  &  &  \\
     01/2008 - 12/2016   &  &  &  \\
    \midrule
    $^{8}$B$_{HER-I}$ & 0.136$^{+0.013+0.003}_{-0.013-0.013}$ & (5.77$^{+0.56+0.15}_{-0.56-0.15}$)  & 5.46(1.0 $\pm$ 0.12)  \text{(HZ)} \\ [3pt]
    & & & 4.50(1.0 $\pm$ 0.12)  \text{(LZ)}\\ [3pt]
    $^{8}$B$_{HER-II}$ & 0.087$^{+0.080+0.005}_{-0.010-0.005}$ & (5.56$^{+0.52+0.33}_{-0.64-0.33}$) & 5.46(1.0 $\pm$ 0.12)  \text{(HZ)} \\ [3pt]
    & & & 4.50(1.0 $\pm$ 0.12) \text{(LZ)} \\ [3pt]
    $^{8}$B$_{HER}$ & 0.223$^{+0.015+0.006}_{-0.016-0.006}$ & (5.68$^{+0.39+0.03}_{-0.41-0.03}$) & 5.46(1.0 $\pm$ 0.12)  \text{(HZ)} \\ [3pt]
    & & & 4.50(1.0 $\pm$ 0.12) \text{(LZ)} \\ [3pt]
    \midrule
    Phase-I (part) + II + III (part)~\cite{ImprovedB8} &  &  &     \\
     11/2009 - 10/2017   &  &  &    \\
     \midrule
    \emph{hep} & $\textless$0.002 (90\% C.L.) & $\textless$180  (90\% C.L.) & 7.98(1.0 $\pm$ 0.30)  \text{(HZ)} \\ [3pt]
    & & & 8.25(1.0 $\pm$ 0.12)  \text{(LZ)} \\ [3pt]
    \midrule
     Phase-III~\cite{CNOpap} &  &  &  \\
     07/2016 - 02/2020   &  &  &  \\
     \midrule
     CNO & 7.2 $^{+3.0}_{-1.7}$ & (7.0 $^{+3.0}_{-2.0}$) & 4.88(1.0 $\pm$ 0.11)  \text{(HZ)} \\ [3pt]
    & & & 3.51(1.0 $\pm$ 0.10) \text{(LZ)} \\ [3pt]
    \bottomrule
    \end{tabular}
}
\end{table}

The analysis strategies for the LER and HER were explained in the previous Section. The results from these analyses are discussed below.

\paragraph{\emph{LER Analysis}}

The result of the multivariate MC fit in the LER is shown in Figure~\ref{fig:fit_LER_Phase2}, which illustrates the four fits, namely the TFC-subtracted, TFC-tagged, pulse shape, and radial distributions, as defined in Equation~\ref{eq:likelihood}. The fit results in terms of interaction rates of solar neutrinos in counts per day per $100$\,ton (cpd/100\,ton) are given in Table~\ref{tab:phase2-3-res} including systematic errors. The fact that the fit is repeated with the CNO rate constrained to HZ- and LZ-SSM predictions influences only the resulting \emph{pep} neutrino rate and is thus given separately with the label HZ and LZ. In both cases, the absence of the \emph{pep} reaction in the Sun is rejected with $>5\sigma$ significance, which makes this measurement the  discovery of solar \emph{pep} neutrinos. 

In spite of the remarkable understanding of the detector response throughout the scintillator volume and in a large energy range (Section~\ref{subsec:calib-MC}), an extensive study of the possible sources of systematic errors has been performed. The results of these studies are summarised in Table~\ref{tab:syst_ler_p2}, which lists the various contributions to the systematic error individually for the \emph{pp}, $^7$Be, and \emph{pep} measurements.
The main contribution to the systematic error comes from the fit model, that is, possible residual inaccuracies in the modelling of the detector response (energy scale, uniformity of the energy response, pulse-shape discrimination shape) and uncertainties in the theoretical energy spectra used in the fit. The second source of systematics is related to the fit method, i.e. eventual differences between the MC-based and analytical fit approach. Further systematic effects arise from the choice of the energy estimator, the details of the implementation of the pile-up, different fit energy ranges and binning, the inclusion of an independent constraint on $^{85}$Kr, and the estimation of the FV. This last uncertainty is determined with calibration data, using sources deployed in known positions throughout the detector volume.

\begin{table}[t] 
\centering
\caption{Relevant sources of systematic uncertainties and their contribution to the measured interaction rates of \emph{pp} chain solar neutrinos in the Borexino Phase-II analysis in LER. The Table is from \cite{NusolAfterNature}. Further details are discussed in the text.}
\label{tab:syst_ler_p2}
\begin{tabular}{lcccccc}
\toprule
&\multicolumn{2}{c}{\emph{pp}}& \multicolumn{2}{c}{$^7$Be} & \multicolumn{2}{c}{\emph{pep}}  \\ 
\cline{2-7}							    
Source of uncertainty & $-\%$	& $+\%$	& $-\%$	& $+\%$	& $-\%$	& $+\%$ \\
\midrule

Fit method (AL/MC)                   &-1.2    &1.2              &-0.2         & 0.2     &-4.0  &4.0\\
Choice of energy estimator                        &-2.5   &2.5                 &-0.1        & 0.1     &-2.4 &2.4 \\
Pile-up modeling		                         &-2.5   &0.5                 &0        & 0     &0 & 0 \\
Fit range and binning                                &-3.0   &3.0                &-0.1       & 0.1      &1.0 &1.0 \\
Fit models (see text)                                 &-4.5  & 0.5                     &-1.0     &  0.2    &-6.8  &2.8\\
Inclusion of  $^{85}$Kr constraint 	                       &-2.2   &2.2               &0       & 0.4      &-3.2  &0\\
Live Time                                       &-0.05     &0.05           &-0.05    & 0.05     &-0.05  &0.05\\
Scintillator density                         &-0.05    &0.05               &-0.05      & 0.05     &-0.05  &0.05\\
Fiducial volume                                        &-1.1    &0.6              &-1.1       &  0.6       &-1.1 &0.6 \\
\midrule               
Total systematics ($\%$)		&-7.1	&4.7                   &-1.5      &0.8     &-9.0 &5.6\\
\bottomrule
\end{tabular}
\end{table}

\paragraph{\emph{HER Analysis}}

The radial fits in HER-I and HER-I ranges to obtain the $^8$B neutrino interaction rate are shown Figure~\ref{fig:b8_radial_fits}. The resulting $^8$B interaction rates are given in Table~\ref{tab:phase2-3-res}. The most important systematic uncertainties arise from the determination of the target mass, that is complicated by the presence of the small leak in the inner vessel. The evolution of the scintillator mass is monitored on a week-by-week basis, by studying the inner vessel shape, which is obtained from the spatial distribution of its surface contamination. Additional sources of systematic error include the energy scale uncertainty, and the application of the $z$-cut in HER-I. The uncertainties from the live-time determination and the knowledge of the scintillator density are almost negligible.

\begin{table}[t] 
\centering
\caption{Relevant sources of systematic uncertainties and their contribution to the measured interaction rate of $^8$B solar neutrinos in the Borexino Phase-I+II analysis in HER-I, HER-II, and combined HER = HER-I + HER-II energy ranges.}
\label{tab:syst_b8_p2}
\begin{tabular}{lccc}
\toprule
Source					&HER-I	&HER-II & HER  \\
&$\sigma[\%]$	&$\sigma[\%]$ &$\sigma[\%]$	\\
\midrule
Active Mass		            	& 2.0  & 2.0 & 2.0 \\
Energy Scale	            	& 0.5  & 4.9 & 1.7 \\
$z$-cut 		            	& 0.7  & 0.0 & 0.4 \\
Live time 			            & 0.05  & 0.05 & 0.05 \\
Scintillator density 			& 0.05  & 0.05 & 0.05 \\
\midrule
Total                           & 2.2   & 5.3 & 2.7\\
\bottomrule
\end{tabular}
\end{table}

To complete the \emph{pp} chain analysis, a search for \emph{hep} neutrinos has been performed. Its flux expectation is two orders of magnitudes smaller than that of $^8$B neutrinos. Even if the endpoint energy for \emph{hep} neutrinos is high, it falls in the energy region containing cosmogenic $^{11}$Be decays and $^8$B neutrinos. Taking into account the whole dataset corresponding to an exposure of 0.745\,kt\,$\times$\,yr, i.e. from November 2009 until October 2017, and considering only the energy interval of 11-20\,MeV, $(12.8\pm2.3)$ events were found, consistent with the background expectation. An upper limit of $0.002$ cpd/100\,t at 90\%\,C.L. has been set. The analysis periods used for $^8$B and \emph{hep} neutrinos overlap but, they are different (see Tables~\ref{tab:SolAnal} and~\ref{tab:phase2-3-res}).

\subsubsection{Implications for solar and neutrino physics}
\label{sec:implications}

The measured interaction rates of solar neutrinos, as discussed in the previous Section, can be used to test our understanding of both the Sun and the basic neutrino properties. Assuming that the physics of neutrino interactions and oscillation are known, the measured rates can be converted to neutrino fluxes, to be then compared individually with the HZ and LZ SSM predictions, which is important for constraining the solar metallicity. In addition, one can quantify the relative intensity of the two primary terminations of the \emph{pp} chain (\emph{pp-I} and \emph{pp-II} in Figure~\ref{fig:pp_cno_sketch}) and evaluate the solar neutrino luminosity. On the other hand, assuming the SSM predictions for the neutrino fluxes, one can evaluate the electron neutrino survival probability $P_{ee}$ for different energies and compare them with the standard prediction of the 3-flavour neutrino oscillations including the MSW effect. The following paragraphs discuss these points in more detail.

\paragraph{\emph{pp-chain solar neutrino fluxes }}

Considering solar neutrino oscillations, the expected
neutrino interaction rate in Borexino $R_{\nu}$ is~\cite{LongPaperPhaseI}:
\begin{equation}
    R_{\nu} = N_{e}\Phi_{\nu}
    \int{dE_{\nu}dT_{e}\frac{d\lambda}{dE_{\nu}}}[\frac{d\sigma_{e}(E_{\nu},T_e)}{dT_{e}}\,P_{ee}(E_{\nu}) + \frac{d\sigma_{\mu,\tau}(E_{\nu},T_e)}{dT_{e}}\,(1-P_{ee}(E_{\nu}))].
    \label{eq:master_formula_rate}
\end{equation}
where $N_e$ is the number of target electrons (see Section~\ref{subsec:detection}), $\Phi_e$ is the solar neutrino flux, $d\lambda/dE_{\nu}$ is the differential energy spectrum of solar neutrinos, $P_{ee}$ is the electron neutrino survival probability (see Section~\ref{sec:nuosc_msw} and \cite{deHolanda:2004fd}), and $d\sigma_{e,\mu,\tau}/dT_{e}$ are the differential cross sections for the scattering reaction discussed in Section~\ref{subsec:detection}. The spectrum of solar neutrinos, normalised according to SSM predicted fluxes (last column of Table~\ref{tab:SolarFluxes}) is shown in the bottom part of Figure~\ref{fig:pp_cno_sketch}. The cross-sections of elastic scattering for different neutrino flavours were discussed in Section~\ref{subsec:detection}. We remind that Borexino has no sensitivity to distinguish between the shapes of recoiled electron spectra from $\nu_e$ and $\nu_{\mu,\tau}$. However, for the same neutrino energy, the $\sigma_e$ is about 4-5 times larger than $\sigma_{\mu,\tau}$. Thus, in order to convert the measured interaction rate to flux, it is important to know the relative proportion of the flavours in the measured flux and therefore, $P_{ee}$. This conversion is relatively simple for mono-energetic neutrinos, such as $^7$Be and \emph{pep}. For solar neutrinos with a continuous energy spectrum and analysis performed in a restricted energy interval of scattered electrons, the situation is more complicated. One has to take into account the energy dependent detector response and assume energy dependence of $P_{ee}$. This procedure is described in Appendix of~\cite{ImprovedB8}. The solar neutrino fluxes converted following this procedure are given in the third column of Table~\ref{tab:SolarFluxes}. In the particular case of $^8$B neutrinos, it is useful to also provide a flux assuming no-flavour conversion: $2.57^{+0.17+0.07}_{-0.18-0.07}\times10^6$\,cm$^{-2}$\,s$^{-1}$. This conversion assumes that all interacting neutrinos are of electron flavour, which has a higher probability to interact with respect to other neutrino flavours. Thus, in order to comply with the measured rate, a smaller flux is sufficient. This number does not depend on $P_{ee}$ and is therefore very useful to compare the results among different experiments. The Borexino result is compatible with the high-precision result of Super-Kamiokande $2.345\pm0.014$\,(stat.)$\pm{+0.036}$\,(syst.)$\times10^6$\,cm$^{-2}$\,s$^{-1}$~\cite{SK16_solar,ImprovedB8}.

\paragraph{\emph{Metallicity}}

\begin{figure}[t]
\centering
    \includegraphics[width=0.6\textwidth]{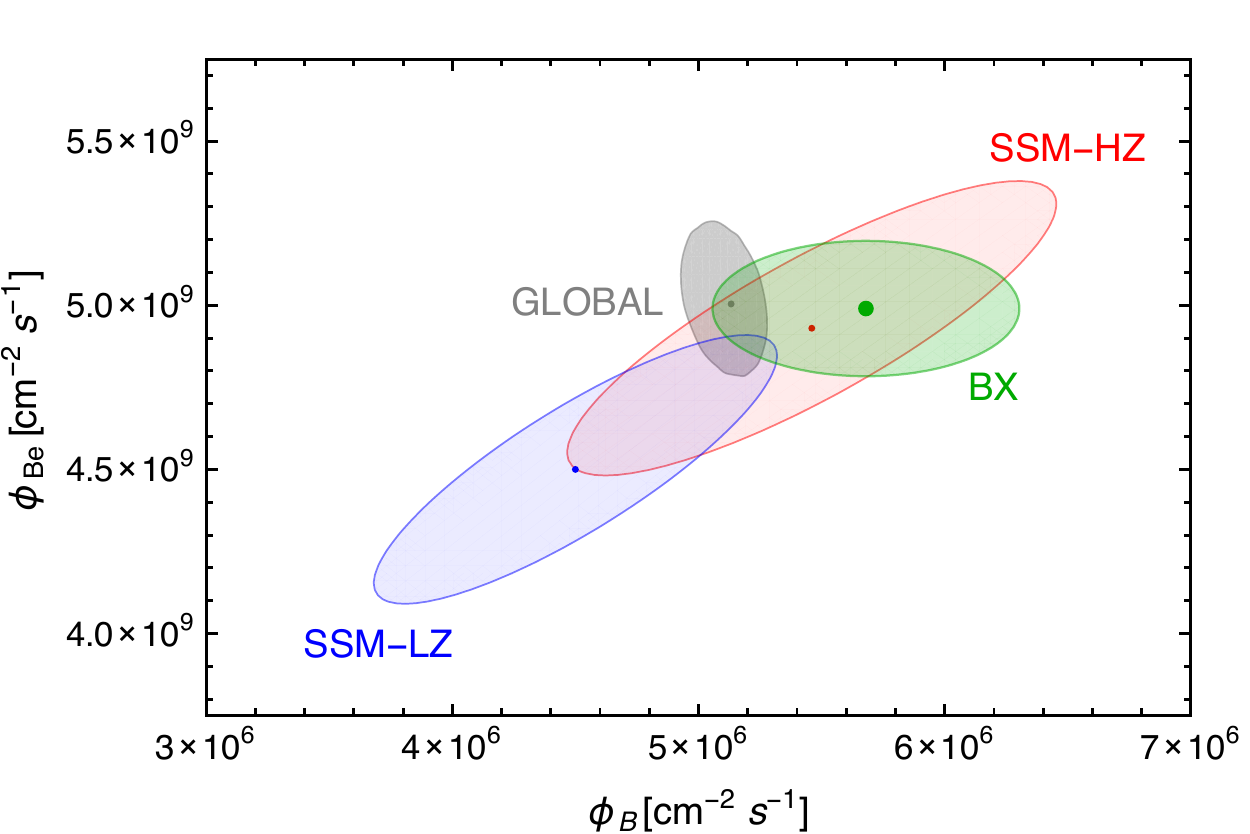}
    \caption{Borexino results for $^7$Be and $^8$B neutrino fluxes (green dot and area) compared to HZ-SSM (red area) and LZ-SSM (blue area) predictions. The allowed contours shown in gray are obtained by combining the Borexino results with all solar and KamLAND data in a global analysis leaving free the oscillation parameters $\theta_{12}$ and $\Delta{m}_{12}^{2}$. All contours represent $68.27\%$ C.L. From~\cite{PPchainNature}.}
\label{fig:metal}
\end{figure}

As discussed in Section~\ref{sec:ssm_metal}, the SSM prediction for 
solar neutrino fluxes depends on the assumption of solar metallicity. For \emph{pp} chain neutrinos, the difference between the HZ- and LZ-SSM predictions is largest for $^7$Be and $^8$B neutrinos, 8.9\% and 17.6\%, respectively, as it is reported in Table\,$\ref{tab:SolarFluxes}$. When comparing the measured $^7$Be and $^8$B neutrino fluxes with these theoretical SSM predictions, as shown in Figure~\ref{fig:metal}, it is possible to evaluate the agreement between the data and the HZ- and LZ-SSM predictions. Note that the errors in the Borexino measurements are in both cases smaller than the theoretical uncertainties.

The Borexino results are compatible with the temperature profiles predicted by both HZ- and LZ-SSMs. However, the $^7$Be and $^8$B solar-neutrino fluxes measured by Borexino provide an interesting hint in favour of the HZ-SSM prediction. A frequentist hypothesis test based on a likelihood-ratio test statistics (HZ versus LZ) was performed by computing the probability distribution functions with a MC approach. Assuming HZ to be true, the LZ is disfavoured at 96.6\% C.L.  Additionally, a Bayesian analysis gives a Bayes factor of 4.9 showing a mild preference towards HZ-SSM. However, this hint weakens when the Borexino data are combined with data of all the other solar neutrino experiments + KamLAND reactor antineutrino data.

\paragraph{{\emph{pp-chain branches}}}

From the measured solar neutrino fluxes, it is possible to evaluate the ratio $R_{I/II}$ between the $^{3}$He–$^4$He and the $^3$He–$^3$He fusion rates, which quantifies the relative intensity of the two primary terminations of the \emph{pp} chain, a critical probe of the solar fusion. Neglecting the $^8$B neutrino contribution, this ratio can be expressed as:
\begin{equation}
R_{I/II} = \frac{^3\textrm{He}+^4\textrm{He}}{^3\textrm{He}+^3\textrm{He}} = 2\cdot \frac{\Phi(^7\textrm{Be})}{\Phi(pp)-\Phi(^7\textrm{Be})}.
\end{equation}
The result obtained with Borexino is $R_{I/II}=0.1780^{+0.027}_{-0.023}$. This value is consistent with both the HZ- and LZ-SSM predictions, $0.180\pm0.011$ and $0.161\pm0.010$, respectively.

\paragraph{\emph{Solar luminosity and thermal stability}}

The neutrino fluxes determined experimentally can be used to derive the total power generated by nuclear reactions in the Sun’s core~\cite{Bergstrom_2016}. Using the measured Borexino fluxes from Table~\ref{tab:SolarFluxes}, the obtained luminosity $L_{\odot}=(3.89^{+0.35}_{-0.42})\times10^{33}$ erg s$^{-1}$ is in agreement with the luminosity calculated using the photon output~\cite{EncycPlanetNature34,SunIrradianceNature35}, $L_{\odot}=(3.846\pm0.015)\times10^{33}$ erg s$^{-1}$. This is a robust and direct evidence of the nuclear origin of the solar power. While neutrinos provide a real time picture of the solar core, it takes around $10^5$ years for the photons to reach the solar photosphere, from where they are free to escape. The comparison of the two luminosities then also proves that the Sun has been in thermodynamic equilibrium over this timescale.

\paragraph{\emph{Electron neutrino survival probability }}

\begin{figure}[t]
\centering
    \includegraphics[width=0.6\textwidth]{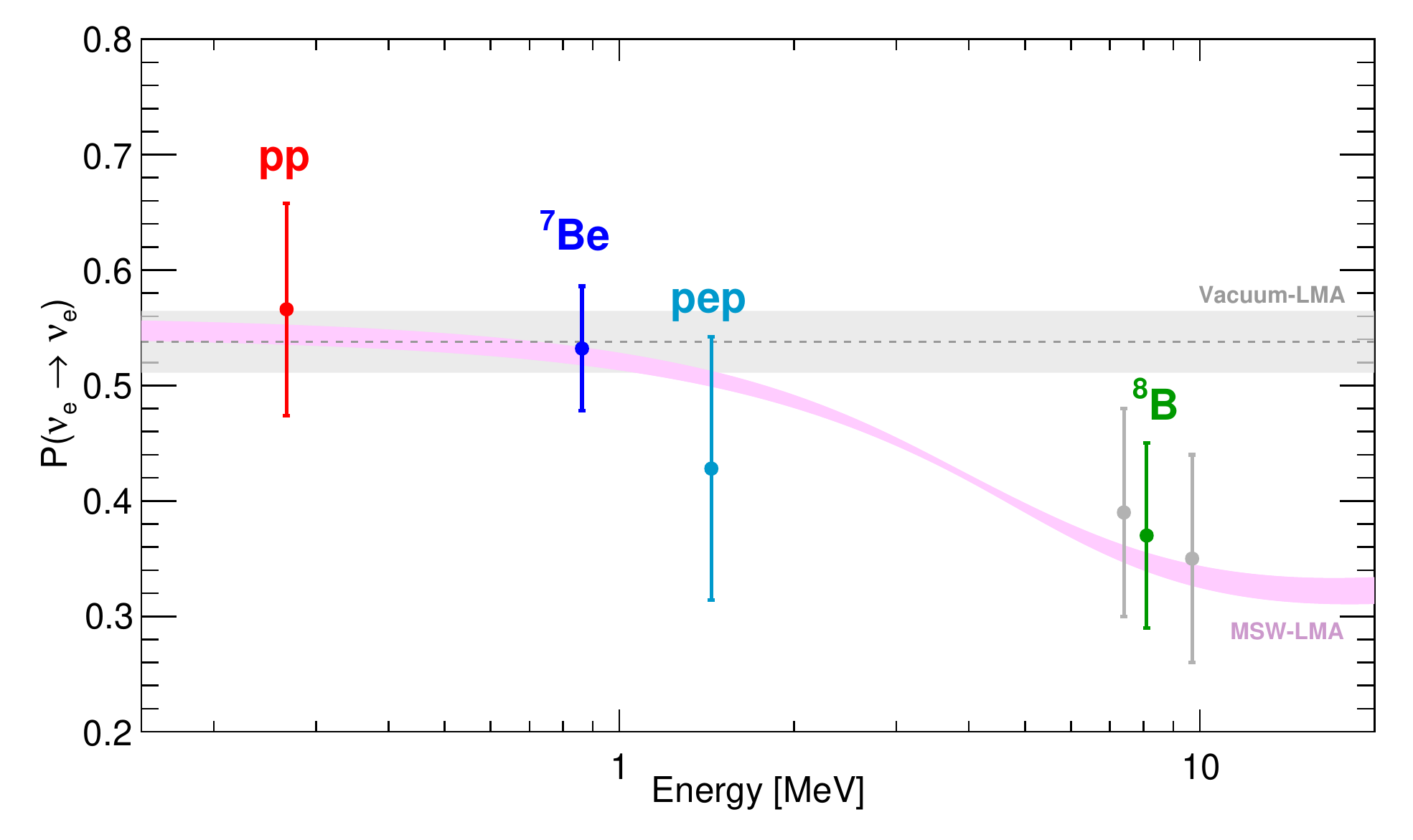}
    \caption{Electron neutrino survival probability $P_{ee}$ as a function of neutrino energy. The data points show the Borexino results, obtained assuming HZ-SSM flux predictions~\cite{B16SSM}, for \emph{pp} (red), $^7$Be (blue), \emph{pep} (cyan), and $^8$B (grey for the separate HER-I and HER-II sub-ranges and green for the combined HER range). The error bars include experimental and theoretical uncertainties. The gray band corresponds to the $P_{ee}$ predicted by the Vacuum-LMA scenario, while the pink band represents the MSW-LMA solution. The width of the bands is $\pm1\sigma$. More details in text. From~\cite{PPchainNature}.}
    \label{fig:pee_p2}
\end{figure}

The measured interaction rates of solar neutrinos can be used to extract the electron neutrino survival probability at different energies. This can be done using already discussed  Equation~\ref{eq:master_formula_rate}, assuming standard neutrino interactions and, in this case, SSM fluxes. Figure~\ref{fig:pee_p2} shows the extracted $P_{ee}$ as a function of the neutrino energy for each measured solar neutrino species. The obtained neutrino survival probabilities are $P_{ee}$(\emph{pp}, 0.267\,MeV)\,=\,$0.57\pm0.09$, $P_{ee}(^7\textrm{Be}, 0.862\,\textrm{MeV})$\,=\,$0.53\pm0.05$, $P_{ee}$(\emph{pep}, 1.44\,MeV)\,=\,$0.43\pm0.11$, $P_{ee}(^8\textrm{B}_{HER}, 8.1\,\textrm{MeV})$\,=\,$0.37\pm0.08$, $P_{ee}(^8\text{B}_{HER-I}, 7.4\,\textrm{MeV})$\,=\,$0.39\pm0.09$, and \\ $P_{ee}(^8\textrm{B}_{HER-II},  9.7\,\textrm{MeV})$\,=\,$0.35\pm0.09$. For continuous neutrino spectra, i.e. for \emph{pp} and $^8$B, the $P_{ee}$ is quoted for the average energy of neutrinos that produce scattered electrons in the given energy range. The quoted errors include the uncertainties on the SSM solar-neutrino flux predictions.

Borexino is the only experiment that can simultaneously test neutrino flavour conversion both in the vacuum and in the matter-dominated regime, providing the most precise measurement of the $P_{ee}$ in the LER. In HER, where the flavour conversion is dominated by matter effects in the Sun, the Borexino results are in agreement with the high-precision measurements of Super-Kamiokande~\cite{bib:Kamiokande,bib:SK} and SNO~\cite{SNO_1,SNO_2}. The Vacuum-LMA\footnote{LMA stands historically for Large Mixing Angle solution of the best parameter space allowed for $\theta_{12}$ mixing angle.} prediction is shown as a grey band in Figure~\ref{fig:pee_p2}, calculated with Equation~\ref{eq:Pee_vac} using $\theta_{12}$ and $\theta_{13}$ values based on measurements of KamLAND~\cite{Gando:2013nba} and Daya Bay~\cite{DayaBay2017}, respectively and obtained without Borexino data. The pink band instead shows the MSW-LMA solution~\cite{deHolanda:2004fd} with the oscillation parameters indicated in~\cite{Esteban2017}. Borexino data disfavours the vacuum-LMA hypothesis at 98.2\% C.L. and are in excellent agreement with the expectations from the MSW-LMA paradigm.

\subsubsection{$^7$Be flux seasonal modulation} \label{sec:be7_seasonal}

The flux of solar neutrinos at Earth is not constant in time, but is instead modulated due to Earth's movement around the Sun on an elliptical trajectory. Given the variation of the solid angle covered by Earth during the year, a variation in the solar neutrino flux is predicted to exist. The net flux variation between the maximum and the minimum is estimated to be around 6.7\%.

\begin{figure}[t]
    \centering
    \includegraphics[width=0.7\textwidth]{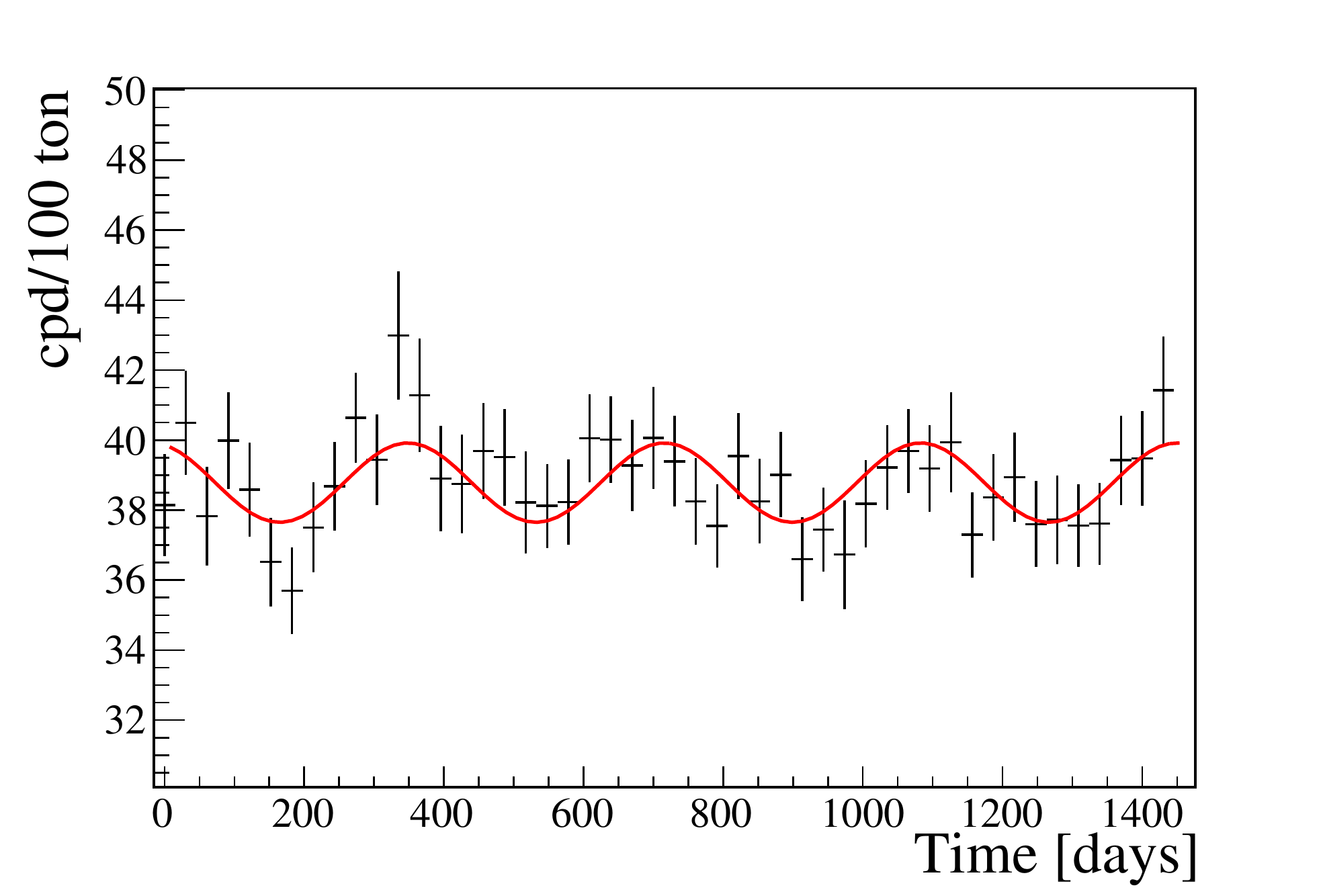}
    \caption{Borexino event rate in Phase-II versus time in the $^7$Be edge region, with associated modulation extraction in red. Distribution is plotted in time bins of $\sim$30 days. From~\cite{Be7mod}.}
    \label{fig:Be7_mod}
\end{figure}

Borexino was able to measure the annual modulation of solar neutrinos with high significance~\cite{LongPaperPhaseI, Be7mod}, confirming the solar origin of the measured $^7$Be signal. In order to maximize the signal-to-background ratio, events were selected in the $^7$Be shoulder energy range. In the latest analysis~\cite{Be7mod}, the energy window has been tuned to be [0.215,0.715]\,MeV and the events have been selected inside a large FV of 98.6\,ton. The chosen energy region is also rich in $^{210}$Po $\alpha$-decays, which are efficiently suppressed by means of the MLP cut (see Section \ref{subsec:event-reco}). In order to extract the modulation signal, three different analytical approaches were used: an analytical fit to event rate, a Lomb-Scargle periodogram, and an Empirical Mode Decomposition analysis. Figure~\ref{fig:Be7_mod} shows the $\beta^-$ event rate in the energy region of interest, in $\sim$30-day time bins. All methods yield compatible results, confirming the observation of solar neutrino flux modulation. The fit values for the modulation periodicity and its amplitude are well consistent with the expectations. Borexino was able to reject the hypothesis of no modulation with a confidence level of 99.99\%. 

\subsection{Detection of CNO neutrinos}
\label{sec:CNO_ana}

This Section reports the details about the analysis performed on Borexino data to extract the CNO signal. Section~\ref{sec:CNO_strategy} describes the challenges of this analysis and the main strategy towards the first observation of the CNO neutrino signal on Phase-III data~\cite{CNOpap}. The achieved 5$\sigma$ significance of this results is compatible with the expected sensitivity~\cite{CNOsens} described in Section~\ref{CNO_sens_corr}. Section~\ref{sec:CNO_LPoF} describes the method used to evaluate the rate of the $^{210}$Bi background contaminating the scintillator. This was used as a constraint in the spectral fit as well as in the counting analysis, leading to the final results discussed in Section~\ref{sec:CNO_results}.

\subsubsection{Analysis strategy}
\label{sec:CNO_strategy}

The energy of recoiling electrons after CNO neutrino interactions, shows a continuous distribution with an endpoint at 1.517\,MeV. The main sensitivity to CNO neutrinos comes from the energy region above the $^7$Be shoulder, i.e. from 0.8 to 1.0\,MeV~\cite{CNOsens}. This energy region is populated by other backgrounds, which limit the sensitivity towards the CNO neutrino events. Exactly as in the \emph{pp} chain LER neutrino analysis (Section~\ref{sec:pp_ana}), the cosmogenic $^{11}$C($\beta^{+}$) background is treated via the TFC algorithm (Section~\ref{sec:solar_nu_ana}). In the TFC-subtracted spectrum, the most important backgrounds are the $\beta^-$ particles emitted by the $^{210}$Bi contaminant and the \emph{pep}-neutrino recoil electrons. These two species have spectral features very similar to that of CNO neutrinos and their absolute rate must then be constrained by means of independent inputs.

In the Phase-II LER analysis, it was not possible to set an independent constraint on $^{210}$Bi. The rate of \emph{pep} neutrinos was constrained indirectly, by constraining the ratio of \emph{pp} and \emph{pep} neutrino fluxes to the SSM predictions of $47.7\pm0.8$ (HZ-SSM) and $47.5\pm 0.8$ (LZ-SSM), while the absolute \emph{pp} and \emph{pep} rates were free fit parameters. This corresponds to an effective constraint of 10\% precision for the \emph{pep} neutrino rate, a value dominated by the precision with which Borexino can measure \emph{pp} neutrinos. The \emph{pp}-\emph{pep} ratio is known very well from nuclear physics due to the fact that both reactions have the same nuclear matrix element. By performing a $\Delta\chi^{2}$-profiling, an upper limit on CNO interaction rate of $8.1$ cpd/100\,ton $(95 \% \textrm{C.L.})$ was obtained.

In the Phase-III analysis, the energy threshold was increased above the end-point of \emph{pp} neutrinos. This was due to a worsened resolution at low energies as a consequence of dying PMTs. The rate of \emph{pep} neutrinos was then constrained directly to a value of $(2.74 \pm 0.04)$\,cpd/100\,ton, corresponding to a precision of 1.4\%~\cite{Bergstrom_2016,CNOsens}. This value results from a combination of robust theoretical assumptions and a global fit to the solar neutrino data, excluding Borexino Phase-III. The other main background for the CNO measurement, consisting of $^{210}$Bi decays, was also constrained in the Phase-III analysis. $^{210}$Bi  has a short lifetime of 7.2\,days, and its overall rate in the detector can be assumed to be in secular equilibrium with its parent nucleus $^{210}$Pb. The decay chain of $^{210}$Pb is summarised in Equation \ref{eq:Pb210chain}. $^{210}$Pb decays are well below the analysis threshold and can therefore be considered invisible. The $^{210}$Po, on the other hand, is an unstable isotope, which produces mono-energetic $\alpha$ particles. $\alpha$-decays are efficiently detected in Borexino on an event-by-event basis, by means of the MLP selection (see Section~\ref{subsec:event-reco}). Provided that the secular equilibrium in Equation~\ref{eq:Pb210chain} is maintained, the measured $^{210}$Po corresponds to the $^{210}$Bi rate. 

\subsubsection{Low Polonium Field}
\label{sec:CNO_LPoF}

As discussed previously, the rate of $^{210}$Bi decays can be constrained via its link with the $^{210}$Po decay rate, with the assumption that this latter term is only supported by in-equilibrium $^{210}$Pb decay chain. Data collected by Borexino since its start, however, indicate that an out-of-equilibrium component of $^{210}$Po is present in the detector. The source of this component is likely the surface of the Inner Vessel, from which $^{210}$Po is detached into the scintillator. The mean free path of $^{210}$Po atoms is calculated to be very small in stable conditions. However, the presence of convective motions in the Borexino scintillator allow $^{210}$Po to spread throughout the scintillator volume. Under this conditions, the measured value of $^{210}$Po decay rate would be much higher than the $^{210}$Bi decay rate, spoiling any possible constraint.

To limit convective motions in the scintillator volume, the Borexino Collaboration pursued a long-lasting effort, culminated in the detector thermal insulation in 2015 and the subsequent installation of active temperature controls (Section~\ref{subsec:detector-setup}). This way, $^{210}$Po mixing has been strongly suppressed since 2016, leading to the formation of a very clean region around the centre of the detector, called the \emph{Low Polonium Field} (LPoF). The in-equilibrium $^{210}$Po decay rate in the LPoF region can be then measured. However, there might be still some residual contribution of convective $^{210}$Po in this region and the measured $^{210}$Po rate can therefore only be translated into an upper limit for the $^{210}$Bi rate.

The $^{210}$Po events in the LPoF have been chosen in the energy range 0.30-0.54\,MeV and selected by means of the MLP (see Section \ref{subsec:event-reco}). A paraboloid equation in 2D, assuming rotational symmetry along the $x-y$ plane, has been used to fit the data. The minimum $^{210}$Po rate ($R(^{210}Po_{min})$) has been then obtained through the fit function:
\begin{equation}\label{eq:bi_fit}
\frac{d^{2}R_{Po}}{d(\rho^{2})dz} = [R(^{210}Po_{min})\varepsilon_{E}\varepsilon_{MLP} + R_{\beta}]\bigg(1 + \frac{\rho^{2}}{a^{2}} + \frac{(z - z_{0})^{2}}{b^{2}} \bigg). 
\end{equation}
Here, $\rho^{2} = x^{2} + y^{2}$, $z_{0}$ is the minimum position of the LPoF along the $z$-axis, $a$ and $b$ are shape parameters along the respective axes, $\varepsilon_{E}$ and $\varepsilon_{MLP}$ are the efficiency of the energy and MLP cuts, respectively, applied to select $^{210}$Po events, and $R_{\beta}$ is the residual rate of $\beta$ events after the selection of $^{210}$Po events.
Since the LPoF slowly moves along the $z$-axis due to residual convective motions, data in the LPoF needs to be aligned along the $z$-direction before performing the fit on the full dataset. This has been done by ``blindly'' aligning the data in the LPoF every month (or every two months) using the centre $z_{0}$ obtained by fitting the data of the previous month. Monthly fits have been performed in large volumes of 70 or 100\,ton. After the blind alignment using the centres of every month, the final fit has been performed on the aligned dataset in around 20\,ton ($\sim$5000 events) using either a simple paraboloid in Equation~\ref{eq:bi_fit} with four free parameters ($R(^{210}Po_{min}), a, b, z_{0}$) or with more free parameters, depending on the method. The simple paraboloid fit can be performed either as a likelihood fit with ROOT~\cite{ROOT} or with the MultiNest Bayesian tool~\cite{MultiNest08,MultiNest09,MultiNest19}. The assumption of the rotational symmetry has been also verified. In addition to the 2D paraboloid fit, 3D ellipsoidal fits have been also performed with MultiNest without assuming rotational symmetry along the $x-y$ plane, resulting in statistically compatible results. In order to account for the complexity of the LPoF along the $z$-axis, a cubic spline function was implemented along the $z$-axis in equation~\eqref{eq:bi_fit} and the fit was performed with MultiNest. Despite its better fit on the LPoF data, the method was statistically compatible with the simple paraboloid fit and both the methods were finally used for the $^{210}$Bi upper limit. 
In addition to the statistical uncertainty of the fit, the other sources of uncertainties considered in this analysis are:
\begin{itemize}
    \item Systematic uncertainties from the fit: mass of the fit region, and binning of the data histogram.
    \item Uncertainty on the $\beta$-leakage estimation, i.e. $R_{\beta}$ in equation~\ref{eq:bi_fit}.
    \item Homogeneity of $\beta$-events: Since the $^{210}$Bi upper limit is estimated from the LPoF, which is only 20\,ton, it is necessary to study the homogeneity of $\beta$-events in the entire FV of the CNO analysis and in the energy region of $^{210}$Bi. The radial homogeneity has been studied by dividing the FV into 25 iso-volumetric shells. The angular homogeneity was studied by extending Fourier decomposition over a sphere surface, by projecting the spatial co-ordinates of the selected events on a sphere.
\end{itemize}
The final $^{210}$Bi upper limit obtained through the estimation of the minimum $^{210}$Po rate in the LPoF of Borexino, including both statistical and systematic contribution, is:
\begin{equation}
    R(\textrm{$^{210}$Bi}) \leq (11.5\pm1.3)\,\textrm{cpd/100\,ton}~(1\sigma).
\end{equation}

\subsubsection{Sensitivity}
\label{CNO_sens_corr}

The strength of the constraint on $^{210}$Bi rate directly affects the precision of the CNO measurement. In order to evaluate quantitatively the precision of the CNO measurement of Borexino as a function of the $^{210}$Bi rate constraint, a MC-based sensitivity study has been performed~\cite{CNOsens}. To assess the expected discovery potential to CNO neutrinos, a frequentist hypothesis test has been performed. Within this framework, two hypotheses have been considered: the \emph{null} hypothesis $H_0$, meaning that no CNO is assumed to exist, and the \emph{alternative} hypothesis $H_1$, that includes the presence of CNO. By indicating with $L(H_0)$ and $L(H_1)$ the two respective maximum likelihood values, the following likelihood ratio $q_0$ can be used as a test statistic~\cite{Lkl_q}:
\begin{equation}
    q = -2\,\textrm{ln}\frac{L(H_0)}{L(H_1)}.
\end{equation}

Two sets of toy data have been produced, one with CNO injected one without CNO. Each dataset has been fit twice, i.e. with the $H_0$ and the $H_1$ hypotheses. The distribution of the test statistics $q$ for the dataset without CNO injected, is called $q_0$. The median of the $q$ distribution for the dataset with CNO injected is called $q_{\mathrm{med}}$. Then, the \emph{p} value of the $q_0$ distribution with respect to $q_{\mathrm{med}}$ defines the discovery potential.

\begin{figure}[t]
\centering
    \includegraphics[width=0.55\textwidth]{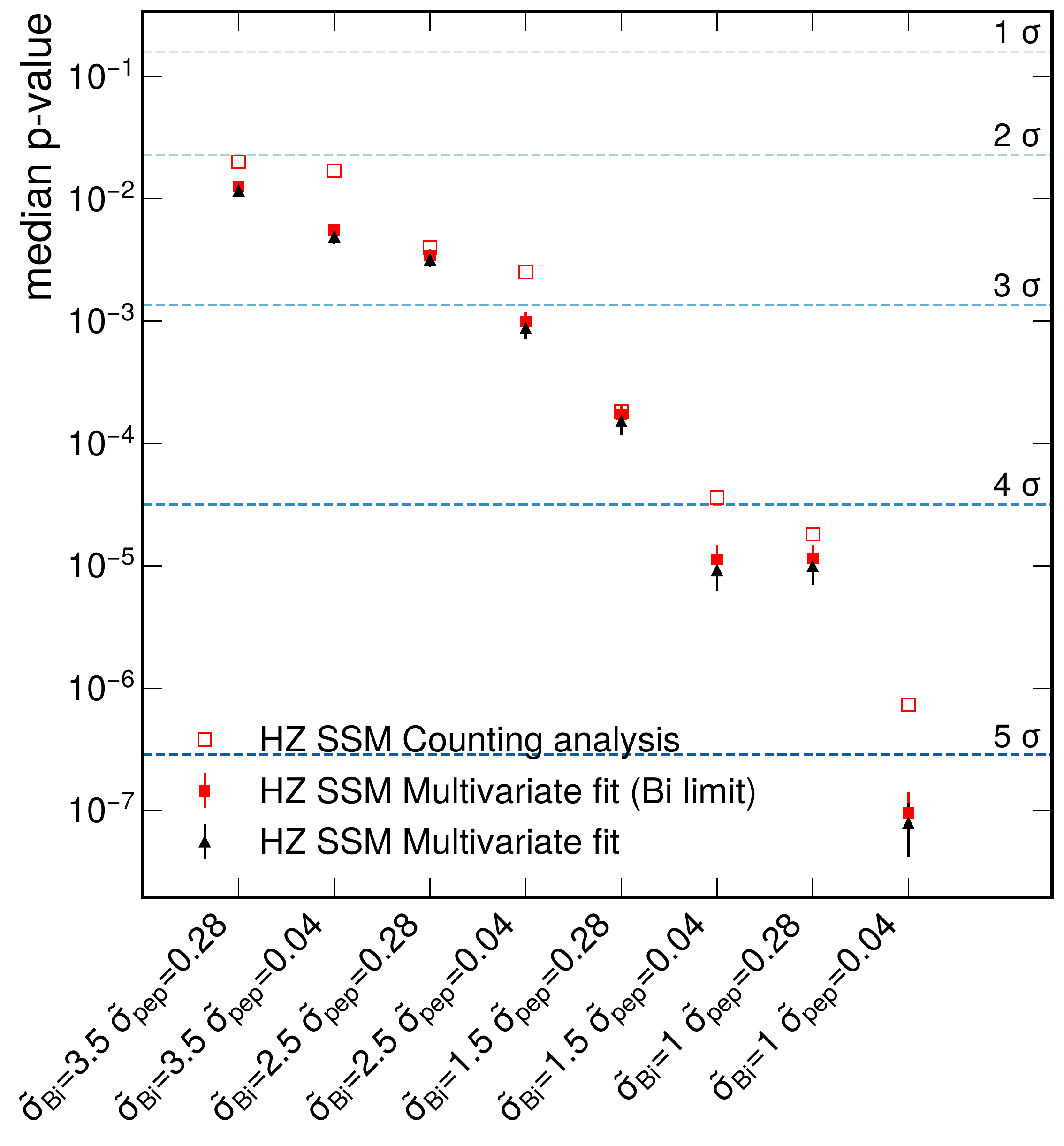}
    \caption{Borexino median discovery significance for the HZ-SSM hypothesis on the CNO rate for the exposure of (1000 $\times$ 71.3)\,days $\times$\,ton (93\% of the Phase-III exposure) for the multivariate spectral fit (filled markers) and counting analysis (empty square). Different scenarios for the penalties on \emph{pep} solar neutrino rate and on $^{210}$Bi rate are considered. The uncertainty $\overset{\sim}{\sigma}_{\mathrm{Bi}}$ and $\overset{\sim}{\sigma}_{pep}$ are in
cpd/100\,ton. For \emph{pep}, 0.04\,cpd/100\,ton corresponds to a 1.4\% constraint. For $^{210}$Bi, the LPoF fit returns an uncertainty of 1.3\,cpd/100\,ton (Section \ref{sec:CNO_LPoF}). From \cite{CNOsens}.  
}
    \label{fig:CNO_sens}
\end{figure}

Figure~\ref{fig:CNO_sens} reports the CNO median discovery significance assuming the HZ-SSM hypotheses and under different assumptions of \emph{pep} and $^{210}$Bi rate constraints, and assuming an exposure of 1000\,days $\times$ 71.3\,ton (93\% of the Phase-III exposure).
Neutrino interaction rates have been chosen according to HZ-SSM
predictions. Background rates have been extracted from a MV fit on Phase-III data with the exception of the $^{210}$Bi rate that has been set to 10\,cpd/100\,ton, a value similar to the upper limit estimated in the Phase-III data (see previous Section). This small difference does not influence the sensitivity, that is dominated by the precision of the constraint. The strong dependence of the sensitivity to CNO neutrinos on the strength of the external constraints on \emph{pep} and $^{210}$Bi rates is evident in Figure~\ref{fig:CNO_sens}. By assuming the uncertainties similar to the one obtained in Phase-III data (Sections \ref{sec:CNO_strategy} and \ref{sec:CNO_LPoF}), a 5$\sigma$ significance on the CNO neutrino signal can be reached by Borexino.

\paragraph{Spectral analysis}

To extract the CNO neutrino interaction rate from data, a multivariate analysis has been performed, by simultaneously fitting the energy spectra between 0.320\,MeV and 2.640\,MeV and the radial distribution of events (Table~\ref{tab:SolAnal}), after all the selection procedure as already introduced in Section $\ref{sec:solar_nu_ana}$. The Phase-III dataset has been considered for the analysis. Apart from the constrained \emph{pep}-neutrino and $^{210}$Bi rates, and the fixed $^8$B rate, all other species rates (including CNO neutrinos) have been left free to vary in the fit. The reference PDFs used to fit both the energy and radial distributions have been built by means of a complete Monte Carlo simulation. Considering only statistical uncertainty, the best fit returns  a CNO interaction rate of  $R = 7.2^{+2.9}_{-1.7}$\,cpd/100\,ton (68\% confidence interval) with Borexino~\cite{CNOpap}.

The effect of the fit configuration (such as the fit ranges) has been found negligible in the analysis. Since the fit heavily relies on the simulated MC PDFs for signal and backgrounds, a mismatch between data and simulations could potentially affect the CNO result and introduce a bias. To take this effect into account, as a possible source of systematics, a toy MC-based study has been performed. By generating several millions of pseudo-datasets, deforming signals and backgrounds every time and fitting with the same non-deformed PDFs, the impact on the CNO measurement has been evaluated. As possible sources of deformation, the following contributions have been considered:

\begin{itemize}
    \item Detector energy response, in terms of the scintillator energy scale (0.23\%), non-uniformity (0.28\%) and non-linearity (0.4\%). The size of the deformation has been chosen based on the allowed values from calibration data and the `standard candles' namely, $^{210}$Po and $^{11}$C.
    \item Deformation of the spectral shape of the cosmogenic $^{11}$C isotope, induced by noise cuts not fully reproduced by MC (2.3\%).
    \item Spectral shape uncertainty of $^{210}$Bi (18\%). The uncertainty has been quoted through comparison of the reference $^{210}$Bi spectrum~\cite{Bi210_ref} with alternative spectra~\cite{Bi210_def1,Bi210_def2}.
\end{itemize}
The final systematic contribution has been found to be $^{+0.6}_{-0.5}$\,cpd/100\,ton, evaluated by comparing the CNO output from toy MC with and without injecting systematic distortions.

    \begin{figure}[t]
    \centering
    \includegraphics[width=0.95\textwidth]{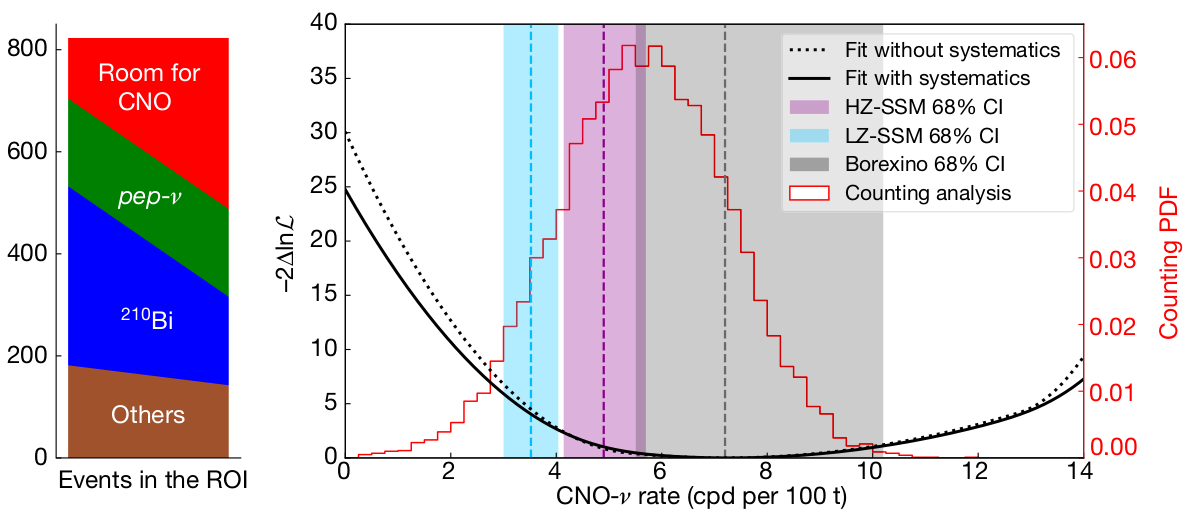}
    \caption{Summary of CNO counting and spectral analysis results. Left, counting analysis bar chart. Right, CNO-neutrino rate negative log-likelihood (ln $\mathcal{L}$) profiles, together with PDF from counting analysis and models predictions. From~\cite{CNOpap}.}
    \label{fig:cno-final}
    \end{figure}

\paragraph{Counting analysis}

As a cross-check of the spectral analysis, a counting analysis has been performed on the same data sample. Data events have been counted inside an energy region where the CNO signal-to-background ratio is maximised, corresponding to [0.780, 0.885]\,MeV. The \emph{pep}-neutrino and $^{210}$Bi numerical rate constraints, used in the spectral analysis, have been used in the counting analysis as well. The rate of $^{210}$Bi has been symmetrically constrained and signals and backgrounds have been described by means of analytical functions. The CNO interaction rate has been extracted by subtracting all the background contributions, evaluated within a certain uncertainty, and then propagating those uncertainties. The final measured rate is (5.6$\pm$1.6)\,cpd/100\,ton, confirming the presence of CNO at 3.5$\sigma$ level. The quoted uncertainty considers both statistical and systematic terms. The uncertainty related to the energy response, in particular, is the dominant contribution in the overall balance.

\subsubsection{Final result}
\label{sec:CNO_results}

Figure~\ref{fig:cno-final} summarizes the Borexino result on the measured CNO neutrino interaction rate. The result of the spectral fit is reported in terms of log-likelihood profiles, with statistical uncertainty (dotted black line) and also with folded systematic contributions (solid black line). The best fit value is $R = 7.2^{+3.0}_{-1.7}$ cpd/100\,ton (68\% confidence interval), including systematic uncertainties. The inferred flux of CNO neutrinos at Earth is $\Phi = 7.0^{+3.0}_{-2.0} \times 10^8 ~ \mathrm{cm}^{-2} ~ \mathrm{s}^{-1}$ (68\% confidence interval). The probability density function obtained from the counting analysis is also reported (solid red line). From the profiling of the log-likelihood, an exclusion of no-CNO hypothesis is achieved with 5.1$\sigma$ significance. A further hypothesis test with 13.8 million pseudo-datasets excludes the no-CNO hypothesis with 5.0$\sigma$ at 99\% confidence level.

The CNO neutrino measurement performed by Borexino is compatible with both HZ-SSM (0.5$\sigma$) and LZ-SSM (1.3$\sigma$) metallicity scenarios. A combined hypothesis test, including $^7$Be and $^8$B solar neutrino fluxes measured by Borexino, shows a preference for the HZ-SSM hypothesis at 2.1$\sigma$ level.

\subsection{Search for BSM physics}
\label{sec:bsm}

This section deals with searches beyond the Standard Model of particle physics with Borexino. Section~\ref{sec:bxnsi} is dedicated to the search for Non-Standard neutrino interactions while Section~\ref{sec:bxnmm} describes the procedure for the determination of the neutrino magnetic moment. 

\subsubsection{Non-standard neutrino interactions}
\label{sec:bxnsi}

\begin{figure}[t]
\centering
\includegraphics[width = 0.45\textwidth]{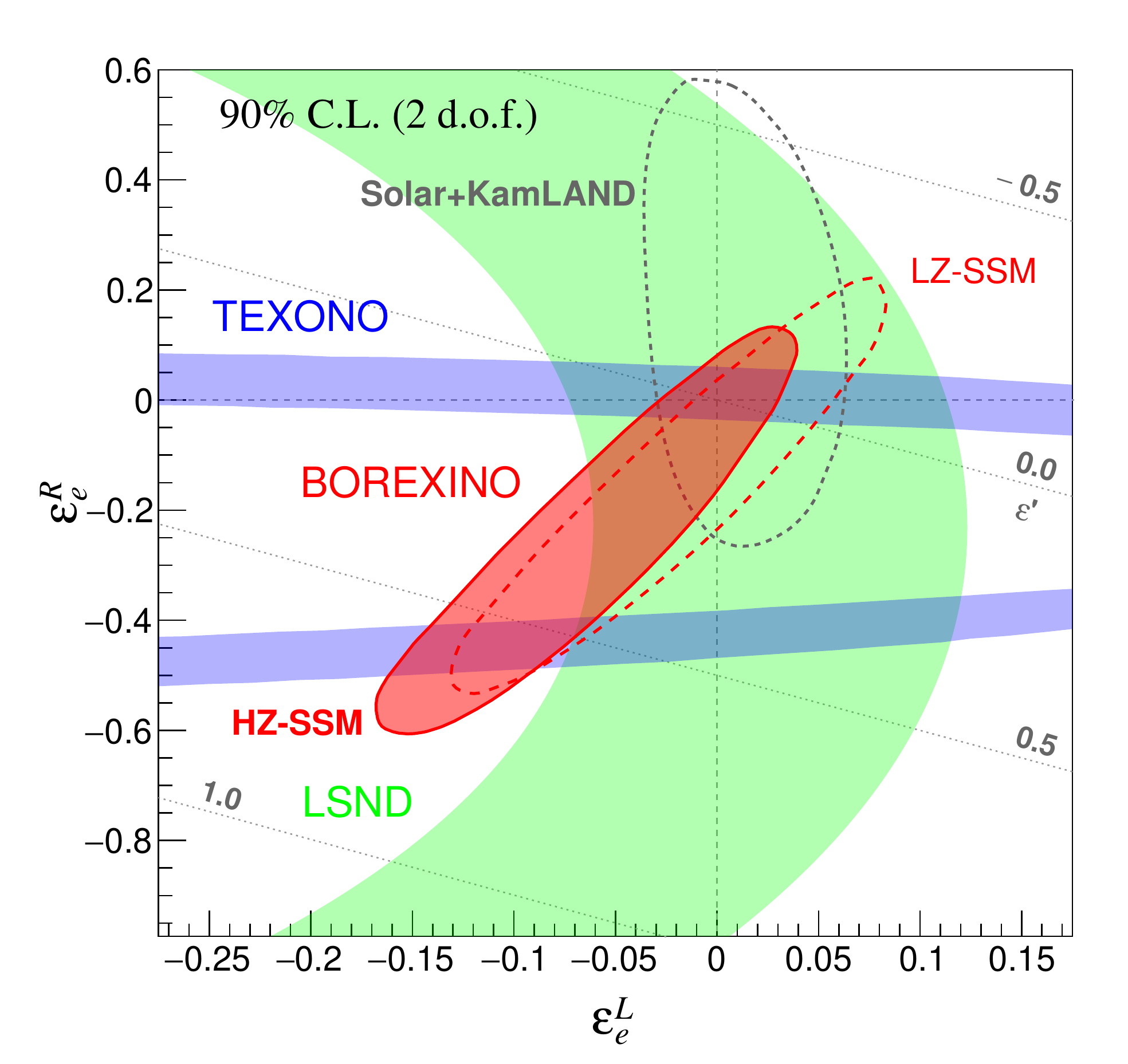}
\includegraphics[width = 0.45\textwidth]{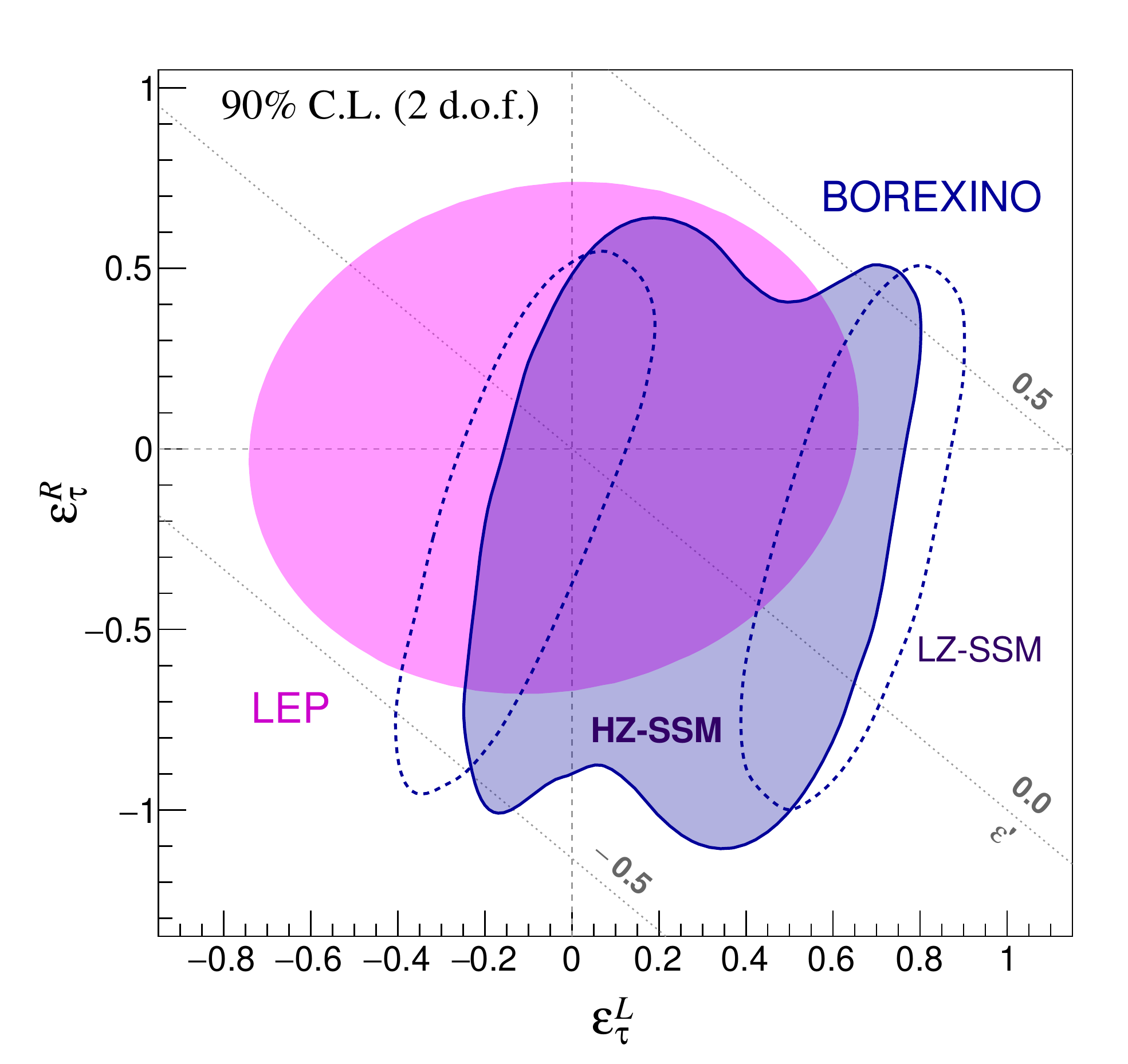}
\caption{Borexino limits on NSI, using Phase-II dataset: (a) Allowed region for the NSI parameters $\varepsilon_e^R$ and $\varepsilon_e^L$. (b) Allowed region for the NSI parameters $\varepsilon_\tau^R$ and $\varepsilon_\tau^L$. Contours are reported for both HZ-SSM and LZ-SSM metallicity scenarios. From \cite{Bx-NSI20}.}
\label{fig:nsi}
\end{figure}

Events selected in Borexino data for the solar neutrino analysis can be used to search for interactions not predicted by the Standard Model of Particle Physics. This class of phenomena is usually referred to as Non-Standard Interactions (NSI). The effect of NSI, regarding the solar neutrino flux, is to modify the electron neutrinos survival probability $P_{ee}$, as well as the couplings between neutrinos and the scattered electrons. By analyzing Borexino Phase-II data \cite{Bx-NSI20}, possible presence of NSI in the $\nu_e e$ and $\nu_\tau e$ couplings have been studied. In particular, the analysis has looked for flavor-diagonal NC interactions. By considering the predictions from SSM, including oscillation effects, as well as both HZ-SSM and LZ-SSM metallicity scenarios, deviations in the measured spectrum have been searched for. Figure~\ref{fig:nsi} reports the allowed regions for NSI extracted from Borexino data, considering $\nu_e e$ and $\nu_\tau e$ couplings. Both regions in Figure~\ref{fig:nsi} are compatible with null values of the $\varepsilon$ parameters, which indicate the absence of NSI. In any case, the limits set by Borexino represent a significant reduction in the allowed parameter space, with respect to other experiments.

The same Phase-II dataset and analysis approach have been used to measure the value of the square sine of the Weinberg angle $\sin^2\theta_W$, by considering it as a free parameter in the analysis. From the likelihood profile, the best fit value results in:
\begin{equation}
\sin^2 \theta_W = 0.229 \pm 0.026~ \mathrm{(stat+syst)},
\end{equation}
which is in agreement with results obtained from other neutrino-electron scattering experiments.

\subsubsection{Neutrino magnetic moment}
\label{sec:bxnmm}

\begin{figure}[t]
\centering
    \includegraphics[width=0.65\textwidth]{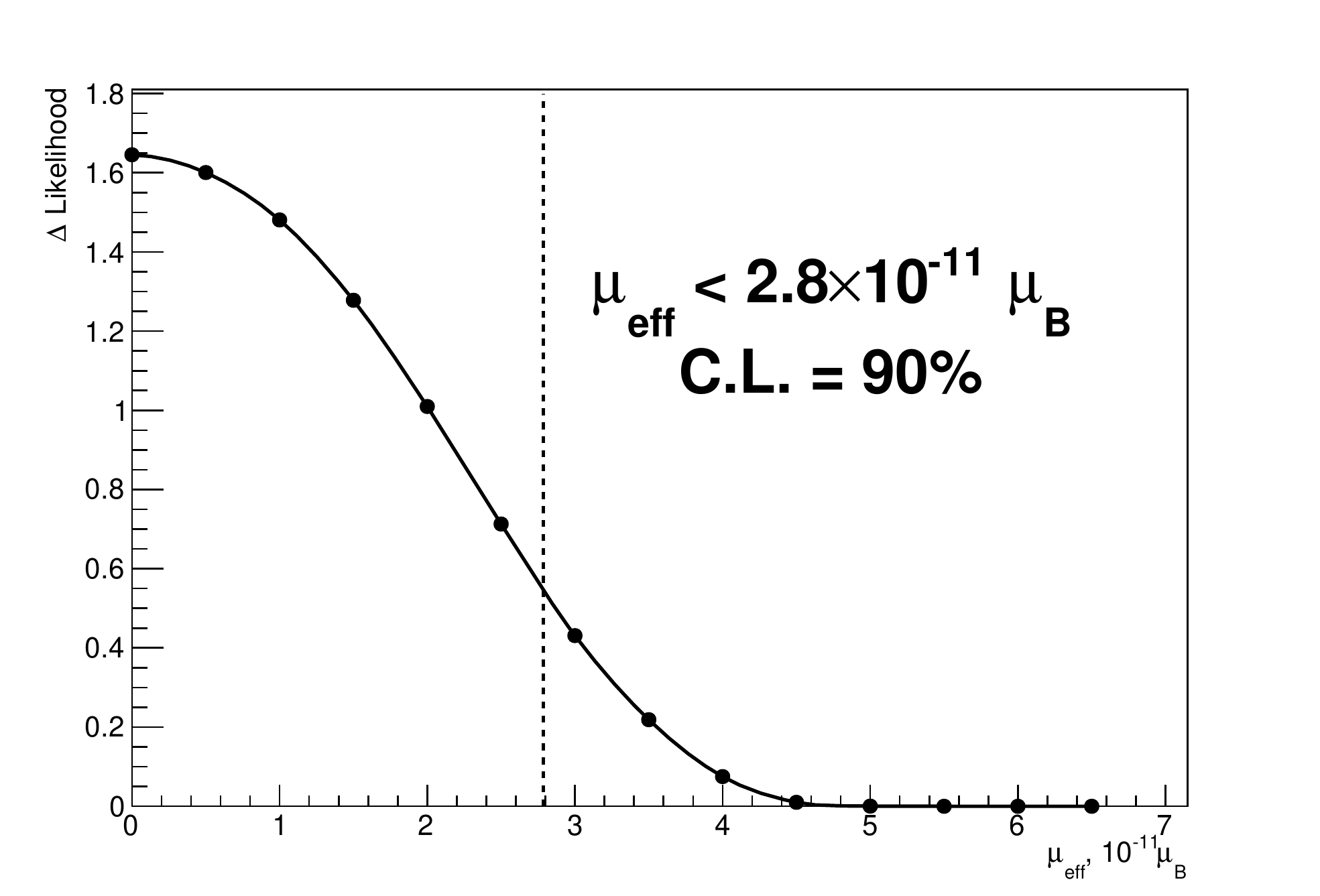}
    \caption{Limit on a possible anomalous effective neutrino magnetic moment with Borexino Phase-II solar neutrino data: weighted likelihood profile for the neutrino effective magnetic moment, including systematic effects. The limit corresponds to 90\% of the total area under the curve. From \cite{Bx-NMM17}.}
    \label{fig:nmm}
\end{figure}

The flux of solar neutrinos detected by Borexino can be also used to derive indications of a neutrino anomalous magnetic moment. The presence of this condition would have the effect of modifying the neutrino-electron cross section and, consequently, the visible energy spectrum of solar neutrinos.
The analysis of Borexino Phase-II data, including an additional component in the fit originating from an anomalous neutrino magnetic moment, has been performed in~\cite{Bx-NMM17}. Figure~\ref{fig:nmm} shows the likelihood profile used to estimate the limit on the neutrino magnetic moment.

The sum of the solar neutrino fluxes have been constrained using inputs from radiochemical gallium experiments. By including possible systematics effects, the derived limit for the neutrino magnetic moment is $\mu_\nu^{\mathrm{eff}} <$ 2.8 $\times 10^{-11} \mu_B$\footnote{$\mu_B$ = $e\hslash/2m_e$ = 5.788 381 8060(17) $\times 10^{-11}$\,MeV T$^{-1}$ is the Bohr magneton~\cite{PDG2020}}. at 90\% confidence level. The result is also free from uncertainties associated with predictions from the SSM neutrino flux. The limit on the effective value can be translated into a limit for magnetic moments of individual flavors, according to the measured values of flavor oscillation parameters. Limits on the value of the neutrino magnetic moment can be obtained from neutrino-electron elastic scattering experiments, either using solar~\cite{SK_NMM04,Bx_NMM08} or reactor neutrinos~\cite{TEXONO07,GEMMA13}. The value obtained by Borexino is numerically the most stringent constraint on the neutrino magnetic moment.

\section{Geoneutrinos} 
\label{sec:geo} 

Geoneutrinos are mostly antineutrinos emitted as byproducts of radioactive decays inside the Earth. The fundamental quests of Earth sciences involve answering long-standing questions about the thermal, geodynamical, and geological evolutions of the Earth. Although geoneutrinos cannot answer all these questions, their measurement can serve as a unique tool in understanding the abundance of radioactive elements inside the Earth, especially in the inaccessible mantle. This can thereby help to determine the contribution of radiogenic heat to the total heat flux measured on Earth - a key parameter in many aspects of geosciences. This part of the review is divided into the following sections: Section~\ref{sec:earth-geo} describes the structure, composition, and the heat flow of the Earth and it also presents geoneutrinos as a new tool for geoscience. The different components required for the geoneutrino analysis in Borexino are described in Section~\ref{sec:geo-analysis}. The final results and geological interpretations of Borexino's recent geoneutrino measurement~\cite{Agostini:2019dbs} are discussed in Section~\ref{sec:geo-results}.

\subsection{The Earth and geoneutrinos}
\label{sec:earth-geo}

This section is dedicated to the detailed structure of the Earth and its heat budget (Section~\ref{subsec:earth}) and the different models predicting the composition of the silicate part of the primordial Earth and thus, the Earth's radiogenic heat (Section~\ref{subsec:bse}). Geoneutrinos and their role in geosciences, in particular in the determination of the Earth's radiogenic heat, are discussed in Section~\ref{subsec:geonu-intro}.

\subsubsection{The Earth's structure and heat budget}
\label{subsec:earth}

The Earth was formed by the accumulation of matter from the solar nebula over time~\cite{RN1398, RN607}. All bodies with sufficient mass undergo the process of differentiation, i.e., the transformation of a homogeneous body into a layered structure. Primitive homogeneous Earth went through the first differentiation through segregation of metals, when the {\it core} separated from the silicate {\it primitive mantle} or {\it bulk silicate earth} (BSE). The BSE further differentiated into the present {\it mantle} and {\it crust}. The metallic core has Fe-Ni chemical composition and is expected to reach temperatures up to about 6000\,K in its central parts. The inner core ($\sim$1220\,km radius) is solid due to high pressure, while the 2263\,km thick outer core is liquid. The outer core has an approximate 10\% admixture of lighter elements and plays a key role in the geodynamo process, which generates the Earth's magnetic field. The {\it core-mantle boundary} (CMB), a seismic discontinuity divides the core from the mantle. The mantle reaches temperature of about 3700\,K at its deeper part. It is solid but viscous on long time scales, leading to mantle convection processes. This drives the movement of tectonic plates at the speed of few cm per year. At a depth of 400-700\,km, the mantle is characterised by a {\it transition zone}, where a weak seismic-velocity heterogeneity is measured. The upper portion of the mantle contains the viscous {\it asthenosphere} on which the {\it lithospheric tectonic plates} are floating. These comprise the uppermost, rigid part of the mantle (i.e. the {\it continental lithospheric mantle} (CLM)) and the two types of crust: {\it oceanic crust} (OC) and {\it continental crust} (CC). The CLM is a portion of the mantle beneath the CC at a typical depth of $\sim$175\,km~\cite{RN367}. The CC has a thickness of (34 $\pm$ 4)\,km~\cite{Huang2013} and is the most differentiated and heterogeneous layer, due to its complex history. It consists of igneous, metamorphic, and sedimentary rocks. The OC with (8 $\pm$ 3)\,km~\cite{Huang2013} thickness is created along the mid-oceanic ridges, where the basaltic magma differentiates from the partially melting mantle, up-welling towards the ocean floor.

The heat flow from the Earth's surface to the space results from a large temperature gradient across the Earth. The current best estimate for the total surface heat flux on Earth is $H_{\mathrm{tot}}$ = (47 $\pm$ 2)\,TW~	\cite{RN593}. Neglecting the small contribution ($<$0.5\,TW) from tidal dissipation and gravitational potential energy released by the differentiation of crust from the mantle, the $H_{\mathrm{tot}}$ is typically expected to originate from two main processes: ({\it i}) {\it secular cooling} $H_{\mathrm {SC}}$ of the Earth, i.e. cooling from the time of the Earth's formation when gravitational binding energy was released due to matter accretion, and ({\it ii}) {\it radiogenic heat} $H_{\mathrm {rad}}$. The {\it radiogenic heat} of the present Earth arises mainly from the decays of isotopes with lifetimes comparable to, or longer, than the Earth's age (4.543\,$\cdot$\,10$^{9}$\,years): $^{232}$Th ($\tau$ = 2.02\,$\cdot$\,10$^{10}$\,years), $^{238}$U ($\tau$ = 6.447\,$\cdot$\,10$^{9}$\,years), $^{235}$U ($\tau$ = 1.016\,$\cdot$\,10$^{9}$\,years), and $^{40}$K ($\tau$ = 1.801\,$\cdot$\,10$^{9}$\,years)~\cite{IAEA}. All these isotopes are labeled as {\it heat-producing elements} (HPEs). The relative contribution of radiogenic heat to the $H_{\mathrm{tot}}$ is crucial in understanding the thermal conditions that existed during the formation of the Earth and the energy that is now available to drive the dynamical processes such as the mantle and outer-core convection. The {\it convective urey ratio} ($UR_{\mathrm{CV}}$) quantifies the ratio of internal heat generation in the mantle over the mantle heat flux and is expressed as follows~\cite{RN404}:
\begin{equation}
UR_{\mathrm{CV}} = \frac {H_{\mathrm {rad}} - H_{\mathrm {rad}}^{\mathrm{CC}}} {H_{\mathrm{tot}} - H_{\mathrm {rad}}^{\mathrm{CC}}},
\label{eq:URCV}
\end{equation}
where $H_{\mathrm {rad}}^{\mathrm{CC}}$ is the radiogenic heat produced in the continental crust, that is relatively well known to be 6.8$^{+1.4}_{-1.1}$\,TW~\cite{Huang2013}. By adding the contribution from the oceanic crust and continental lithospheric mantle, the radiogenic heat from the lithosphere results to be 8.1$^{+1.9}_{-1.4}$\,TW~\cite{Agostini:2019dbs}. The mantle radiogenic heat is  poorly constrained and ranges between 1.2 and 39.8\,TW~\cite{Agostini:2019dbs}. This is discussed further in Section~\ref{subsec:exp-signal}. No radiogenic heat is expected to be produced in the metallic core. Preventing dramatically high temperatures during the initial stages of Earth  formation, the present-day $UR_{\mathrm{CV}}$ must be in the range between 0.12 to 0.49~\cite{RN1395}.  

\subsubsection{Bulk Silicate Earth models}
\label{subsec:bse}

The Bulk Silicate Earth (BSE) models define the original chemical composition of the primitive mantle, including the abundances of HPE's and thus, the respective radiogenic heat. The elemental composition of BSE is obtained assuming a common origin for celestial bodies in the solar system. It is supported, for example, by the strong correlation observed between the relative (to Silicon) isotope abundances in the solar photosphere and in the CI chondrites, a special group of rare stony meteorites belonging to the carbonaceous chondrites~\cite{Lodders_2003}. Such correlations can be then assumed for the material from which the Earth was created. The BSE models agree with each other in the prediction of major elemental abundances (e.g. O, Si, Mg, Fe) within 10\%. Uranium and Thorium abundances are assumed based on relative abundances in chondrites, and dramatically differ between different models. There are three main classes of BSE models: the \emph{cosmochemical}, \emph{geochemical}, and \emph{geodynamical} models, as defined in~\cite{BELLINI20131,RN630}. The \emph{cosmochemical} (CC) model~\cite{RN367} is characterised by a relatively low amount of U and Th, producing a total $H_{\mathrm{rad}}$ = (11 $\pm$ 2)\,TW. This model assumes that the Earth is composed of enstatite chondrites. The \emph{geochemical} (GC) model predicts intermediate HPE abundances for primordial Earth. It adopts the relative abundances of refractory lithophile elements as in CI chondrites, while the absolute abundances are constrained by terrestrial samples~\cite{RN1380, RN361}. The \emph{geodynamical} (GD) model shows relatively high U and Th abundances. It is based on the energy dynamics of mantle convection and the observed surface heat loss~\cite{RN368}. Additionally, an extreme \emph{fully radiogenic} (FR) model can be obtained following the approach described in~\cite{RN356}, where the terrestrial heat $H_{\mathrm {tot}}$ of 47\,TW is assumed to be entirely from radiogenic heat production $H_{\mathrm {rad}}$. Apart from these four main classes of models, there are also individual geological models used for comparison in the geological interpretations of Borexino's measurement discussed in Section~\ref{sec:geo-results}. The predictions of these models are discussed in detail in~\cite{Agostini:2019dbs}.

\paragraph{\emph{Th/U ratio}}
A global assessment of the Th/U mass ratio of the primitive mantle could provide information about the early evolution of the Earth and its differentiation. The most precise estimate of the planetary Th/U mass ratio has been refined to a value of $M_{\mathrm{Th}}$/$M_{\mathrm {U}}$ = (3.876 $\pm$ 0.016)~\cite{RN1377} and is relevant for the geoneutrino analysis as it will be discussed in the later sections. Significant deviations from this average value can be found locally, in the heterogeneous continental crust surrounding the detectors. The area surrounding the Borexino detector is characterised by a Th/U mass ratio ranging from $\sim$0.8 (carbonatic rocks) to $\sim$3.7 (terrigenous sediments)~\cite{coltorti}.

\paragraph{\emph{ K/U ratio}}

Potassium is the only semivolatile HPE. The bulk mass of Potassium predicted by different Earth models varies widely. Due to different possible scenarios, the K/U ratio predicted by different BSE models differs from 9700 to 16000~\cite{RN630}. According to these ratios, the mantle radiogenic heat from $^{40}$K varies between 2.6 and 4.3\,TW. This results in an average contribution of 18\% to the total mantle radiogenic power and is used in the determination of radiogenic heat from Borexino's geoneutrino measurement in Section~\ref{subsec:results-heat-urey}.

\subsubsection{Probing the Earth with geoneutrinos}
\label{subsec:geonu-intro}

Geoneutrinos are (anti)neutrinos emitted during the decay of the long-lived HPEs, discussed in the previous subsection, that can be summarised by the following equations:
\begin{align}
^{238}\mathrm{U} \rightarrow& \, ^{206}\mathrm{Pb} + 8\alpha + 8 e^{-} + 6 \bar{\nu}_e + 51.7\,\mathrm{MeV}\label{Eq:geo1}\\
^{235}\mathrm{U} \rightarrow& \, ^{207}\mathrm{Pb} + 7\alpha + 4 e^{-} + 4 \bar{\nu}_e + 46.4\,\mathrm{MeV} \label{Eq:geo2}\\
^{232}\mathrm{Th} \rightarrow& \, ^{208}\mathrm{Pb} + 6\alpha + 4 e^{-} + 4 \bar{\nu}_e + 42.7\,\mathrm{MeV} \label{Eq:geo3} \\
^{40}\mathrm{K} \rightarrow& \, ^{40}\mathrm{Ca}  +  e^{-} +  \bar{\nu}_e + 1.31\,\mathrm{MeV}~\mathrm{(89.3\%)} \label{Eq:geo4}\\
^{40}\mathrm{K} + e^{-} \rightarrow& \, ^{40}\mathrm{Ar}  +  \nu _{e} +  1.505\,\mathrm{MeV}~\mathrm{(10.7\%)} \label{Eq:geo5}
\end{align}
In the $\beta$ decays of $^{238/235}$U and $^{232}$Th chains and that of $^{40}$K, all geoneutrinos are antineutrinos. Neutrinos are emitted only in the electron-capture decays of $^{40}$K, having 10.7\% branching ratio. In the radioactive decays of HPEs, the amount of released geoneutrinos and radiogenic heat are in a well-known ratio [equations~\eqref{Eq:geo1} to~\eqref{Eq:geo5}]. Thus, a direct measurement of the geoneutrino flux provides useful information about the composition of the Earth's interior. Consequently, it also provides an insight into the radiogenic heat contribution to the measured Earth's surface heat flux. Even though their small interaction cross section limits our ability to detect them, it makes them a unique probe of inaccessible innermost parts of the Earth. The first ideas to detect geoneutrinos started in the 1960s~\cite{EDER1966657,Marx1960, Marx1969}, further developed in 1984~\cite{KraussGlashowSchramm}, and the potential of liquid scintillator detectors to measure geoneutrinos was suggested in the 1990s~\cite{Rothschild:1997dd, Raghavan:1997gw}. It took several years to prove the methods feasible and the first investigation was performed by KamLAND in 2005~\cite{Araki:2005qa}. So far, only KamLAND~\cite{Gando:1900zz, Gando:2013nba} and Borexino~\cite{Bellini:2010geo, Bellini:2013geo, Agostini:2015cba, Agostini:2019dbs} have been able to measure geoneutrinos.

The determination of the radiogenic component of Earth's internal heat budget has proven to be a difficult task, since an exhaustive theory is required to satisfy geochemical, cosmochemical, geophysical and thermal constraints, which are often based on indirect arguments. Therefore, direct geoneutrino measurements can be of great help. Geoneutrinos have also the potential to determine the mantle radiogenic heat, the key unknown parameter. This can be done by constraining the relatively-well known lithospheric contribution, as it will be shown in Section~\ref{subsec:mantle}. Since the oceanic crust is depleted of HPEs and has a very small lithospheric contribution that can be easily determined, it would make the ocean floor an ideal environment for geoneutrino detection. Geoneutrino measurements can also contribute to the discussion \textcolor{blue}{} about possible additional heat sources, whproposed by some authors. For example, stringent limits can be set on the power of a hypothetical Uranium natural georeactor suggested in~\cite{herndon1993feasibility,herndon1996substructure,rusov2007geoantineutrino,de2008feasibility} and discussed in Section~\ref{subsec:exp-antinu-bckg} and~\ref{subsec:georea}. 

\subsection{Geoneutrino analysis}
\label{sec:geo-analysis}

The various components required for the geoneutrino analysis with Borexino are discussed in this section. These include: the expected geoneutrino signal at LNGS (Section~\ref{subsec:exp-signal}), the evaluation of other antineutrino signals (Section~\ref{subsec:exp-antinu-bckg}) and non-antineutrino backgrounds (Section~\ref{subsec:exp-nonanti-bckg}), the data selection cuts (Section~\ref{subsec:geo-selcuts}), the spectral fit used to extract the signal (Section~\ref{subsec:geo-fit}), and the evaluation of various systematic uncertainties (Section~\ref{subsec:syst-geo}).

\subsubsection{Expected geoneutrino signal}
\label{subsec:exp-signal}

The Earth shines in a flux of antineutrinos with a luminosity $L\sim 10^{25}$\,s$^{-1}$. Due to the IBD threshold discussed in Section~\ref{subsec:detection}, only $^{238}$U and $^{232}$Th geoneutrinos can be measured by LS detectors. For a detector placed on the continental crust, the expected $^{238}$U and $^{232}$Th geoneutrino flux is of the order of $10^6$\,cm$^{-2}$\,s$^{-1}$ and is typically dominated by the crustal contribution. The differential flux of geoneutrinos emitted from isotope $i$ = ($^{238}$U, $^{232}$Th) and expected at LNGS location $\vec{r}$ is calculated using the following expression:
 \begin{eqnarray}
\frac{d\Phi (i;E_{\bar{\nu}},\vec{r})}{dE_{\bar{\nu}}} &=&  \varepsilon_{\nu} (i) \frac{dn(i;E_{\bar{\nu}})}{dE_{\bar{\nu}}} \\
\nonumber
    &\times&  \int_{V}d\vec{r \prime} P_{ee} \left( E_{\bar{\nu}}, \lvert\vec{r} - \vec{r \prime}\rvert\right) \frac{a(i; \vec{r \prime}) \cdot \rho (\vec{r \prime})}{4 \pi \lvert \vec{r} - \vec{r \prime}\rvert^2},
 \label{eq:geonuS}
 \end{eqnarray}
where $\varepsilon_{\nu} (i)$ is the {\it specific antineutrino production rate} for isotope $i$ per 1\,kg of naturally occurring element and $E_{\bar{\nu}}$ is geoneutrino energy. The \emph{geoneutrino energy spectra} $\frac{dn(i;E_{\bar{\nu}})}{dE_{\bar{\nu}}}$ is normalised to one. The electron-flavour survival probability $P_{ee}$ after the propagation of geoneutrinos from a geological reservoir located at $\vec{r \prime}$ to the detector is calculated considering oscillations in vacuum. The matter effect is estimated to be of the order of 1\%~\cite{baldoncini2015reference}, i.e., much less than other uncertainties involved in the geoneutrino signal prediction. The average survival probability $\left \langle P_{ee} \right \rangle$ is 0.55. Despite the dependence of $\left \langle P_{ee} \right \rangle$ w.r.t. energy, the geoneutrino energy spectrum is unchanged since the geoneutrinos undergo several oscillations before reaching the detector. The $\rho(\vec{r \prime})$ is the density of the voxel emitting geoneutrinos and it is taken from geophysical models of lithosphere~\cite{Huang2013} and mantle~\cite{Dziewonski:1981xy}. The abundances $a(i; \vec{r \prime})$ of isotope $i$ are expressed per mass unit of rock. The integration is done over the whole volume of the Earth, considering geological constraints of the main HPE reservoirs.

The antineutrino signal can be conveniently expressed in \emph{Terrestrial Neutrino Units} (TNU). 1\,TNU corresponds to 1 antineutrino event detected via IBD (Section~\ref{subsec:detection}) over 1\,year by a detector with 100\% detection efficiency containing 10$^{32}$ free target protons (roughly corresponding to 1\,kton of LS). To convert the differential geoneutrino flux $\frac{d\Phi (i;E_{\bar{\nu}},\vec{r})}{dE_{\bar{\nu}}}$ to geoneutrino signal $S (i)$ expressed in TNU, it is necessary to perform an integration over the geoneutrino energy spectra, considering the energy dependence of the IBD cross section~\cite{strumia2003precise}, and $10^{32}$ target protons in 1\,year measuring time. The geoneutrino energy spectrum that can be observed by Borexino extends from 1.8\,MeV (IBD threshold) to 3.7\,MeV. Note that for a reference oscillated flux of $10^6$\,cm$^{-2}$\,s$^{-1}$, the geoneutrino signals from $^{238}$U and $^{232}$Th are $S$(U) = 12.8\,TNU and $S$(Th) = 4.04\,TNU, respectively. Considering the specific antineutrino production rates $\varepsilon_{\nu} (\mathrm {U, Th})$\footnote{ $\varepsilon_{\nu} (\mathrm {U})$  = $7.41 \cdot 10^{7}$\,kg$^{-1}$\,s$^{-1}$, $\varepsilon_{\nu} (\mathrm  {Th})$  = $1.62 \cdot 10^{7}$\,kg$^{-1}$\,s$^{-1}$} one can calculate the signal ratio $R_S$ for a homogeneous reservoir characterised by a fixed $a$(Th)/$a$(U) ratio (where $a$ denotes the abundance):
\begin{equation}
    \begin{aligned}
        R_s=\frac{S(\mathrm{Th})}{S(\mathrm{U})}=0.069\frac{a(\mathrm{Th})}{a(\mathrm{U})}.
    \end{aligned}
    \label{eq:RS}
\end{equation}
This signal ratio thus depends on the composition of the reservoir. Adopting the CI chondrites $a$(Th)/$a$(U) = 3.9 for the bulk Earth, the signal ratio $R_s$ = 0.27 can be extracted. Geophysical and geochemical observations of the lihtosphere constrain the $a$(Th)/$a$(U) = 4.3~\cite{Agostini:2019dbs}, implying a signal ratio of 0.29. As a consequence of maintaining the global chondritic ratio of 3.9 for the bulk Earth, the inferred mantle ratio $a$(Th)/$a$(U) results to be 3.7, which corresponds to a $R_S$ = 0.26. This is used for the extraction of mantle signal explained in Section~\ref{subsec:mantle}.

\begin{figure}[t]
\centering
    \includegraphics[width = 0.85\textwidth]{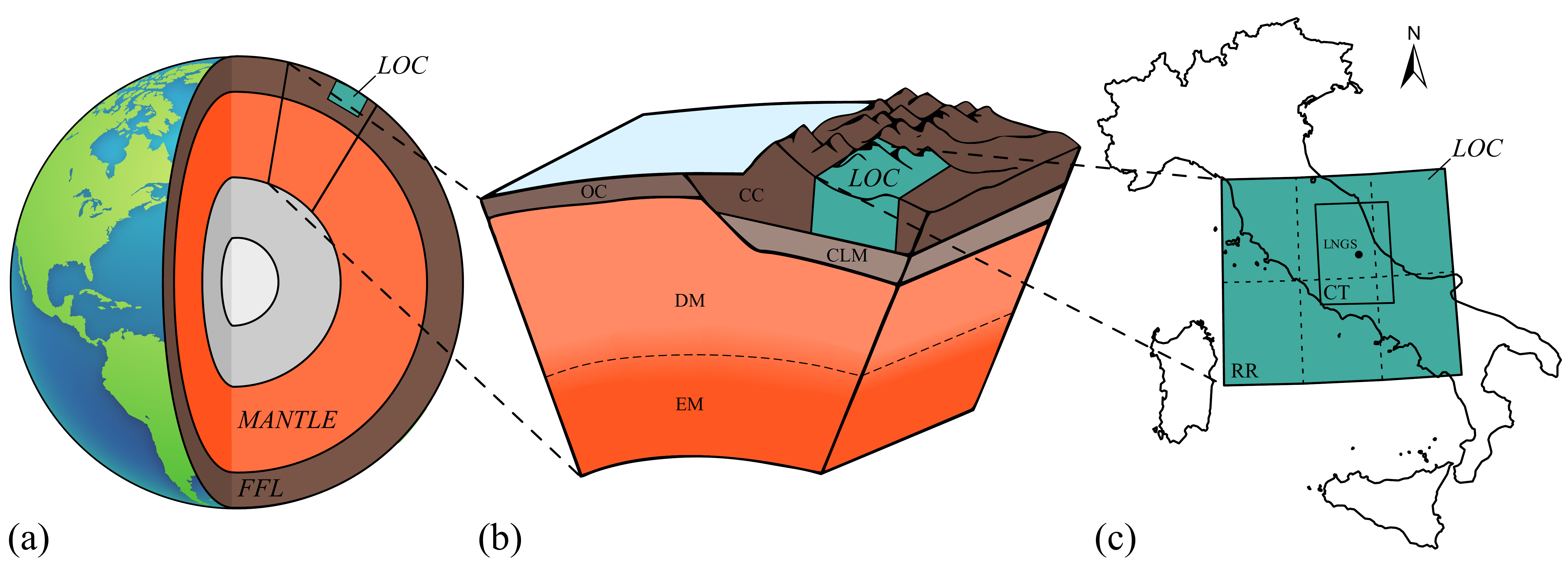}
        \caption{(a) Schematic drawing of the Earth’s structure, showing the three units contributing to the expected geoneutrino signal at LNGS: (i) the \emph{local crust} (LOC), (ii) the \emph{far field lithosphere} (FFL), and (ii) the \emph{mantle}. The inner and outer portions of the \emph{core} (in grey) do not contribute to the geoneutrino signal. Not to scale. (b) Schematic section detailing the components of the BSE. The \emph{lithosphere} includes the LOC and the FFL. The latter comprises the rest of the \emph{continental crust} (CC), the \emph{oceanic crust} (OC), and the \emph{continental lithospheric mantle} (CLM). In the mantle, two portions can be distinguished: a lower \emph{enriched mantle} (EM) and an upper \emph{depleted mantle} (DM). Not to scale. (c) Simplified map of the LOC. The \emph{central tile} (CT) of the $\ang{2}$ $\times$ $\ang{2}$ centreed at LNGS is modelled separately from the remaining six tiles which represent the \emph{rest of the region} (RR). Figure taken from~\cite{Agostini:2019dbs}.}
       \label{fig:GeolContrib} 
\end{figure}

The expected geoneutrino signal in Borexino $S$(U+Th) can be expressed as the sum of three components and this is shown in detail in Figure~\ref{fig:GeolContrib}.
\begin{itemize}
    \item $S_{\mathrm{LOC}}$(U+Th), the \emph{local crust} (LOC) signal produced from the 492\,km $\times$ 444\,km crustal area surrounding LNGS, 
    \item $S_{\mathrm{FFL}}$(U+Th), the signal from the \emph{far field lithosphere} (FFL), which includes the continental lithospheric mantle (CLM), i.e. the brittle portion of the mantle underlying the CC, and the remaining crust obtained after the removal of the LOC.
    \item $S_{\mathrm{mantle}}$(U+Th), the signal from the mantle. 
\end{itemize}
The sum of $S_{\mathrm{LOC}}$(U+Th) and $S_{\mathrm{FFL}}$(U+Th) is the expected signal from the bulk lithosphere (U + Th) and is evaluated to be 25.9$^{+4.9}_{-4.1}$\,TNU. This constraint is used for the extraction of mantle signal at LNGS, described in Section~\ref{subsec:mantle}. The mantle geoneutrino signal and the mantle radiogenic heat are highly debated topics in Earth sciences. The distribution of HPEs in the mantle is not known and the assumed different scenarios influence the strength of the mantle signal. There is a \emph{high scenario}, which assumes homogeneous HPE distribution, an \emph{intermediate scenario}, that assumes HPEs distributed in two layers, a \emph{lower enriched mantle} (EM) and an upper \emph{depleted mantle} (DM) separated at 2180\,km of depth, and a \emph{low scenario}, where the HPEs are placed just above the CMB. Assuming these scenarios, the predictions of the mantle geoneutrino signal $S_{\mathrm{mantle}}$(U+Th) vary between 0.9 and 33.0\,TNU for different BSE models. Using the relatively well-known contribution from the bulk lithosphere, and the various predictions for the mantle signal, the total geoneutrino signal at LNGS $S$(U+Th) varies between 28.5 and 55.3\,TNU. The individual predictions for the different BSE models discussed in Section~\ref{subsec:bse} are: Cosmochemical CC = 28.5$^{+5.5}_{-4.8}$\,TNU, Geochemical GC = 34.6$^{+5.5}_{-4.8}$\,TNU, Geodynamical GD = 45.6$^{+5.6}_{-4.9}$\,TNU, Fully Radiogenic FR = 55.3$^{+5.7}_{-5.0}$\,TNU~\cite{Agostini:2019dbs}.

\subsubsection{Other antineutrino signals}
\label{subsec:exp-antinu-bckg}

The antineutrino backgrounds relevant for the geoneutrino analysis (Section~\ref{sec:geo-analysis}) are described below. The main source of background in geoneutrino detection is the production of electron antineutrinos by nuclear power plants, the strongest man-made antineutrino source. In addition, atmospheric neutrinos can have a small impact. This has been treated as a systematic source of uncertainty in the recent measurement~\cite{Agostini:2019dbs}. Another source of antineutrinos can be a hypothetical georeactor inside the Earth, searched for in a separate analysis (Section~\ref{subsec:georea}).

\paragraph{\emph{Reactor antineutrinos}}
Many nuclei, produced in the fission process of nuclear fuel, decay through $\beta$-processes, with the consequent emission of electron antineutrinos, the so-called \emph{reactor antineutrinos}. Their energy spectrum extends up to $\simeq$10\,MeV, well beyond the end point of the geoneutrino spectrum (3.27\,MeV). As a consequence, in the geoneutrino energy window (1.8 - 3.27\,MeV), there is an overlap between geoneutrino and reactor antineutrino signals. At present, there are approximately 440 nuclear power reactors in the world. A reactor with a typical thermal power of 3\,GW emits $5.6 \times 10^{20} \,  \bar{\nu}_e$ s$^{-1}$, considering $\sim$200\,MeV average energy released per fission and 6 $\bar{\nu}_e$ produced along the $\beta$-decay chains of the neutron-rich unstable fission products. The nominal thermal power and the monthly load factors of the reactors are taken from the \emph{Power Reactor Information System} (PRIS), developed and maintained by the International Atomic Energy Agency (IAEA)~\cite{IAEA}. An accurate determination of the expected signal and spectrum of reactor antineutrinos requires a wide set of information, spanning from the characteristics of nuclear cores to neutrino properties and is discussed in detail in~\cite{Agostini:2019dbs}.  Various experimental results from reactor antineutrino experiments show that the IBD positron energy spectrum deviates significantly from the spectral predictions of Mueller et al. 2011~\cite{mueller2011improved} in the energy range between 4 - 6\,MeV, i.e. the so-called  ``5\,MeV excess'' and also an overall deficit with respect to the prediction. These features are treated as a systematic uncertainty (Section~\ref{subsec:syst-geo}) using the Daya Bay high precision measurement~\cite{an2016measurement}. The expected signal from reactor antineutrinos can be expressed in TNU, assuming 100\% detection efficiency for a detector containing 10$^{32}$ target protons and operating continuously for 1\,year. The uncertainties related to reactor antineutrino production, propagation, and detection processes are estimated using a Monte Carlo-based approach discussed in~\cite{baldoncini2015reference}. The expected reactor antineutrino signal with and without the ``5\,MeV excess'' at LNGS for the period December 2007 to April 2019 are 79.6$^{+1.4}_{-1.3}$\,TNU and 84.5$^{+1.5}_{-1.4}$\,TNU, respectively.

\paragraph{\emph{Atmospheric neutrinos}}
Atmospheric neutrinos can act as a potential background source for the geoneutrino measurement and this was studied for the first time in~\cite{Agostini:2019dbs}. Atmospheric neutrinos originate in sequential decays of $\pi^{\pm}$ and K$^{\pm}$ mesons and $\mu^{\pm}$ leptons produced in cosmic rays' interactions with atmospheric nuclei. The flux of atmospheric neutrinos contains both neutrinos and antineutrinos, and the muon flavour is roughly twice abundant with respect to the electron flavour. The process of neutrino oscillations then alters the flavour composition of the neutrino flux passing through the detector. Atmospheric neutrinos interact in many ways with the nuclei constituting the Borexino scintillator. The most copious isotopes in the Borexino scintillator are $^1$H ($6.00\times 10^{31}$/kton), $^{12}$C ($4.46\times 10^{31}$/kton), and $^{13}$C ($5.00\times 10^{29}$/kton). Besides the IBD reaction itself, there are many reactions with $^{12}$C and $^{13}$C atoms that may, in some cases, mimic the IBD interactions. They have the form of $\nu + A \to \nu(l) + n + \dots + A'$, where $A$ is the target nucleus, $A'$ is the nuclear remnant, $l$ is charged lepton produced in CC processes, $n$ is the neutron, and dots are for other produced particles like nucleons (including additional neutrons) and mesons (mostly $\pi$ and $K$ mesons). A dedicated simulation code was developed to precisely calculate this background in Borexino~\cite{Agostini:2019dbs}. The expected number of IBD-like interactions due to atmospheric neutrinos in the energy window used for the analysis was evaluated via MC to be 6.7 $\pm$ 3.7 events in the analysed period.

\paragraph{\emph{Georeactor}}

A possible existence of a georeactor, i.e. a natural nuclear fission reactor in the Earth's interior, was first suggested by Herndon in 1993~\cite{ herndon1993feasibility}. Since then, several authors have discussed its possible existence and its characteristics. Different models suggest the existence of natural nuclear reactors at different depths~\cite{herndon1996substructure, rusov2007geoantineutrino, de2008feasibility}. New upper bounds on the power of a potential georeactor are discussed in Section~\ref{subsec:georea}. In order to be able to set such limits for different hypothetical locations of the georeactor, the expected antineutrino spectra of a 1\,TW point-like georeactor were calculated. It was assumed to operate continuously during the entire analysis period with the constant power fractions of fuel components, as suggested in~\cite{herndon2005background} ($^{235}$U : $^{238}$U $\approx$ 0.76 : 0.23). The energy released and the antineutrino spectra per fission were calculated similar to reactor antineutrinos, using only the flux parametrisation of~\cite{mueller2011improved}. Three different depths were considered for the georeactor location: 1) GR1: the Earth's centre ($d$ = $R_{\mathrm{Earth}}$), 2) GR2: the CMB at just below the LNGS site ($d$ = 2900\,km), and 3) GR3: the CMB on the opposite hemisphere ($d$ = 2$R_{\mathrm{Earth}}-2900$ = 9842\,km). In the calculation of the survival probability, the matter effect is taken into account by assuming a constant Earth density (density variation in between the crust and the core causes about 3\% signal variation). Unfortunately, Borexino's energy resolution does not allow to distinguish the spectral differences due to oscillations. The errors were calculated via the method similar to that used for reactor antineutrinos. The expected georeactor signals for a 1\,TW georeactor located at GR1, GR2, and GR3 are 43.1$\pm$ 1.3\,TNU, 8.9 $\pm$ 0.3\,TNU, and 3.7 $\pm$ 0.1\,TNU, respectively.

\subsubsection{Non-antineutrino backgrounds}
\label{subsec:exp-nonanti-bckg}

There are various non-antineutrino backgrounds that can mimic an IBD and should therefore be estimated for the geoneutrino analysis. The three important non-antineutrino backgrounds that are constrained in the final spectral fit (Section~\ref{subsec:geo-fit}) -  cosmogenic $^{9}$Li, accidental, and ($\alpha$, n) backgrounds - are described in more detail below. In addition to these, there are various other minor backgrounds that are also estimated. These include: backgrounds due to untagged muons, fast neutrons from the surrounding rocks and the water tank, ($\gamma, n$) reactions, fission reactions in PMTs, and $^{214}$Bi-$^{214}$Po coincidences due to $^{222}$Rn contamination during the water extraction period. Table~\ref{tab:backg_summary} summarizes the expected number of events from all non-antineutrino backgrounds passing the IBD selection cuts discussed in Section~\ref{subsec:geo-selcuts}.

\paragraph{\emph{Cosmogenic $^{9}$Li}}

Cosmic muons reaching Borexino can produce many spallation products that can in turn produce ($\beta^{-}$\,+\,n) pairs as shown in Table~\ref{tab:cosmo_bckg}. Such decays are indistinguishable from the IBD events, since the prompt is represented by the $\beta^{-}$ and the delayed is due the neutron, exactly as in the IBD. The hadronic background for geoneutrino analysis is dominated by the $^{9}$Li isotope. It has a \emph{Q}-value of 13.6\,MeV, a lifetime of 257.2\,ms~\cite{Bellini:2013cosmo} and decays according to the following process:
    \begin{equation}
      	^{9}\text{Li} \rightarrow e^{-} + \bar{\nu}_{e} + n + 2\alpha.
      \label{eq:Li9}
   \end{equation}
The $^{9}$Li background can be estimated by studying the events that follow immediately after a cosmic muon, using the same IBD selection cuts described in Section~\ref{subsec:geo-selcuts}. The residual background after the time and space muon vetoes can then be estimated using the measured lifetime of the $^{9}$Li \emph{prompts} and their distance to the muon, respectively.
In order to account for other cosmogenic isotopes such as $^{8}$He and $^{9}$B, which can also be present in the muon daughter sample, the measured lifetime from Borexino data (260 $\pm$ 21)\,ms~\cite{Agostini:2019dbs} is used to evaluate this background instead of the $^{9}$Li lifetime.

\paragraph{\emph{Accidental background}}
Accidental coincidences happen due to the random coincidences of prompt-like and delayed-like events in Borexino. This is an important background for the geoneutrino analysis and depends on the selection cuts applied. They are mostly due to external backgrounds, since their reconstructed positions are mostly near the IV. They are usually searched for in an off-time window of 2\,s to 20\,s after the prompt and scaled back to the IBD time window of 1.28\,ms (Section~\ref{subsec:geo-selcuts}). Such a long time is used to maximize the number of selected accidental coincidences and thus to reduce the error with which the shape and rate of this background can be constrained.

\paragraph{\emph{($\alpha, n$) coincidences}}
Decays along the chains of $^{238}$U, $^{235}$U, and $^{232}$Th can produce $\alpha$-particles, which in turn can give rise to ($\alpha, n$) interactions that can mimic IBD coincidences. In Borexino, this is only due to the $^{210}$Po isotope; while it is a part of the $^{238}$U chain, it is found fully out of the secular equilibrium and its contamination is several orders of magnitudes higher than the rest of the chain. The $\alpha$ particle can then interact with other isotopes, mostly $^{13}$C in the LS:
 \begin{equation}
    ^{13}\text{C} + \alpha \longrightarrow \: ^{16}\text{O} + n.
    \label{eq:alpha_n}
 \end{equation}
The cross section of this interaction is 200\,mb~\cite{Mohr:2018alphaNBgr} and the produced neutron can have energies up to 7.3\,MeV, almost indistinguishable from the IBD \emph{delayed}. There are three different possibilities for a prompt-like interaction, which are described in detail in~\cite{Agostini:2019dbs}. This background is estimated using the amount of $^{210}$Po events (tagged on an event-by-event basis via pulse shape discrimination techniques, see Section~\ref{subsec:event-reco}) in the entire analysed exposure, the probability of an ($\alpha, n$) interaction to produce IBD-like coincidences passing the selection cuts, and the neutrino-yield of the reaction in pseudocumene.

\begin{table} [t]
\centering
		\caption{Summary of the expected number of events from non-antineutrino backgrounds in the antineutrino candidate sample (exposure $\mathcal{E}_p$ = (1.29 $\pm$ 0.05) $\times 10^{32}$\, protons $\times$ yr). Upper limits are given at 95\% C.L. From~\cite{Agostini:2019dbs}}. \label{tab:backg_summary}
	\begin{tabular}{cc}
		\toprule
		Background Type & Events \\
        \midrule
	    $^9$Li background& 3.6 $\pm$ 1.0 \\
		Untagged muons & 0.023 $\pm$ 0.007 \\
		Fast n's ($\mu$ in water tank) & $<$0.013 \\
		Fast n's ($\mu$ in rock) & $<$1.43  \\
		Accidental coincidences & 3.846 $\pm$ 0.017 \\
		($\alpha$, n) in scintillator & 0.81 $\pm$ 0.13 \\ 
		($\alpha$, n) in buffer & $<$2.6  \\
		($\gamma$, n) & $<$0.34 \\
		Fission in PMTs & $<$0.057   \\
		$^{214}$Bi-$^{214}$Po & 0.003 $\pm$ 0.001\\
		\midrule
		Total & 8.28 $\pm$ 1.01 \\			
		\bottomrule
	\end{tabular}
	
\end{table}

\subsubsection{Selection cuts}\label{subsec:geo-selcuts}

The various criteria used to select IBD candidates for the geoneutrino analysis in Borexino are described below. The detection efficiency of geoneutrinos for these selection cuts was estimated through MC to be (87.0 $\pm$ 1.5)\%.

    \paragraph {\emph{Muon vetoes}} Muon vetoes are applied after both internal and external muons in Borexino, in order to reduce the cosmogenic background. Generally, a 2\,ms veto is applied after external muons, similarly to the solar neutrino analysis, in order to remove fast neutrons. In the geoneutrino analyses performed until 2015, a conservative 2\,s veto (8 times the lifetime of $^{9}$Li) was applied after internal muons to eliminate muon daughters. In the improved analysis performed in 2019~\cite{Agostini:2019dbs}, 4 different kinds of more complex space-time vetoes were applied, depending on the type of muon. This reduced the exposure loss from 10-11\% in the previous analyses to 1.2\% in the latest analysis.
    
    \paragraph {\emph{Energy cuts}} Energy cuts are applied on both \emph{prompt} and \emph{delayed}, after the consideration of the IBD threshold and the $n$-capture energy peaks described in Section~\ref{subsec:detection}. The energy spectrum of \emph{prompt} starts at $\sim$1\,MeV, which corresponds to the two 0.511\,MeV gammas. The threshold energy of the \emph{prompt} is chosen as 408\,p.e., corresponding to about 0.8\,MeV. There is no upper limit on the energy of the \emph{prompt}. The energy of the delayed can either be due to the neutron capture on $^{1}$H (2.2\,MeV gamma) or on $^{12}$C with 1.1\% probability (4.95\,MeV gamma). However, at large radii, gammas can partially deposit their energy in the buffer, decreasing the visible energy. Consequently, the gamma-peak develops a low-energy tail and even the peak position can be shifted to lower values. Therefore, in the latest geoneutrino analysis, the energy ranges 700-1300\,p.e and 1300-3000\,p.e. were used for IBD selection, to account for the n-captures on $^{1}$H and $^{12}$C, respectively. However, the lower energy threshold was increased to 860\,p.e. for the water extraction period between 2010 and 2011, because of increased $^{222}$Rn contamination.

    \paragraph {\emph{Time space correlation}} 
    The cuts applied on the time and distance between the \emph{prompt} and the \emph{delayed} are important background suppression cuts. The IBD events can either have both the \emph{prompt} and \emph{delayed} clusters in the same 16\,$\mu$s DAQ window (double cluster events) or have them in separate DAQ windows as single cluster point-like events. A time coincidence window of 20-1280\,$\mu$s is used for IBD selection of single cluster events. The lower threshold guarantees that the delayed event can trigger after the DAQ deadtime of 3-4\,$\mu$s after the prompt event and the higher threshold corresponds to 5 times the measured neutron capture time of (254.5 $\pm$ 1.8)\,$\mu$s~\cite{Bellini:2011yd}. In the latest geoneutrino analysis, the time window 2.5-12.5\,$\mu$s was also considered, in order to include double cluster events. The lower threshold was tuned on reactor antineutrino MC and guarantees that even for the highest energy \emph{prompt}, the light from the \emph{prompt} cannot alter the delayed event after 2.5\,$\mu$s. The higher threshold was chosen by studying the variable position of the prompt event inside the DAQ window in the entire analysed period. In addition, it also allows the delayed event to have a cluster duration up to 2.5\,$\mu$s before the end of the DAQ window.
    
    The reconstructed distance between the \emph{prompt} and the \emph{delayed} is larger than the distance between their points of production due to different reasons~\cite{Agostini:2019dbs}. The important factors to be considered before choosing the spatial cut are the MC efficiency and the accidental background. The efficiency drops quickly below 1\,m and the accidental background steadily increases with increasing distance. This spatial cut was optimised to be 1.3\,m, using MC-based sensitivity studies.
    
    \paragraph {\emph{Pulse-shape discrimination}}
    
    Pulse shape discrimination is applied to the \emph{delayed} in order to distinguish it from other $\alpha$-like coincidences. The MLP parameter mentioned in Section~\ref{subsec:event-reco} was used for the latest geoneutrino analysis. A cut $\textgreater$ 0.8 was applied on the \emph{delayed} to distinguish it from ($\alpha$+$\gamma$) decays of $^{214}$Po~\footnote{In 1.04 $\times$ $10^{-4}$ and in 6 $\times$ $10^{-7}$ of cases, the $^{214}$Po decays to excited states of $^{210}$Pb and the $\alpha$ is accompanied by the emission of prompt gammas of 0.7997\,MeV and of 1.0977\,MeV energy, respectively}, arising from $^{222}$Rn contamination during the water extraction period.

    \paragraph {\emph{Dynamical Fiducial Volume cut}}
    
    The vessel shape of Borexino changes in time due to the presence of a small leak and it can be reconstructed, as discussed in Section~\ref{subsec:detector-setup}. The Dynamical Fiducial Volume (DFV) cut is applied in the geoneutrino analysis based on the distance of the \emph{prompt} from the IV. The cut was optimised to be 10\,cm, using MC-based sensitivity studies in the latest geoneutrino analysis. This distance is sufficient enough to suppress background events near the IV, and to account for the uncertainty due to the IV shape reconstruction. The enlarged fiducial volume in the recent analysis resulted in a 15~\% increase in exposure when compared to the previous analyses.
    
    \paragraph {\emph{Multiplicity cut}} 
    
    The multiplicity cut is applied to suppress neutron-neutron or buffer muon-neutron pairs that can enter the IBD selection due to undetected muons. This cut requires that there are no high energy events ($\textgreater$ 400\,p.e.) present in a $\pm$2\,ms time window (about 8 times neutron capture time) around the \emph{prompt} or \emph{delayed}. The corresponding exposure  loss  due  to  accidental  coincidences  of  IBD candidates with $\textgreater$400 p.e. events within 2\,ms is of the order of 0.01\%, which is negligible for this analysis.

\subsubsection{Spectral fit}
\label{subsec:geo-fit}

After the selection of IBD candidates, the geoneutrino signal is extracted from the unbinned-likelihood spectral fit of the charges of all $prompts$:
\begin{equation}
  L = (\vec{\theta}; \vec{N}_{pe}^{\mathrm{p}}) = \prod_{i=1}^{N_{\mathrm {IBD}}} L(\vec{\theta}; N_{pe}^{\mathrm{p_i}}),
  \label{eq:Lkl}
\end{equation}
where $\vec{N}_{pe}^{\mathrm{p}}$ is the vector of individual \emph{prompt} charges $N_{pe}^{\mathrm{p_i}}$, and index $i$ runs from 1 to $N_{\mathrm IBD}$, i.e. the total number of IBD candidates. The symbol $\vec{\theta}$ indicates the set of the variables with respect to which the function is maximised, namely the number of events of signal and backgrounds. The shapes of all spectral components used in the fit are taken from the Probability Distribution Density Functions (PDFs) constructed using MC, with the exception of the accidental background which can be measured with sufficient precision using off-time coincidences in data. The result of the fit is the number of events due to each spectral component. 

The spectral fit is generally performed in the range 408-4000\,p.e. ($\approx$0.8-8\,MeV). The number of geoneutrinos is always kept free. One way of doing it is by having one free fit parameter for geoneutrinos, when using the PDF in which the $^{232}$Th and $^{238}$U contributions are summed and weighted according to the chondritic mass ratio of 3.9, corresponding to a signal ratio of 0.27 (Section~\ref{subsec:exp-signal}). Alternatively, $^{232}$Th and $^{238}$U contributions can be fit as two independent contributions. Additional combinations are also possible. The number of reactor antineutrino events is typically kept free. This is an important cross-check of the ability of Borexino to measure electron antineutrinos and can be verified by comparing the fit results to the well-known prediction of reactor antineutrinos in Section~\ref{subsec:exp-antinu-bckg}. The three non-antineutrino backgrounds (Section~\ref{subsec:exp-nonanti-bckg}) are constrained using additional multiplicative Gaussian pull terms in the likelihood function of Equation~\ref{eq:Lkl}. Using the known exposure and detection efficiency, the number of detected geoneutrinos and reactor antineutrinos can then be expressed in the units of TNU.	

\subsubsection{Systematic sources of uncertainty}
\label{subsec:syst-geo}

The systematic uncertainty on the measured geoneutrino signal is very small compared to the statistical uncertainty. The total uncertainties on the geoneutrino and reactor antineutrino signals were evaluated to be $^{+5.2}_{-4.0}$\% and $^{+5.1}_{-5.5}$\%, respectively. The different systematic sources are discussed below and summarised in Table~\ref{tab:geo-summary_sys}.
\begin{itemize}
     \item  \emph{Atmospheric neutrinos:} 
    Atmospheric neutrinos as a source of background for geoneutrinos were discussed in Section~\ref{subsec:exp-antinu-bckg}. The uncertainty on the expected number of atmospheric neutrino events is estimated to be around 50\%. Therefore, the atmospheric neutrino PDF was included in the standard spectral fit, in order to evaluate the impact of this systematic source.  
     \item  \emph{Shape of reactor antineutrino spectra:}
    The standard spectral fit was performed using the MC PDF of reactor antineutrinos without the ``5\,MeV excess'' which was discussed in Section~\ref{subsec:exp-antinu-bckg}. The effect of this assumption was studied by using PDFs with and without the ``5\,MeV excess'' in the spectral fit.
   \item\emph{Inner vessel shape reconstruction:}
    The systematic uncertainty on the FV definition due to the 5\,cm error on the IV shape reconstruction is negligible. However, there is a systematic uncertainty associated with the selection of the IBD candidates using the DFV cut, which was evaluated by smearing the distance-to-IV of each IBD candidate with a Gaussian function ($\sigma$ = 5\,cm). Then, the spectral fit was performed on the newly selected candidates with the DFV cut and repeated 50 times. 
    \item  \emph{MC efficiency:}
    The major source of uncertainty for the MC efficiency arises from the event losses close to the IV edges, especially near the south pole, because of the combined effect of a large number of broken PMTs and the IV deformation.
     \item  \emph{Position reconstruction:}
    Since the events are selected inside the DFV based on the reconstructed position, the uncertainty in the position reconstruction of events affects the error on the fiducial volume, and thus, on the resulting exposure.   
\end{itemize}

\begin{table} [t]
\centering
	\caption{Summary of the different sources of systematic uncertainty in the geoneutrino and reactor antineutrino measurement. Different contributions are summed up as uncorrelated. From~\cite{Agostini:2019dbs}.}
	\label{tab:geo-summary_sys}
	\begin{tabular}{ccc}
		\toprule
		Source & Geo &  Reactor \ \\
               & Error &  Error \\
               &  [\%] &  [\%]  \ \\
		\midrule
		Atmospheric neutrinos & {\Large $^{+0.00}_{-0.38}$} & {\Large $^{+0.00}_{-3.90}$} \Large \\ [8pt]
		Shape of reactor spectrum &  {\Large  $^{+0.00}_{-0.57}$} & {\Large $^{+0.04}_{-0.00}$} \\ [8pt]
		Vessel shape & {\Large $^{+3.46}_{-0.00}$} & {\Large $^{+3.25}_{-0.00}$} \\ [8pt]
		Efficiency & 1.5 & 1.5 \\ [8pt]
		Position reconstruction & 3.6  & 3.6 \ \\
		\midrule
		Total &  {\Large $^{+5.2}_{-4.0}$} & {\Large $^{+5.1}_{-5.5}$} \Large\Large \\
		\bottomrule
	\end{tabular}

\end{table}

\subsection{Results and geological interpretations}
\label{sec:geo-results}

In the period between December 9, 2007 and April 28, 2019, corresponding to 3262.74 days of data acquisition, 154 IBD candidates were found~\cite{Agostini:2019dbs}. The exposure of (1.29 $\pm$ 0.05) $\times$ 10$^{32}$ protons $\times$ year represents an increase by a factor of two with respect to the previous analysis in 2015~\cite{Agostini:2015cba}. The time, spatial, and energy distributions of the IBD candidates were compatible with the expectations. The following subsections discuss the geoneutrino signal measured by Borexino at LNGS (Section~\ref{subsec:results-geo-signal}), the extraction of the mantle geoneutrino signal (Section~\ref{subsec:mantle}), the calculation of the radiogenic heat and Urey ratio from Borexino measurement and its comparison to different BSE model predictions (Section~\ref{subsec:results-heat-urey}), and the upper limits on a hypothetical georeactor (Section~\ref{subsec:georea}).

\subsubsection{Geoneutrino signal at LNGS} \label{subsec:results-geo-signal}

\begin{figure}[t]
     \centering
    \subfigure[]{\includegraphics[width = 0.45\textwidth]{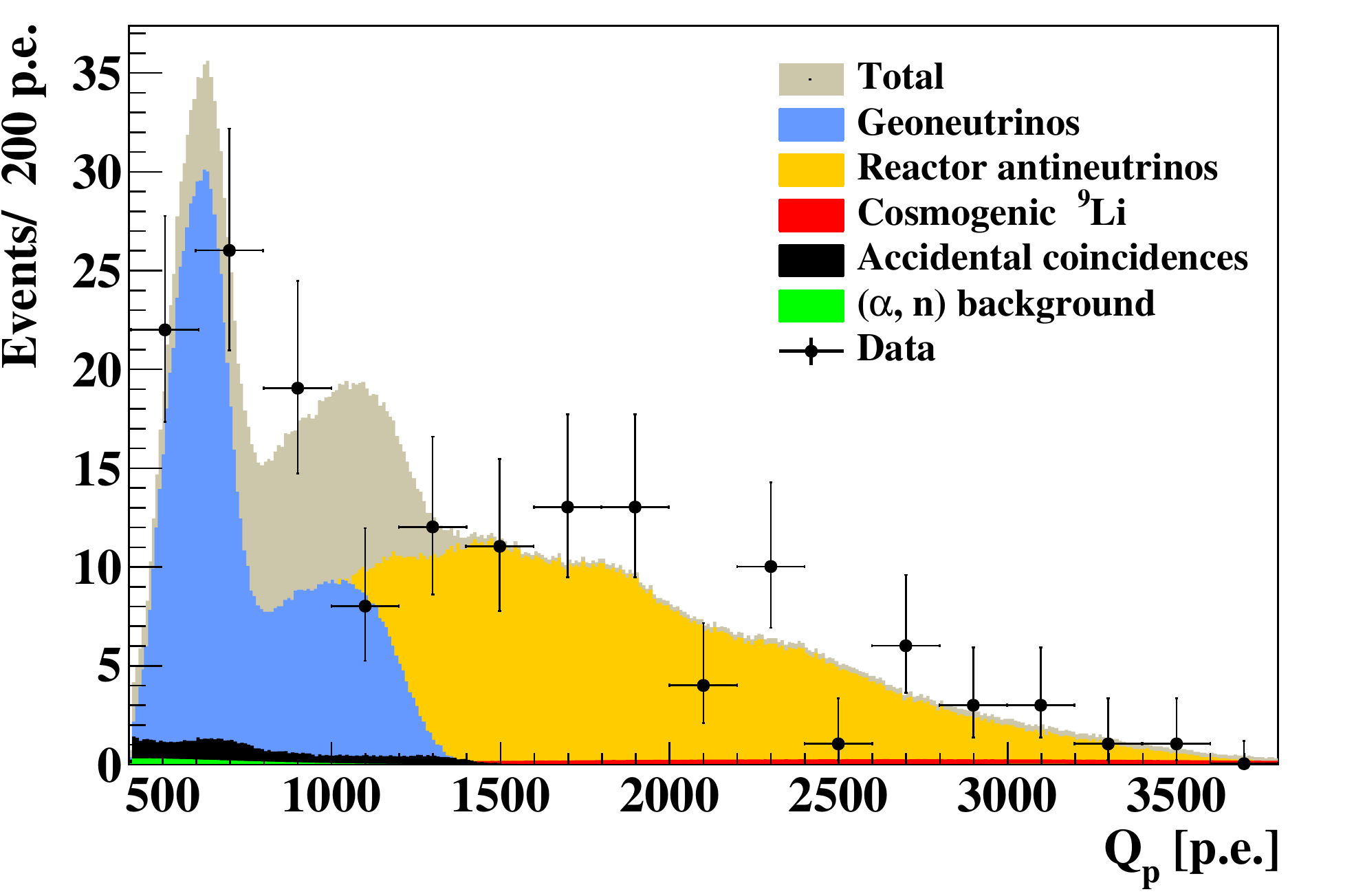}}
    \subfigure[]{\includegraphics[width = 0.45\textwidth]{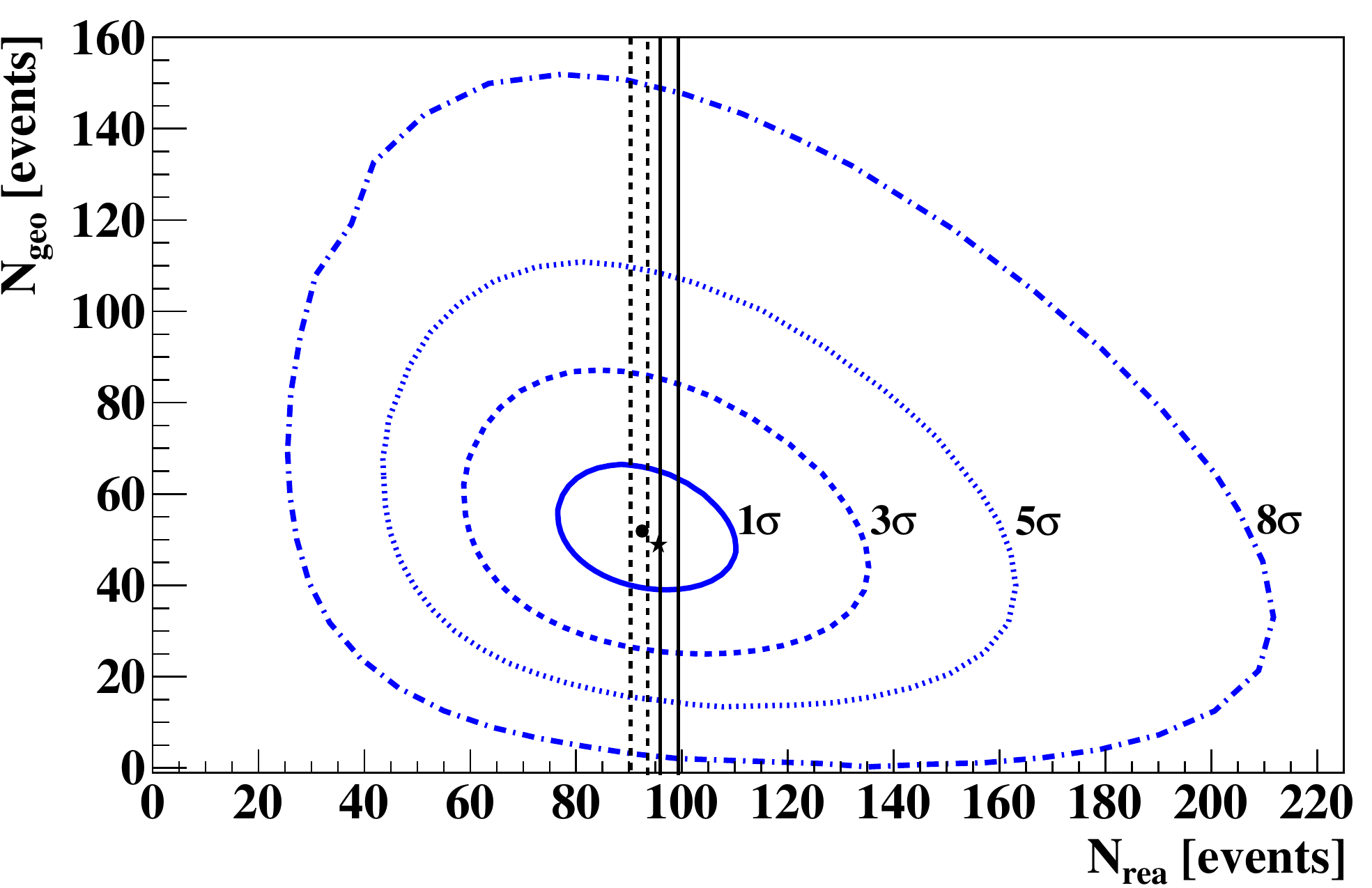}}
     \caption{Results of the analysis of 154 golden IBD candidates. (a) Spectral fit of the data (black points with Poissonian errors) assuming the chondritic Th/U ratio. The total fit function containing all signal and background components is shown in brownish-grey. Geoneutrinos (blue) and reactor antineutrinos (yellow) were kept as free fit parameters. Other non-antineutrino backgrounds were constrained in the fit. (b) The best fit point (black dot) and the contours for the 2D coverage of 68, 99.7, (100 - $5.7 \times 10^{-5})$\%, and (100 - $1.2 \times 10^{-13})$\%, (corresponding to 1, 3, 5, and 8$\sigma$, respectively), for number of geoneutrinos  (N$_{\mathrm{geo}}$) versus number of reactor antineutrinos (N$_{\mathrm{rea}}$), assuming Th/U chondritic ratio. The vertical lines mark the 1$\sigma$ bands of the expected reactor antineutrino signal (solid - without ``5\,MeV excess", dashed - with ``5\,MeV excess"). For comparison, the star shows the best fit performed assuming the $^{238}$U and $^{232}$Th contributions as free and independent fit components. From~\cite{Agostini:2019dbs}.}
     \label{fig:fits_contour}
    \end{figure}

\begin{figure}[t]
\centering
\subfigure[]{\includegraphics[width = 0.50\textwidth]{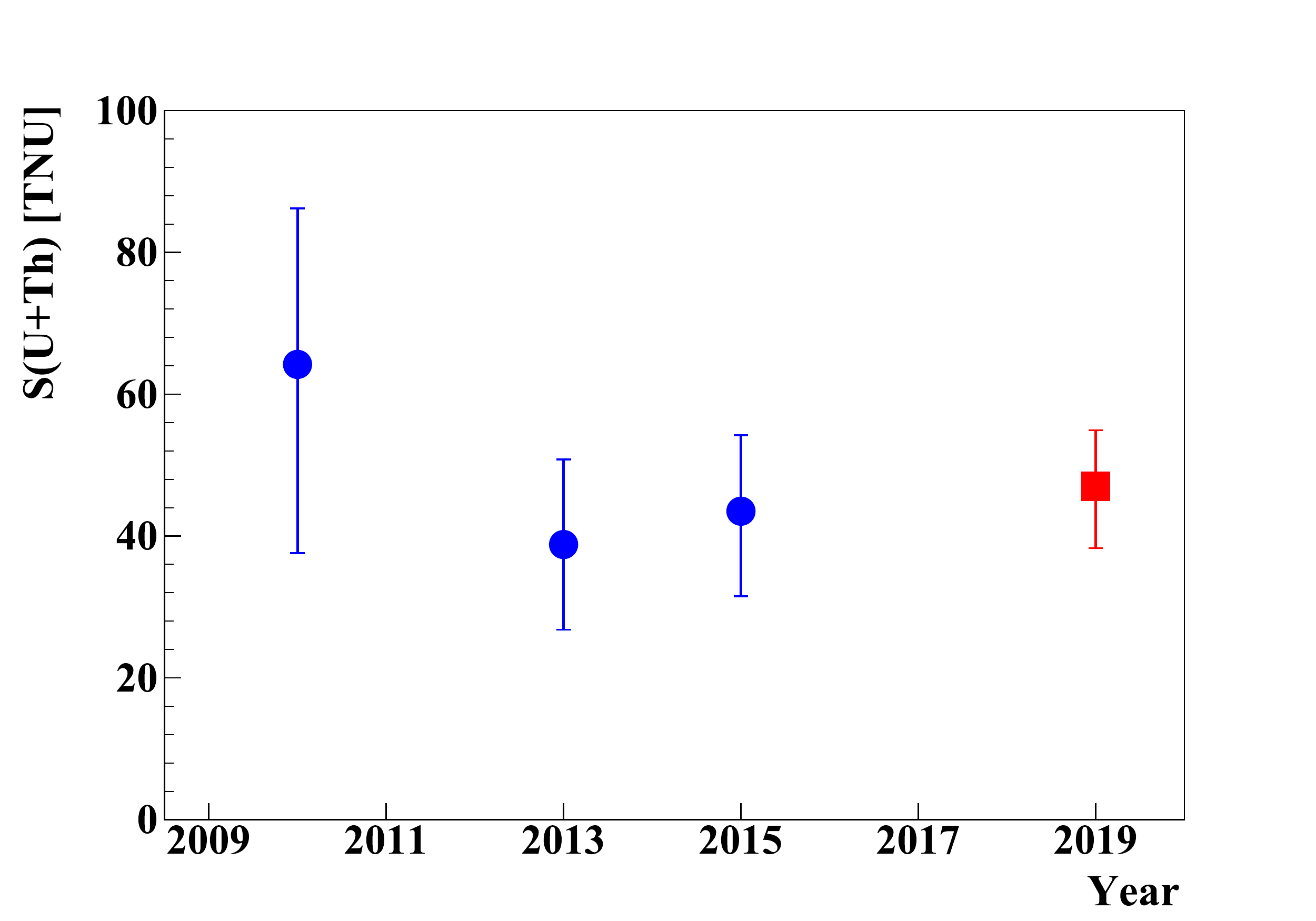}\label{fig:geo_signal_LNGS}}
\subfigure[]{\includegraphics[width = 0.45\textwidth]{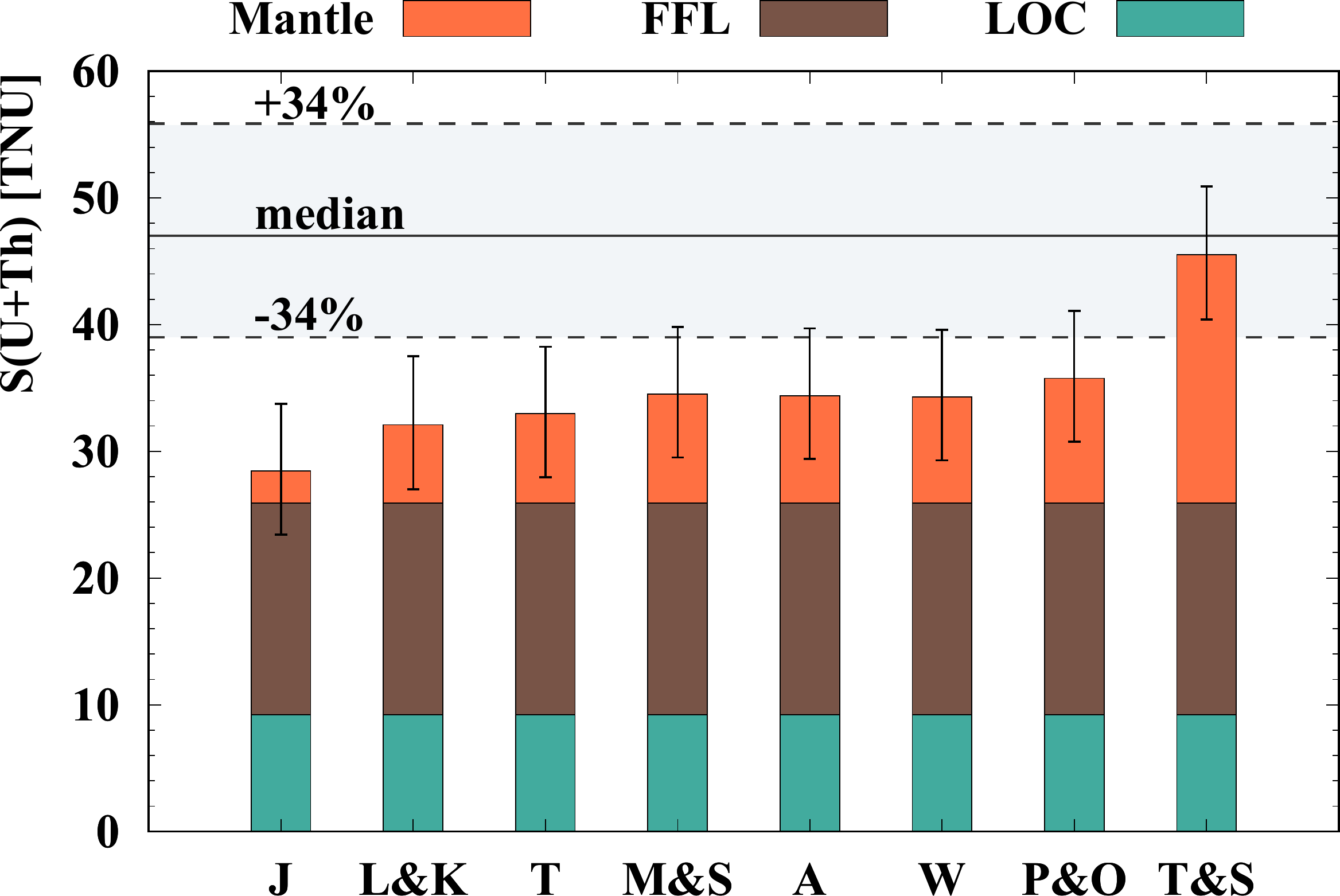}\label{fig:Sgeo_Bx_vs_model}}
\caption{(a) Comparison of the geoneutrino signal $S_{\mathrm{geo}}$(U+Th) at LNGS as measured by Borexino. Blue circles indicate the results from 2010~\cite{Bellini:2010geo}, 2013~\cite{Bellini:2013geo}, and 2015~\cite{Agostini:2015cba}, while the red square represents the most recent result from 2019~\cite{Agostini:2019dbs}. (b) Comparison of the expected geoneutrino signal $S_{\mathrm{geo}}$(U+Th) at LNGS (calculated according to different BSE models) with the Borexino measurement~\cite{Agostini:2019dbs}. For each model, the bulk lithosphere contribution (LOC+FFL) is the same, while the mantle signal, obtained considering an intermediate scenario, varies. The horizontal solid back line represents the median geoneutrino signal $S_{\mathrm {geo}}^{\mathrm{med}}$, while the grey band represents the 68\% coverage interval. From~\cite{Agostini:2019dbs}.}
\end{figure}

The spectral fit described in Section~\ref{subsec:geo-fit} was performed on the \emph{prompt} charge\textcolor{magenta}{\st{s}} of the 154 IBD candidates.  The three major non-antineutrino backgrounds, namely, the cosmogenic $^{9}$Li background, the ($\alpha$, n) background from the scintillator, and accidental coincidences were included in the fit and were constrained according to values in Table~\ref{tab:backg_summary} with Gaussian pull terms. The geoneutrino and reactor antineutrino contributions were left free in the fit and the geoneutrino MC PDF was constructed assuming a Th/U chondritic ratio of 3.9. This resulted in 52.6$_{-8.6}^{+9.4}$ (stat) $_{-2.1}^{+2.7}$ (sys) geoneutrinos from $^{238}$U and $^{232}$Th, corresponding to a geoneutrino signal of $S_{geo}$(U+Th) = 47.0$_{-7.7}^{+8.4}$ (stat)$_{-1.9}^{+2.4}$ (sys)\,TNU at LNGS. The resulting reactor antineutrino signal from the spectral fit was 80.5$_{ -9.3}^{+9.8}$ (stat) $_{-4.4}^{+4.1}$ (sys)\,TNU and is compatible with the expectations discussed in Section~\ref{subsec:exp-antinu-bckg}, confirming the ability of Borexino to measure antineutrinos. The resulting spectral fit and the  1, 3, 5, and 8$\sigma$ contours for the number of geoneutrinos versus reactor antineutrinos are shown in Figure~\ref{fig:fits_contour}. Compatible results were obtained when the spectral fit was performed by leaving the $^{238}$U and $^{232}$Th contributions free, without fixing them to the chondritic ratio. Unfortunately, Borexino does not have the sensitivity to measure the Th/U ratio. 
The geoneutrino signal measured by Borexino over the years is compared with the latest measurement in Figure~\ref{fig:geo_signal_LNGS}. The signal along with the 68\% C.I. is compared to different BSE models in Figure~\ref{fig:Sgeo_Bx_vs_model}. The bulk lithosphere contribution for all these models is the same as discussed in Section~\ref{subsec:exp-signal}, while the mantle signal varies according to the different model predictions.

\subsubsection{Extraction of mantle signal}
\label{subsec:mantle}

\begin{figure}[t]
  \centering
 \includegraphics[width = 0.6\textwidth]{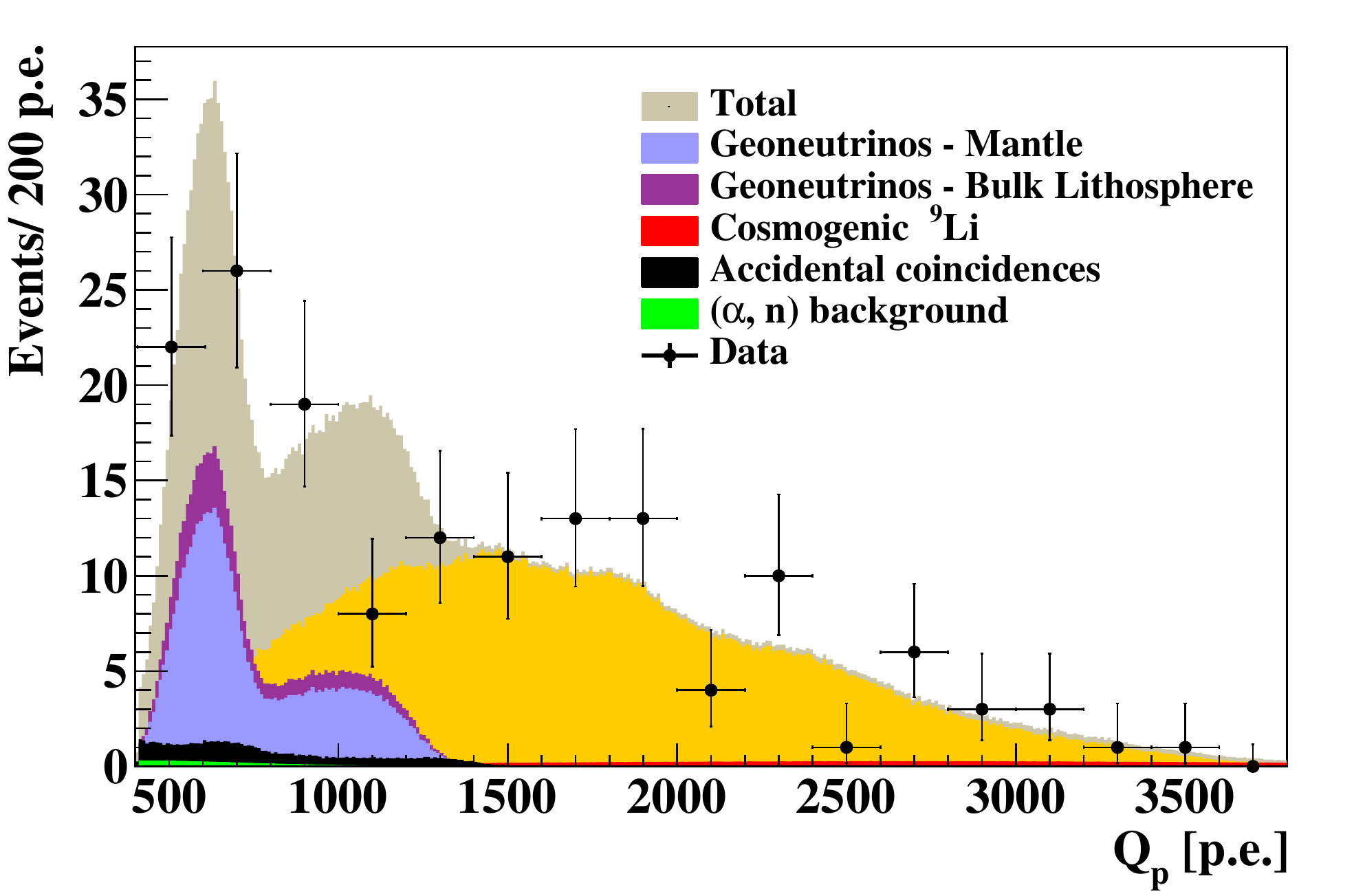} 
    \caption{Spectral fit of the charge spectrum of \emph{prompt} IBD candidates to extract the mantle signal after constraining the contribution of the bulk lithosphere. The grey shaded area shows the summed PDFs of all the signal and background components. From~\cite{Agostini:2019dbs}.}
      \label{fig:fit_mantle}
  \end{figure}

The mantle signal was extracted from the spectral fit after constraining the contribution from the bulk lithosphere to be 28.8$^{+5.5}_{-4.6}$\,events corresponding to the expected signal of 25.9$^{+4.9}_{-4.1}$~TNU mentioned in Section~\ref{subsec:exp-signal}. The corresponding MC PDF was constructed from the PDFs of $^{232}$Th and $^{238}$U geoneutrinos. They were scaled with the lithospheric Th/U signal ratio equal to 0.29, that is based on geological observations. The MC PDF used for the mantle was also constructed from the $^{232}$Th and $^{238}$U PDFs, but the applied Th/U signal ratio was 0.26. This procedure maintains the chondritic Th/U mass ratio of 3.9 for the bulk Earth and infers this ratio in the mantle to be 3.7 (Section~\ref{subsec:exp-signal}). The mantle signal, as well as the reactor antineutrino contribution were kept free in the fit. The data points and the different PDFs of the fit are shown in Figure~\ref{fig:fit_mantle}. After considering the systematic uncertainties, the final mantle signal obtained was $21.2 ^{+9.5}_{-9.0}\,({\rm stat}) ^{+1.1}_{-0.9}\,({\rm sys})$\,TNU. The statistical significance of the mantle signal was studied using MC pseudo-experiments with and without a generated mantle signal. This way, the null-hypothesis of the mantle signal has been rejected with 99.0\%\,C.L. (corresponding to 2.3$\sigma$ significance).

\subsubsection{Radiogenic heat and convective Urey ratio}
\label{subsec:results-heat-urey}
\begin{figure}[t]
 \centering
\includegraphics[width = 0.6\textwidth]{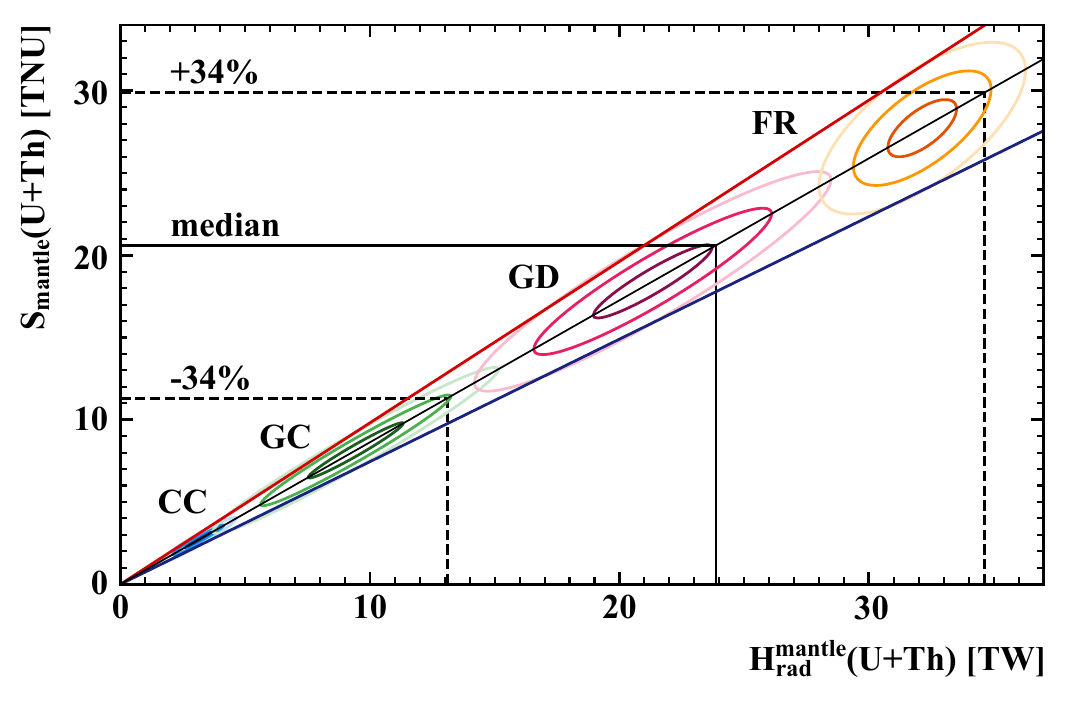}
\caption{Mantle geoneutrino signal expected in Borexino as a function of $^{238}$U and $^{232}$Th mantle radiogenic heat: the area between the red and blue lines denotes the full range allowed between a homogeneous mantle (high scenario) and a unique rich layer just above the CMB (low scenario). The slope of the central inclined black line ($\beta_{\mathrm{centr}}$ = 0.86\,TNU/TW) is the average of the slopes of the blue and red lines. The blue, green, red, and yellow ellipses are calculated using the $^{238}$U and $^{232}$Th mantle radiogenic power (with 1$\sigma$ error) according to different BSE models: Cosmochemical (CC) model, Geochemical (GC) model, Geodynamical (GD) model, and Fully radiogenic (FR) model. For each model darker to lighter shades of respective colours represent 1, 2, and 3$\sigma$ contours. The black horizontal lines represent the mantle signal measured by Borexino: the median mantle signal (solid line) and the 68\% coverage interval (dashed lines). From~\cite{Agostini:2019dbs}.}
\label{fig:SUTh_vs_Heat} 
\end{figure}

The mantle geoneutrino signal $S_{\mathrm{mantle}}$(U+Th) is linearly proportional to the mantle radiogenic power $H_{\mathrm{rad}}^{\mathrm {mantle}}$(U+Th), that in turn scales with the $^{238}$U and $^{232}$Th masses in the mantle $M_{\mathrm{mantle}}$(U) and $M_{\mathrm{mantle}}$(Th), respectively. This relationship is expressed by the following equation:
\begin{eqnarray}
\label{eq:S_vs_H}
\nonumber
S_{\mathrm {mantle}} (\mathrm {U+Th}) &=& 
\beta \cdot H_{\mathrm{rad}}^{\mathrm{mantle}}(\mathrm{U + Th})  \\
&=& \beta \cdot \left [ h(\mathrm{U})+3.7 \cdot h(\mathrm{Th}) \right ] \cdot M_{\mathrm{mantle}}(\mathrm{U}),
\end{eqnarray}
where $h$(U) = 98.5\,$\mu$W/kg and $h$(Th) = 26.3\,$\mu$W/kg~\cite{Dye:2011mc} are the $^{238}$U and $^{232}$Th specific heats. The value 3.7 is the assumed Th/U mass ratio in the mantle as discussed in the previous subsection. The scaling factor $\beta$ depends only on $^{238}$U and $^{232}$Th distributions in the mantle. For Borexino, a central value of $\beta_{\mathrm{centr}} = 0.86$\,TNU/TW represents the average between the assumptions of a homogeneous mantle ($\beta_{\mathrm{high}} = 0.98$\,TNU/TW) and of the HPE-rich layer just above the CMB ($\beta_{\mathrm{low}} = 0.75$\,TNU/TW). This relation is shown in Figure~\ref{fig:SUTh_vs_Heat} for the range of mantle radiogenic power predicted by various groups of BSE models discussed in Section~\ref{subsec:bse}. Thus, the area between the two extreme lines denotes the region allowed by all possible $^{238}$U and $^{232}$Th distributions in the mantle, assuming that the abundances in this reservoir are radial and non-decreasing functions of the depth. The solid black horizontal line represents the median of the Borexino measurement and falls within the prediction of the Geodynamical model (GD). The 68\% C.I. is shown by horizontal black dashed lines, and covers the area of prediction of the GD and the Fully Radiogenic (FR) model. Borexino is least compatible with the Cosmochemical model (CC), whose central value disagrees with the measurement at 2.4$\sigma$ level. Adopting the $\beta_{\mathrm{centr}}$ in Equation~\eqref{eq:S_vs_H}, the Borexino measured mantle signal of  $S_{\mathrm{mantle}}$(U+Th) = $21.2 ^{+9.5}_{-9.0}\,({\rm stat}) ^{+1.1}_{-0.9}\,({\rm sys})$\,TNU from Section~\ref{subsec:mantle} corresponds to the radiogenic heat from the mantle $H^{\mathrm{mantle}}_{\mathrm{rad}}$(U+Th) of $24.6 ^{+11.1}_{-10.4}$\,TW (68\% C.I.). Summing the radiogenic power in the lithosphere $H^{\mathrm{LSp}}_{\mathrm{rad}}$(U+Th) = 6.9$^{+1.6}_{-1.2}$\,TW, the Earth's radiogenic power from $^{238}$U and $^{232}$Th is $H_{\mathrm{rad}}$(U+Th) = 31.7$^{+14.4}_{-9.2}$\,TW. The likelihood profiles of the components are summed for this calculation and the median value is taken as the central value.

\begin{figure}[t]
\centering
\subfigure[]{\includegraphics[width = 0.45\textwidth]{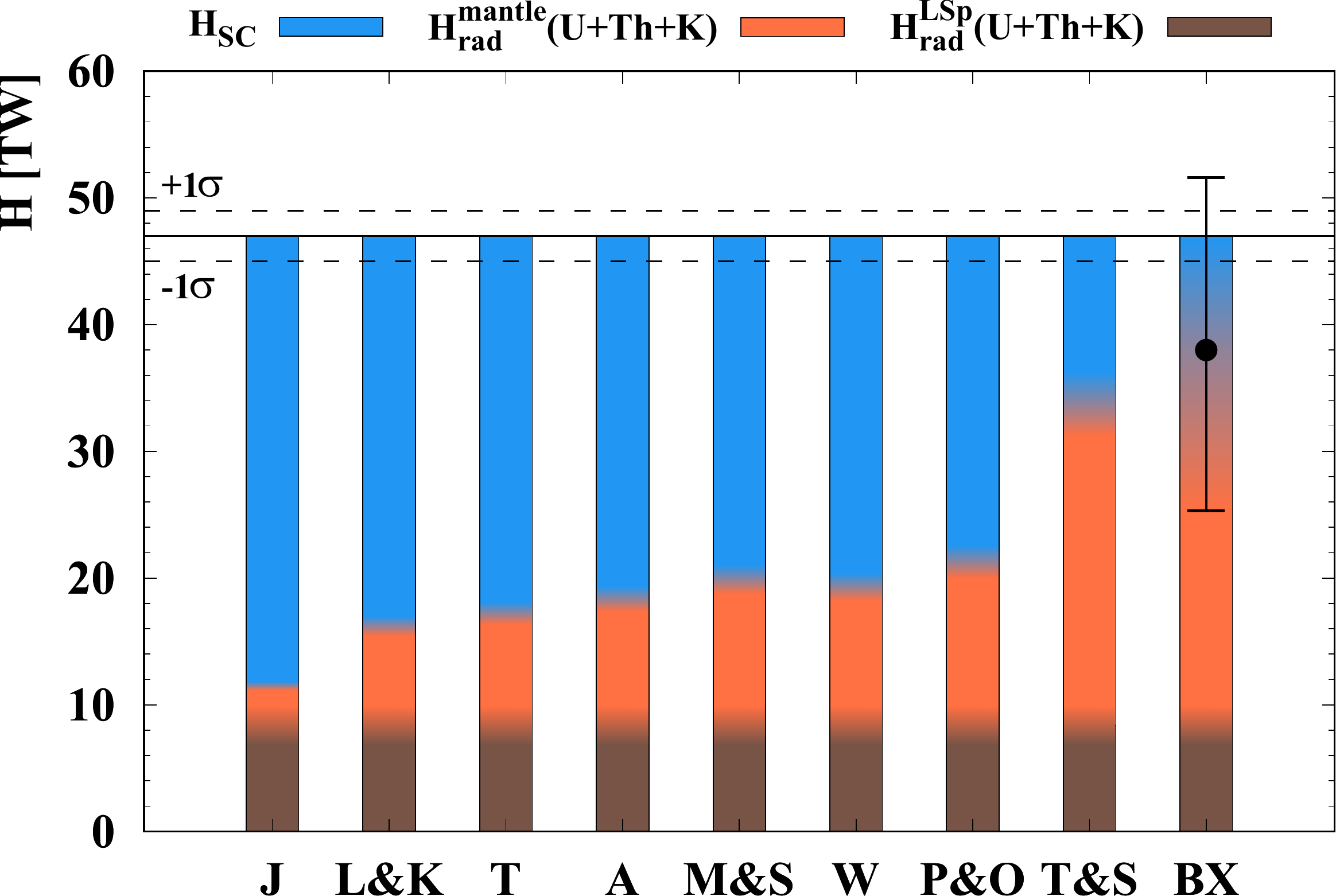}\label{fig:H_vs_BSE}}
\subfigure[]{\includegraphics[width = 0.46\textwidth]{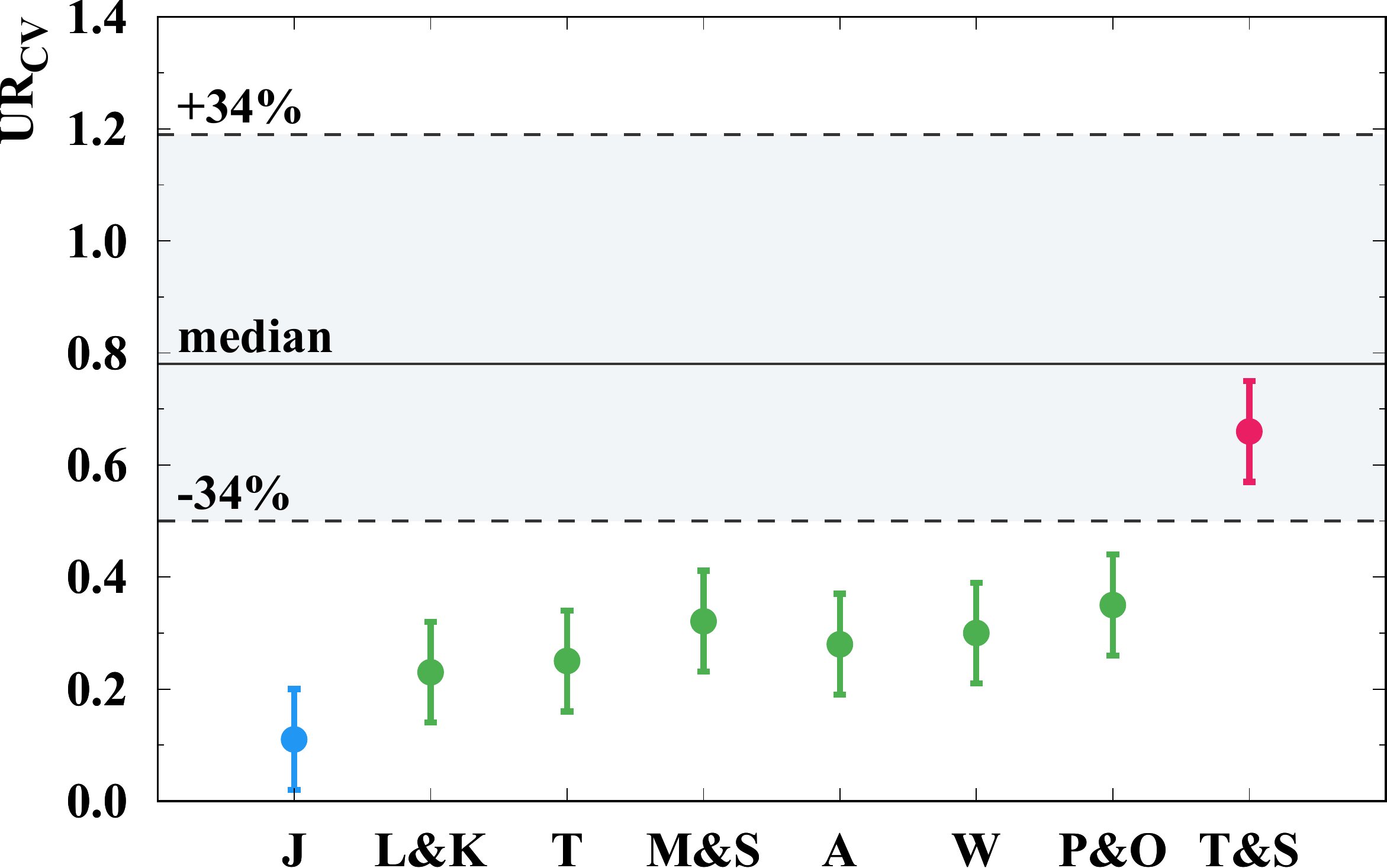}\label{fig:UR_vs_BSE}}
\caption{(a) Decomposition of the Earth's total surface heat flux $H_{\mathrm{tot}}$ = $(47\pm2)$\,TW (horizontal black lines) into its three major contributions: lithospheric radiogenic heat $H_{\mathrm{rad}}^{\mathrm{LSp}}$ (brown), mantle radiogenic heat $H_{\mathrm{rad}}^{\mathrm{mantle}}$  (orange), and secular cooling $H_{\mathrm{SC}}$ (blue). The labels on the $x$ axis identify different BSE models, while the last bar labeled BX represents the Borexino measurement. The lithospheric contribution is the same for all bars. (b) Comparison of Borexino constraints (horizontal band) with predictions of the BSE models (points with $\pm$3$\sigma$ error bars) for the convective Urey ratio $UR_{\mathrm{CV}}$ (Equation~\ref{eq:URCV}), assuming the total heat flux $H_{\mathrm{tot}}$ = $(47\pm 2)$\,TW. The blue, green, and red colours represent different BSE models (CC, GC, and GD, respectively). From~\cite{Agostini:2019dbs}.}
\end{figure}

In order to compare the Borexino estimate of the radiogenic power with the Earth's total surface heat flux of $H_{\mathrm{tot}}$ = $(47\pm2)$\,TW~\cite{RN593}, the contribution from $^{\mathrm{40}}$K must be considered. Firstly, assuming the contribution from $^{\mathrm{40}}$K to be 18\% of the total mantle radiogenic heat (Section~\ref{subsec:earth}), the total radiogenic mantle signal can be expressed as $H^{\mathrm {mantle}}_{\mathrm {rad}}(\mathrm{U+Th+K})$ = 30.0$^{+13.5}_{-12.7}$\,TW. Through the further addition of the lithospheric contribution $H_{\mathrm{rad}}^{\mathrm{LSp}}$(U+Th+K) = 8.1$^{+1.9}_{-1.4}$\,TW, the 68\% C.I. for the Earth's radiogenice heat can be obtained as $H_{\mathrm {rad}}$(U+Th+K) = 38.2$^{+13.6}_{-12.7}$\,TW. Figure~\ref{fig:H_vs_BSE} compares the decomposition of the Earth's total surface heat flux into radiogenic heat and the heat coming from secular cooling $H_{\mathrm{SC}}$, when performed for various BSE models and the Borexino measurement. A preference is found for models with relatively high radiogenic power, which indicates a cool initial environment at early formation stages of Earth and assumes the fraction of heat coming from secular cooling $H_{\mathrm{SC}}$ to be small. However, no model can be excluded at 3$\sigma$ level.

The total radiogenic heat estimated by Borexino can be used to extract the convective Urey ratio, according to Equation~\eqref{eq:URCV} in Section~\ref{subsec:earth}. The resulting value of $UR_{\mathrm{CV}}$ = 0.78$^{+0.41}_{-0.28}$ is compared to the $UR_{\mathrm{CV}}$ predicted by different BSE models in Figure~\ref{fig:UR_vs_BSE}. The Borexino geoneutrino measurement constrains the mantle radiogenic heat power at a 90(95)\% C.L. to be $H_{\mathrm{rad}}^{\mathrm{mantle}}$(U+Th) $>$ 10(7)\,TW and $H_{\mathrm{rad}}^{\mathrm{mantle}}$(U+Th+K) $>$ 12.2(8.6)\,TW and the convective Urey ratio to be $UR_{CV}$ $>$ 0.13(0.04).

\subsubsection{Limits on a hypothetical georeactor}
\label{subsec:georea}

Limits on a hypothetical georeactor, mentioned in Section~\ref{subsec:exp-antinu-bckg}, can be obtained by constraining the number of expected reactor antineutrino events in the spectral fit, since both these components have similar spectral shapes. The georeactor PDFs for the fit were generated for three different positions. However, their shapes were practically identical and Borexino does not have any sensitivity to distinguish them. The geoneutrino (Th/U fixed to chondritic mass ratio of 3.9) and georeactor contributions were left free in the fit. This resulted in an upper limit of 18.7\,TNU, at 95\% C.L., on a georeactor signal. Considering the predicted georeactor signal expressed in TNU, for a 1\,TW georeactor in different locations~\cite{Agostini:2019dbs}, Borexino has excluded the existence of a georeactor with a power greater than 0.5/2.4/5.7 TW at 95\% C.L., assuming its distance from the detector to be 2900\,km (CMB below LNGS) /6371\,km (Earth's centre) /9842\,km (CMB on opposite hemisphere), respectively.

\section{Conclusions and outlook}
\label{sec:conclusions}

In this paper the latest Borexino measurements on neutrinos from the Sun and Earth were discussed, from highlighting the key elements of the analyses up to the discussion and interpretation of the results. The success of Borexino in the measurement of low-energy neutrinos is primarily based on the extreme radio-purity of the liquid scintillator.

Borexino measures solar neutrinos via neutrino–electron elastic scattering. The direct result of the solar neutrino analysis are the interaction rates (given for zero-threshold) for neutrinos produced by different nuclear reactions. Borexino has performed a complete spectroscopy of the \emph{pp} chain solar neutrinos~\cite{PPchainNature}. By performing a multivariate spectral fit in the Low Energy Region, the rates of neutrinos from initial proton–proton fusion \emph{pp}($^{+8.7}_{-10.6}$\%), the three-body proton–electron–proton fusion \emph{pep}($^{+16.0}_{-17.4}$\% for HZ-SSM and $^{+14.7}_{-16.3}$\% for LZ-SSM), and the electron-capture decay of $^7$Be($^{+2.4}_{-2.7}$\%) were extracted. The numbers in parentheses indicate the total experimental error summing quadratically the statistical and systematic uncertainties. The constraint on the CNO rate  used in the fit, based on the HZ-SSM and LZ-SSM predictions, influences only the \emph{pep} neutrinos. In both cases however, the absence of the \emph{pep} reaction in the Sun was rejected with $>5\sigma$ significance. The rate of scattered electrons above 3\,MeV due to $^8$B($^{+7.1}_{-7.7}$\%) solar neutrinos interactions was extracted  via a radial fit without any assumption on the form of the energy spectrum and thus, oscillation parameters. Borexino is the only experiment that measured all (with the exception of \emph{hep} neutrinos) neutrinos from the \emph{pp} chain. While the Low Energy Region results are the most precise measurements existing in the world, the $^8$B result is less precise than the measurements of large volume water Cherenkov detectors, but is compatible with them. Borexino measurements provide a direct determination of the relative intensity of the two primary terminations of the \emph{pp} chain (\emph{pp}-I and \emph{pp}-II) and, assuming standard three-neutrino oscillations, an indication that the temperature profile in the Sun is more compatible with SSM models that assume high surface metallicity. Assuming the SSM prediction for the solar fluxes to hold, the survival probability of solar electron neutrinos can be determined from the measured rates: by comparing its values at different energies, Borexino probes simultaneously the neutrino flavour-conversion paradigm, both in vacuum and in matter-dominated regimes. The vacuum-only hypothesis is disfavoured at 98.2\% C.L. Borexino also confirmed the solar origin of the measured signal assigned to $^7$Be solar neutrinos, by observing the expected seasonal variation of the respective rate~\cite{Be7mod}, induced by the Earth elliptical orbit around the Sun. Measurement of solar neutrinos also help in constraining the non-standard neutrino interactions~\cite{Bx-NSI20} and placing a stringent limit on the effective neutrino magnetic moment~\cite{Bx-NMM17}.

Borexino provided the first experimental evidence at 5$\sigma$ significance of neutrinos produced in the CNO fusion cycle in the Sun~\cite{CNOpap}. This was achieved by a multivariate spectral fit performed with a constraint on the rates of \emph{pep} solar neutrinos and $^{210}$Bi internal background. The \emph{pep} rate was constrained with 1.4\% precision based, on theoretical expectations and a global fit to all solar data, excluding the Borexino dataset used in this analysis. An upper limit constraint was placed on $^{210}$Bi rate, obtained via a fit of the $^{210}$Po distribution in the Low Polonium Field. This procedure was made possible, thanks to the thermal stabilisation of the Borexino detector during the Phase III, that minimised the convective currents bringing $^{210}$Po from the nylon vessel holding the scintillator to the fiducial volume of the analysis. Thus, in the region of the scintillator free from convection, $^{210}$Po rate approaches the rate of its parent $^{210}$Bi nuclei. In the CNO cycle, the fusion of Hydrogen is catalysed by Carbon, Nitrogen, and Oxygen, and so its rate as well as the flux of emitted CNO neutrinos depend directly on the abundance of these elements in the solar core. Borexino result is compatible with both HZ-SSM and LZ-SSM predictions, but paves the way towards a direct measurement of the solar metallicity using CNO neutrinos. In addition, Borexino measurements quantify the relative contribution of CNO fusion in the Sun to be of the order of 1\%. In massive stars, however, CNO is the dominant process of energy production and thus, the primary mechanism for the stellar conversion of Hydrogen into Helium in the Universe.

Geoneutrinos, antineutrinos from the decays of long-lived radioactive elements inside the Earth, can be exploited as a new and unique tool to study our planet. Only two experiments in the world, KamLAND and Borexino, have detected geoneutrinos so far. Geoneutrinos are detected via the Inverse Beta Decay (IBD) on protons. The 1.8\,MeV kinematic threshold of this interaction allows to measure the high-energy part of geoneutrinos emitted along $^{238}$U and $^{232}$Th chains, while it leaves $^{40}$K geoneutrinos completely unreachable for the present-day technology. Inverse Beta Decay has about two orders of magnitude higher cross section than the elastic scattering, a huge advantage for the detection of geoneutrinos with the flux about four orders of magnitude lower than the \emph{pp} solar neutrino flux. In addition, it provides a unique background-suppressing topology of the fast space-time coincidences between a prompt and a delayed signal. Borexino has presented its latest geoneutrino measurement in 2019~\cite{Agostini:2019dbs}. The geoneutrino signal was extracted through a spectral fit of the prompt events, related to the energy of incident antineutrinos. Geoneutrino and reactor antineutrino contributions were kept as free fit parameters, while non-antineutrino backgrounds were constrained to the values estimated independently. Thanks to both more acquired data and improved analysis techniques in an enlarged fiducial volume, the updated measurement reached $^{+18.3}_{-17.2}$\% total precision, assuming the same Th/U mass ratio as found in chondritic CI meteorites. Antineutrino background from reactors was fit unconstrained and found compatible with the expectations. The null-hypothesis of observing a geoneutrino signal from the Earth's mantle was excluded at a 99.0\% C.L., when exploiting detailed knowledge of the local crust near the experimental site. The measured mantle signal was then converted to mantle radiogenic heat from decays of Uranium and Thorium, assuming a range of geological models describing the distribution of these elements in the mantle: from a homogeneous distribution up to an assumption of an enriched layer at the core-mantle boundary. The measured mantle signal is compatible with different geological predictions, however there is a $\sim$2.4$\sigma$ tension with those Earth models, which predict the lowest concentration of heat-producing elements in the mantle. In addition, by constraining the number of expected reactor antineutrino events, the existence of a hypothetical georeactor at the centre of the Earth having power greater than 2.4\,TW was excluded at 95\% C.L.

To summarize, this paper reviewed the latest Borexino measurements of solar and geo neutrinos\textemdash results that certainly mark the history of neutrino physics. The Borexino collaboration is analysing the latest data taken with the detector, featuring exceptional radio-purity and thermal stability, promising conditions for an improved CNO solar neutrino measurement. In spite of that, Borexino is expected to stop data taking within 2021. 

\section*{Funding}This work is funded by the recruitment initiative of the Helmholtz Association of German Research Centres (HGF).

\section*{Data availability}Data used to obtain presented results are provided at \url{https://bxopen.lngs.infn.it/}, shared and owned by the Borexino collaboration.

\section*{Acknowledgments}We would like to acknowledge the whole Borexino collaboration for the team work that lead to the results discussed in this paper. All of them were published for the first time elsewhere in original publications, that are referenced in this paper. It is an honour of the authors to be a part of the Borexino collaboration.

\bibliography{main.bib}

\end{document}